\documentclass[11pt]{article}
\usepackage{bbm}

\usepackage{mathrsfs}
\usepackage{amsfonts}
\usepackage{amssymb}
\usepackage{amsfonts,amssymb,mathrsfs,amsmath,amssymb,theorem,float,xypic}
\usepackage{color}

\newtheorem{theorem}{Theorem}[section]
\newtheorem{prop}[theorem]{Proposition}
\newtheorem{cor}[theorem]{Corollary}

\newtheorem{thm}[theorem]{Theorem}
\newtheorem{defn}[theorem]{Definition}
\newtheorem{lem}[theorem]{Lemma}
\newtheorem{exmp}[theorem]{Example}
\newtheorem{rem}[theorem]{Remark}

\floatstyle{ruled}
\newfloat{algorithm}{t}{loa}
\floatname{algorithm}{Algorithm}
\newcommand{\Inp}[1]
  {\noindent\begin{tabular}{@{}p{1.5cm}@{}p{12.0cm}@{}}
   {\bf Input: }&#1 \end{tabular}}
\newcommand{\Outp}[1]
  {\noindent\begin{tabular}{@{}p{1.5cm}@{}p{12.0cm}@{}}
   {\bf Output: }&#1 \end{tabular}}

\def\imply{\Rightarrow}
\def\qed{{\vrule height5pt width2pt depth2pt}}

\def\and{\cap}

\def\bref#1{(\ref{#1})}
\def\proof{{\noindent\em Proof:} }

\newcommand{\SPC}{\hspace*{15pt}}
\newcommand{\qedd}{\hspace*{\fill}$\Box$\medskip}
\newcommand{\bm}[1]{\mbox{\boldmath{$#1$}}}

\floatstyle{ruled}
\newfloat{algorithm}{t}{loa}
\floatname{algorithm}{Algorithm}

\def\st#1{{\em{#1}}}

\def\Y{{\mathbb{Y}}}

\def\V{{\mathbb{V}}}
\def\L{{\mathbb{L}}}
\def\J{{\mathcal {J}}}

\def\FB{{\mathbb{F}}}
\def\RB{{\mathbb{R}}}

\def\AH{{\mathbb{A}}}
\def\IH{{\mathbb{I}}}

\def\A{{\mathcal A}}
\def\B{{\mathcal B}}
\def\C{{\mathcal C}}
\def\D{{\mathcal D}}

\def\MI{{\mathcal{M}}}

\def\D{{\mathbb D}}
\def\G{{\mathbb G}}

\def\Q{{\mathbb Q}}

\def\bu{{\mathbf{u}}}

\def\beps{{\bm\epsilon}}

\def\prem{\hbox{\rm prem}}

\def\grem{\hbox{\rm grem}}

\def\sat{\hbox{\rm{sat}}}

\def\max{\hbox{\rm{max}}}

\def\span{\hbox{\rm{Span}}}
\def\lcm{\hbox{\rm{lcm}}}
\def\gcd{\hbox{\rm{gcd}}}

\def\gcd{\hbox{\rm{gcd}}}
\def\deg{\hbox{\rm{deg}}}

\def\init{\hbox{\rm{I}}}
\def\ord{\hbox{\rm{ord}}}

\def\lead{\hbox{\rm{ld}}}

\def\dim{\hbox{\rm{dim}}}

\def\lv{\hbox{\rm{lvar}}}
\def\mod{\hbox{\rm{mod}}}

\def\ord{\hbox{\rm{ord}}}
\def\rk{\hbox{\rm{rk}}}

\def\I{\mathcal{I}}

\def\LT{{\bf LT}}
\def\c{{\bf c}}
\def\b{{\bf b}}
\def\f{{\bf f}}
\def\g{{\bf g}}
\def\h{{\bf h}}
\def\e{{\bf e}}

\def\a{{\bf a}}

\def\Z{{\mathbb{Z}}}
\def\Q{{\mathbb{Q}}}
\def\N{{\mathbb{N}}}

\def\F{{\mathcal{F}}}

\def\ix{{\rm{i}}}

\def\grem{\hbox{\rm grem}}
\def\and{\cap}

\newcounter{bean}
\def\bl{\begin{list}{Step \arabic{bean}}{\usecounter{bean}}\labelwidth=34pt}
\def\el{\end{list}}

\def\deg{{\rm deg}}
\def\lvar{{\rm lvar}}

\def\init{{\rm I}}

\def\normalization1{{\rm normalization1}}
\def\normalization{{\rm normalization}}

\def\irrfactor1{{\rm irrfactor1}}
\def\irrfactor{{\rm irrfactor}}

\def\sat{{\rm sat}}

\def\gb{Gr\"obner基}

\def\grem{\hbox{\rm{grem}}}

\def\p{\mathfrak{p}}

\def\st{\hbox{ {\rm{s.t.}} }}

\def\Zx{\Z[x]}
\def\Zxn{\Z[x]^n}

\def\fb{{\mathbbm{f}}}
\def\gb{{\mathbbm{g}}}

\def\mb{{\mathbbm{m}}}

\def\lr#1{{\langle{#1}\rangle}}
\def\col{\hbox{\rm{Col}}}

\begin{document}
\title{Binomial Difference Ideals}

\author{Xiao-Shan Gao, Zhang Huang, Chun-Ming Yuan\\
KLMM,  Academy of Mathematics and Systems Science\\
 The Chinese Academy of Sciences, Beijing 100190, China}
\date{}

\maketitle

\begin{abstract}
In this paper, binomial difference ideals are studied.
Three canonical representations for Laurent binomial difference ideals
are given in terms of the reduced Gr\"obner basis of $\Zx$-lattices,
regular and coherent difference ascending  chains, and partial
characters over $\Zx$-lattices, respectively.
Criteria for a Laurent binomial difference ideal to be reflexive,
prime, well-mixed, and perfect are given in terms of their
support lattices.
The reflexive, well-mixed, and perfect closures of a Laurent binomial difference
ideal are shown to be binomial.
Most of the properties of Laurent binomial difference
ideals are extended to the case of difference binomial ideals.
Finally, algorithms are
given to check whether a given Laurent binomial difference ideal
$\I$ is reflexive, prime, well-mixed, or  perfect, and in the
negative case, to compute the reflexive, well-mixed, and perfect
closures of $\I$. An algorithm is given to decompose a finitely
generated perfect binomial difference ideal as the intersection of
reflexive prime binomial difference ideals.


\end{abstract}


\section{Introduction}
\label{sec-intro}

A polynomial ideal is called binomial if it is generated by
polynomials with at most two terms. Binomial ideals were first
studied by Eisendbud and Sturmfels \cite{es-bi},
which were further studied in \cite{Barile1,Dickenstein1,Mart1,Mayr1,Peeva1}
and were applied in
algebraic statistics \cite{pachter1},
chemical reactions \cite{Millan1},
and error-correcting codes \cite{Saleemi1}.

In this paper, we initiate the study of binomial difference ideals
and hope that they will play similar roles in
difference algebraic geometry to their algebraic counterparts.
Difference algebra and difference algebraic geometry were founded by
Ritt \cite{ritt-dd1} and Cohn \cite{cohn}, who aimed to study
algebraic difference equations in the way polynomial
equations were studied in commutative algebra and algebraic geometry \cite{cohn, Hru1, levin, wibmer}.
%

We now describe the main results of this paper.
In Section \ref{sec-zxm}, we prove basic properties of $\Z[x]$-lattices.
By a $\Z[x]$-lattice, we mean a $\Z[x]$-module in $\Z[x]^n$.
$\Z[x]$-lattices play the same role as $\Z$-lattices do in
the study of binomial ideals. Here, $x$ is used
to denote the difference operator $\sigma$. For instance,
$a^3\sigma(a)^2$ is denoted as $a^{2x+3}$.
%
%
Since $\Zx$ is not a PID, the Hermite normal form for a matrix with
entries in $\Zx$ does not exist. In this section, we introduce the
concept of generalized Hermite normal form and show that a matrix is
a generalized Hermite normal form if and only if its columns form a
reduced Gr\"obner basis for a $\Zx$-lattice.
%

In Section \ref{sec-lbi}, we give three canonical representations for Laurent
binomial difference ideals in terms of reduced Gr\"obner bases of $\Zx$-lattices, difference characteristic sets, and partial characters.
Gr\"obner bases play an important role in the study of binomial
ideals \cite{es-bi}. In general, a binomial difference ideal is not
finitely generated and does not have a finite Gr\"obner basis.
Instead, the theory of characteristic set for difference polynomial
systems \cite{gao-dcs} is used for similar purposes.
It is shown that any Laurent binomial difference ideal can be written as
$[\A]$, where $\A$ is a regular and coherent difference ascending chain consisting
of binomial difference polynomials.

Let $\I$ be a proper Laurent binomial difference ideal and
$L=\{\f\in\Zxn\,|\,\Y^\f-c_\f\in\I\}$ the {support lattice} of $\I$,
which is a $\Zx$-lattice.
In Section \ref{sec-lbicr}, we give criteria for a
Laurent binomial difference ideal to be prime, reflexive,
well-mixed, and perfect in terms of its support lattice.
The criterion for prime ideals is similar to the algebraic case, but
the criteria for reflexive, well-mixed, and perfect difference ideals
are unique to difference algebra and are first proposed in this paper.
%
%
Furthermore, it is shown that the reflexive,
well-mixed, and perfect closures of a Laurent binomial difference
ideal $I$ with support lattice $L$ are still binomial, whose support
lattices are the $x$-, $M$-, and the $P$-saturation
of $L$, respectively.
It is further shown that any perfect Laurent binomial difference
ideal $\I$ can be written as the intersection of Laurent reflexive
prime binomial difference ideals whose support lattices are the
$x$-$\Z$-saturation of the support lattice of $\I$.

In Section \ref{sec-bi}, binomial difference ideals are studied. It
is shown that a large portion of the properties for binomial ideals
proved in \cite{es-bi} can be easily extended to the difference case.
We also identify a class of normal binomial difference ideals which
are in a one to one correspondence with Laurent  binomial difference
ideals.
With the help of this correspondence, most properties proved for Laurent
binomial difference ideals are extended to the non-Laurent case.
%

In Section \ref{sec-alg}, algorithms are given to check whether a
$\Zx$-lattice $L$ is $\Z$-, $x$-, M-, or P-saturated, or
equivalently, whether a Laurent binomial difference ideal $\I$ is prime,
reflexive, well-mixed, or perfect. If the answer is negative, we can also
compute the $\Z$-, $x$-, M-, or {\rm{P}}-saturation of $L$
and the reflexive, well-mixed, or perfect closures of $\I$.
Based on these algorithms, we give an algorithm to decompose a
finitely generated perfect binomial difference
ideal as the intersection of reflexive prime binomial difference ideals.
This algorithm is stronger than the general decomposition algorithm
in that for general difference polynomials,
it is still open on how to decompose a finitely generated perfect difference
ideal as the intersection of reflexive prime difference ideals \cite{gao-dcs}.

A distinctive feature of the algorithms presented in this paper is
that problems about difference binomial polynomial ideals are
reduced to problems about $\Zx$-lattices which are pure algebraic
and have simpler structures.

\section{Preliminaries about difference algebra}
\label{sec-pre}
In this section, some basic notations about
difference algebra will be given. For more details about difference algebra,
please refer to \cite{cohn, gao-dcs,Hru1, levin, wibmer}.

\subsection{Difference polynomial and Laurent difference polynomial}

An ordinary difference field, or simply a $\sigma$-field, is a field
$\F$ with a third unitary operation $\sigma$ satisfying:  for any
$a, b\in\F$, $\sigma(a+b)=\sigma(a)+\sigma(b)$,
$\sigma(ab)=\sigma(a)\sigma(b)$, and $\sigma(a)=0$ if and only if
$a=0$.
We call $\sigma$ the {\em transforming operator} of $\F$. If
$a\in\F$,  $\sigma(a)$ is called the transform of $a$ and is denoted
by $a^{(1)}$. For $n\in\Z_{>0}$,
$\sigma^n(a)=\sigma^{n-1}(\sigma(a))$ is called the $n$-th transform
of $a$ and denoted by $a^{(n)}$,  with the usual assumption
$a^{(0)}=a$.
If $\sigma^{-1}(a)$ is defined for each $a\in\F$,  $\F$ is called
{\em inversive}. Every $\sigma$-field has an inversive closure
\cite{cohn}. A typical example of inversive $\sigma$-field is
$\Q(\lambda)$ with $\sigma(f(\lambda))=f(\lambda+1)$.

In this paper, $\F$ is assumed to be inversive and of characteristic
zero. Furthermore, we use $\sigma$- as the abbreviation for
difference or transformally.

We introduce the following useful notation.
Let $x$ be an algebraic indeterminate and $p=\sum_{i=0}^s c_i x^i
\in\Z[x]$. For $a$ in any $\sigma$-over field of $\F$, denote
 $$ a^p = \prod_{i=0}^s (\sigma^i a)^{c_i}.$$
For instance,  $a^{x^2-1} = a^{(2)}/a$. It is easy to check that for
$p, q\in\Z[x]$,  we have
 $$a^{p+q}=a^{p} a^{q}, a^{pq}= (a^{p})^{q}, (ab)^p=a^pb^p.$$
%
By $a^{[n]}$ we mean the set $\{a, a^{(1)}, \ldots, a^{(n)}\}$. If
$S$ is a set of elements, we denote $S^{[n]}=\cup_{a\in S} a^{[n]}$.

Let $S$ be a subset of a  $\sigma$-field $\mathcal{G}$ which
contains $\mathcal {F}$.   We will  denote respectively by $\mathcal
{F}[S]$,  $\mathcal {F}(S)$,  $\mathcal {F}\{S\}$,  and $\mathcal
{F}\langle S\rangle$  the smallest subring,  the smallest subfield,
the smallest $\sigma$-subring,  and the smallest $\sigma$-subfield
of $\mathcal{G}$ containing $\mathcal {F}$ and $S$.  If we denote
$\Theta(S)=\{\sigma^ka|k\geq0, a\in S\}$,  then we have $\mathcal
{F}\{S\}=\mathcal    {F}[\Theta(S)]$ and $\mathcal {F}\langle
S\rangle=\mathcal    {F}(\Theta(S))$.
%
%
%

Now suppose $\Y=\{y_{1},   \ldots,  y_{n}\}$ is a set of
$\sigma$-indeterminates over $\F$.   The elements of $\mathcal
{F}\{\Y\}=\mathcal {F}[y_j^{(k)}:j=1, \ldots, n;k\in \N]$ are called
{\em $\sigma$-polynomials} over $\F$ in $\Y$,  and $\mathcal
{F}\{\Y\}$ itself is called the {\em $\sigma$-polynomial ring } over
$\F$ in $\Y$. A {\em $\sigma$-polynomial ideal}, or simply a
$\sigma$-ideal, $\mathcal {I}$ in $\mathcal {F}\{\Y\}$ is an
ordinary algebraic ideal which is closed under transforming, i.e.
$\sigma(\mathcal {I})\subset\mathcal {I}$. If $\mathcal{I}$ also has
the property that $a^{(1)}\in\mathcal{I}$ implies that
$a\in\mathcal{I}$,  it is called a {\em reflexive $\sigma$-ideal}.
A prime $\sigma$-ideal is a $\sigma$-ideal which is prime as an
ordinary algebraic polynomial ideal. For convenience,  a prime
$\sigma$-ideal is assumed not to be the unit ideal in this paper.
A $\sigma$-ideal $\I$ is called {\em well-mixed} if $fg\in\I$
implies $fg^x\in\I$ for $f,g\in\F\{\Y\}$.
A $\sigma$-ideal $\I$ is called {\em perfect} if for any
$a\in\N[x]\setminus\{0\}$ and $p\in\F\{\Y\}$, $p^a\in\I$ implies
$p\in\I$.
If $S$ is a subset of  $\F\{\Y\}$,  we use $(S)$, $[S]$, $\langle
S\rangle$, and $\{S\}$ to denote the algebraic ideal, the
$\sigma$-ideal, the well-mixed $\sigma$-ideal, and the perfect
$\sigma$-ideal generated by $S$.

An $n$-tuple over $\F$ is an $n$-tuple of the form $\eta=(\eta_1,
\ldots, \eta_n)$ where the $\eta_i$ are selected from a
$\sigma$-overfield of $\F$. For a $\sigma$-polynomial
$f\in\F\{\Y\}$, $\eta$ is called a $\sigma$-zero of $f$ if when
substituting $y_i^{(j)}$ by $\eta_i^{(j)}$ in $f$, the result is
$0$.
%


%
For $\f=(f_1, \ldots, f_n)^\tau\in \Z[x]^{n}$,  we define $\Y^\f =
\prod_{i=1}^n y_i^{f_i}$. $\Y^\f$ is called a {\em Laurent
$\sigma$-monomial} in $\Y$ and $\f$ is called its {\em support}.
A nonzero vector $\f=(f_1, \ldots, f_n)^\tau\in\Z[x]^n$ is said to be {\em
normal} if  the leading coefficient of $f_s$ is positive,  where $s$
is the largest subscript such that $f_s\ne0$.

A {\em Laurent $\sigma$-polynomial} over $\F$ in $\Y$ is an
$\F$-linear combination of Laurent $\sigma$-monomials in $\Y$.
Clearly, the set of all Laurent $\sigma$-polynomials form a
commutative $\sigma$-ring under the obvious sum, product, and the
usual transforming operator $\sigma$, where all Laurent
$\sigma$-monomials are invertible. 
We denote the $\sigma$-ring of Laurent $\sigma$-polynomials with
coefficients in $\mathcal {F}$ by $\F\{\Y^{\pm}\}$.
%
%
Let $p$ be a Laurent $\sigma$-polynomial in $\F\{\Y^{\pm}\}$. An
$n$-tuple $(a_1,$ $\ldots,a_n)$ over $\F$ with each $a_i\neq0$ is
called a {\em nonzero $\sigma$-solution} of $p$  if
$p(a_1,\ldots,a_n)=0.$
%
%

\subsection{Characteristic set for a difference polynomial system}
\label{sec-cs}

Let $f$ be a $\sigma$-polynomial in $\F\{\Y\}$.  The order of $f$
w.r.t. $y_i$ is defined to be the greatest number $k$ such that
$y_{i}^{(k)}$ appears effectively in $f$,  denoted by $\ord(f,
y_{i})$. If $y_{i}$ does not appear in $f$,  then we set $\ord(f,
y_{i})=-\infty$. The {\em order} of $f$ is defined to be $\max_{i}\,
\ord(f, y_{i})$,  that is, $\ord(f)=\max_{i}\, \ord(f, y_{i})$.

The {\em elimination ranking} $\mathscr{R}$ on $\Theta
(\Y)=\{\sigma^ky_i|1\leq i\leq n, k\geq0\}$ is used in this paper:
$\sigma^{k} y_{i}>\sigma^{l} y_{j}$ if and only if $i>j$ or $i=j$
and $k>l$, which is a total order over $\Theta (\Y)$.     By
convention, $1<\theta y_{j}$ for all $\theta y_{j}\in \Theta (\Y)$.

Let $f$ be a $\sigma$-polynomial in  $\mathcal {F}\{\Y\}$.  The
greatest $y_j^{(k)}$ w.r.t.  $\mathscr{R}$ which  appears
effectively in $f$ is called the {\em leader} of $f$,     denoted by
$\lead(f)$ and correspondingly $y_j$ is called the {\em leading
variable }of $f$, denoted by $\lv(f)=y_j$.
%
%
The leading coefficient of $f$  as a univariate polynomial in
$\lead(f)$ is called the {\em initial} of $f$ and is denoted by
$\init_{f}$.

    Let $p$ and $q$ be two $\sigma$-polynomials in $\F\{\Y\}$.
    $q$ is said to be of higher rank than $p$ if
 $\lead(q)>\lead(p)$  or
 $\lead(q)=\lead(p)=y_j^{(k)}$ and $\deg(q, y_j^{(k)})>\deg(p,  y_j^{(k)})$.

Suppose $\lead(p)=y_j^{(k)}$.  $q$ is said to be {\em reduced}
w.r.t. $p$ if  $\deg(q, y_j^{(k+l)})<\deg(p, y_j^{(k)})$  for all
$l\in\N$.

 A finite sequence of nonzero $\sigma$-polynomials $\mathcal
{A}=A_1, \ldots, A_m$  is said to be a
    {\em difference ascending chain}, or simply a {\em $\sigma$-chain}, if
 $m=1$ and $A_1\neq0$ or
 $m>1$,  $A_j>A_i$ and $A_j$ is reduced
    w.r.t. $A_i$ for $1\leq i<j\leq m$.

A $\sigma$-chain $\mathcal{A}$ can be written as the following form
\begin{equation}\label{eq-asc}
\mathcal{A}: A_{11}, \ldots, A_{1k_1},\ldots, A_{p1}, \ldots, A_{pk_p}
\end{equation}
where $\lv(A_{ij})=y_{c_i}$ for $j=1, \ldots, k_i$ and $\ord(A_{ij},
y_{c_i})<\ord(A_{il}, y_{c_i})$ for $j<l$.
%
%
The following are two $\sigma$-chains
\begin{equation}\label{ex-L11}
 \begin{array}{llllll}
 \A_1& =& y_1^{x}-1, &y_1^2y_2^2-1, &y_2^{x}-1& \\
 \A_2& =& y_1^2-1,   &y_1^{x}-y_1, &y_2^2-1, &y_2^{x}+y_2\\
\end{array}
\end{equation}

Let $\mathcal {A}=A_{1},A_{2},\ldots,A_{t}$ be a $\sigma$-chain with
$\init_{i}$ as the initial of $A_{i}$, and $f$ any
$\sigma$-polynomial.
   Then there exists an
   algorithm, which reduces
   $f$ w.r.t. $\mathcal {A}$ to a  polynomial $r$ that is
   reduced w.r.t. $\mathcal {A}$ and satisfies the relation
   \begin{equation}\label{eq-prem}
   \prod_{i=1}^t \init_{i} ^{e_{i}} \cdot f \equiv
   r, \mod \, [\mathcal {A}],\end{equation}
   where the $e_{i}\in\N[x]$ and
$r=\prem(f,\A)$ is called the {\em
$\sigma$-remainder} of $f$ w.r.t. $\A$~\cite{gao-dcs}.

A $\sigma$-chain $\mathcal {C}$ contained in a $\sigma$-polynomial
set $\mathcal {S}$ is said to be a {\em characteristic set} of
$\mathcal {S}$, if  $\mathcal {S}$ does not contain any nonzero
element reduced w.r.t. $\mathcal {C}$. Any $\sigma$-polynomial
set has a characteristic set.
A characteristic set
$\mathcal{C}$ of a $\sigma$-ideal $\mathcal {J}$ reduces to zero all
elements of $\mathcal {J}$.

Let $\A:A_1,\ldots,A_t$ be a $\sigma$-chain, $I_i =\init(A_i)$,
$y_{l_i}^{(o_i)} = \lead(A_i)$.
$\A$ is called {\em regular} if for any $j\in \N$, $I_i^{x^j}$ is
invertible w.r.t $\A$ \cite{gao-dcs} in the sense that
$[A_1,\ldots,A_{i-1},I_i^{x^j}]$ contains a nonzero
$\sigma$-polynomial involving no $y_{l_i}^{(o_i+k)},k=0,1,\ldots$.
%
%
%
To introduce the concept of coherent $\sigma$-chain, we need to
define the {\em $\Delta$-polynomial} first. If $A_i$ and $A_j$ have
distinct leading variables, we define $\Delta(A_i,A_j)=0$. If $A_i$
and $A_j$ ($i<j$) have the same leading variable $y_l$, then
$o_i=\ord(A_i,y_l) < o_j=\ord(A_j,y_l)$.
Define
 \begin{equation}\label{eq-delta}
 \Delta(A_i,A_j) =\prem((A_i)^{x^{o_j-o_i}},A_j).\end{equation}
 Then $\A$ is called {\em
coherent} if $\prem(\Delta(A_i,A_j),\A)=0$ for all $i< j$
\cite{gao-dcs}.

Let $\mathcal {A}$ be a $\sigma$-chain. Denote $\mathbb{I}_{\mathcal
{A}}$ to be  the minimal multiplicative set containing the initials
of elements of $\mathcal{A}$ and their transforms.    The {\em
saturation ideal} of $\A$ is defined to be
 $$\sat(\A)=[\mathcal  {A}]:\mathbb{I}_{\mathcal
 {A}} = \{p\in\F\{\Y\}: \exists h\in \mathbb{I}_{\mathcal {A}},  \, {\text s. t.}\,  hp\in[A]\}.$$
%
%
 The following result is needed in this paper.
\begin{thm}\cite[Theorem 3.3]{gao-dcs}\label{th-rp}
A $\sigma$-chain $\A$ is a characteristic set of $\sat(A)$ if and
only if $\A$ is regular and coherent.
\end{thm}
%
%
%
%

\section{$\Zx$-lattice}
\label{sec-zxm}
In this section, we prove basic properties of $\Z[x]$-lattices,
which will play the role of lattices in the study of binomial ideals.
%

For brevity, a $\Z[x]$-module in $\Z[x]^n$ is called a {\em $\Z[x]$-lattice}.
Since $\Z[x]$ is a Noetherian ring,  any $\Zx$-lattice $L$ has a finite set of
generators $\fb=\{\f_1,\ldots,\f_s\}\subset\Z[x]^n$:
 $$L = \span_{\Zx}\{\f_1,\ldots,\f_s\} \triangleq (\f_1,\ldots,\f_s).$$
A {\em matrix representation} of $\fb$ or $L$ is
 $$M = [\f_1,\ldots,\f_s]_{n\times s},$$
with $\f_i$ to be the $i$-th column of $M$. We also denote $L=(M)$.
The {\em rank} of a $\Zx$-lattice $L$ is defined to be the rank of
any matrix representation of  $L$, which is clearly well defined.

We list some basic concepts and properties of
Gr\"{o}bner bases of modules. For details, please refer to
\cite{cox-1998}.

Denote ${\beps}_i$ to be the $i$-th standard basis vector
$(0,\ldots,0,1,0,\ldots,0)^\tau\in\Z[x]^n$, where $1$ lies in the
$i$-th row of ${\beps}_i$. A {\em monomial} ${\bf m}$ in $\Z[x]^n$
is an element of the form $ax^k{\beps}_i\in\Z[x]^n$, where $a\in\Z$
and $k\in\N$.
The following {\em monomial order} $>$ of $\Z[x]^n$ will be used in
this paper: $ax^\alpha{\beps}_i > bx^\beta{\beps}_j$ if $i
> j$, or $i=j$ and $\alpha > \beta$, or $i=j$, $\alpha = \beta$, and
$|a|>|b|$.

With the above order, any $\f\in\Z[x]^n$ can be written in a unique
way as a  linear combination of monomials,
$ \f = \sum_{i=1}^l \h_i,$
where ${\h_i}\ne 0$ and $\h_1 > \h_2>\cdots > \h_l$. The {\em
leading term} of $\f$  is defined to be $
\LT(\f) = \h_1.$
For any $\G\subset \Zxn$, we denote by $\LT(\G)$ the  set of leading
terms of $\G$.

The order $>$ can be extended to elements of $\Zxn$ as follows: for
$\f,\g\in\Zxn$, $\f <\g$ if and only if $\LT(\f) < \LT(\g)$.

Let $\G\subset\Z[x]^n$ and  $\f\in\Z[x]^n$. We say that $\f$ is {\em
G-reduced} with respect to $\G$ if any monomial of $\f$ is not a
multiple of $\LT(\g)$ by an element in $\Zx$ for any $\g\in\G$.

\begin{defn}
A finite set $\fb = \{\f_1,\ldots,\f_s \}\subset\Z[x]^n$ is called a
{\em Gr\"{o}bner basis} for the $\Zx$-lattice $L$ generated by $\fb$
if for any $\g\in L$, there exists an $i$, such that $\LT(\g) | \LT(\f_i)$.
A Gr\"{o}bner basis $\fb$ is called {\em reduced} if for any
$\f\in\fb$, $\f$ is G-reduced with respect to $\fb\setminus\{\f\}$.
In this paper, it is always assumed that $\f_1<\f_2<\cdots<\f_s$.
\end{defn}

Let $\fb$ be a Gr\"{o}bner basis. Then any $\f\in\Zxn$ can be
reduced to a unique normal form by $\fb$, denoted by
$\grem(\f,\fb)$, which is G-reduced with respect to $\fb$.

\begin{defn}\label{def-sv}
Let ${\bf f,g}\in \Zx^n,~{\bf LT(f)}=ax^k{\bf e}_i,~{\bf LT(g)}=bx^s{\bf e}_j$, $s\leq k$. Then the S-polynomial of ${\bf f}$ and ${\bf g}$ is defined as follows: if $i \neq j$ then $S({\bf f,g})=0$; otherwise $S({\bf f,g})=$
\begin{equation}
\left\{
  \begin{array}{ll}
    {\bf f}-\frac{a}{b}x^{k-s}{\bf g}, & \hbox{ if } b\,|\,a; \\
    \frac{b}{a}{\bf f}-x^{k-s}{\bf g}, & \hbox{ if } a\,|\,b; \\
    u{\bf f}+vx^{k-s}{\bf g}, & \hbox{if $a\nmid b\hbox{ and }~b\nmid a,~\emph{where}~ \gcd(a,b)=ua+vb$.}
  \end{array}
\right.
\end{equation}
\end{defn}
The following basic property for Gr\"{o}bner basis is obviously true
for $\Zx$-lattices and a polynomial-time algorithm to compute G\"obner bases for
 $\Zx$-lattices is given in \cite{GHNF-alg}.
\begin{thm}[Buchberger's Criterion]\label{th-buch}
The following
statements are equivalent.
\begin{description}
\item[1)] $\fb = \{\f_1,\ldots,\f_s \}\subset\Z[x]^n$  is a Gr\"{o}bner basis.

\item[2)] $\grem(S(\f_i,\f_j),G)=0$  for all $i,j$.

\item[3)] $\f\in(\fb)$ if and only if $\grem(\f,\fb)=0$.
\end{description}

\end{thm}

We will study the structure of a Gr\"obner basis for a $\Zx$-lattice
by introducing the concept of generalized Hermite normal form.
First, we consider the case of $n=1$.

\begin{lem}\label{lm-21}
Let  $B = \{b_1,\ldots,b_k\}$ be a reduced Gr\"{o}bner basis of a
$\Zx$-module in $\Z[x]$, $b_1<\cdots<b_k$, and
$\LT(b_i)=c_ix^{d_i}\in\N[x]$. Then

1) $0\le d_1 < d_2 < \cdots<d_k$.

2) $c_k | \cdots | c_2 | c_1 $ and $c_i \ne c_{i+1}$ for $1\le i\le
k-1$.

3) $\frac{c_i}{c_k} | b_i$ for $1\le i< k$. If
$\widetilde{b}_1$ is the primitive part of $b_1$, then
$\widetilde{b}_1 | b_i$ for $1< i\le k$.
%

4) The S-polynomial $S(b_i,b_j)$ can be reduced to zero by $B$ for
any $i,j$.
\end{lem}
\proof  1) and 4) are consequences of Theorem~\ref{th-buch}. To
prove 2), assume that there exists an $l$ such that $c_{l-1} |
\cdots | c_2 | c_1 $ but $c_l \not| c_{l-1}$. Let $r =
\gcd(c_l,c_{l-1}) = p_1 c_l+p_2c_{l-1}$, where $p_1,p_2\in\Z$.
Then $|r| < |c_{l-1}|$ and $|r| < |c_{l}|$. Since $c_{l-1} | \cdots
| c_2 | c_1 $, we have $|r| < |c_{i}|, i=1,\ldots,l$. Let $g =
p_1b_l + p_2x^{d_l-d_{l-1}}b_{l-1}$. Then $\LT(g) = rx^{d_l}$
which is reduced w.r.t. $B$ and $g\in (B)$, contradicting to the
definition of Gr\"{o}bner bases.

We prove 3) by induction on $k$. When $k=2$, let $b_1 = c_1x^{d_1} +
s_{11}x^{d_1-1} + \cdots + s_{1d_1}$ and $b_2 = c_2x^{d_2} +
s_{21}x^{d_2-1} + \cdots + s_{2d_2}$. Then, $c_2 | c_1$ and $d_1 <
d_2$. Let $c_1 = c_2 t$, we need to show $t | b_1$. Since the
S-polynomial $S(b_1,b_2) = tb_2-x^{d_2-d_1}b_1$ can be reduced to
zero by $b_1$, we have $tb_2-x^{d_2-d_1}b_1 = u(x)b_1$, where
$u(x)\in\Zx$ and $\deg(u(x))<d_2-d_1$. Then, $tb_2 = (x^{d_2-d_1} +
u(x))b_1$, and $t | b_1$ follows since $x^{d_2-d_1} + u(x)$ is a
primitive polynomial in $\Zx$. The claim is true. Assume that for
$k=l-1$, the claim is true, then
$\widetilde{b}_1 | b_i$ for $1\le i \le l-1$. We will prove the
claim for $k=l$. Since $S(b_{1},b_l) =
\frac{c_{1}}{c_l}b_l-x^{d_l-d_{1}}b_{1}$ can be reduced to zero by
$B$. We have $\frac{c_{1}}{c_l}b_l-x^{d_l-d_{1}}b_{1} =
\sum_{i=1}^{l-1} f_ib_i$ with $f_i\in\Zx$ and $\deg(f_ib_i) \le
d_l-1$. Then, $\frac{c_{1}}{c_l}b_l = x^{d_l-d_{1}}b_{1} +
\sum_{i=1}^{l-1} f_ib_i$. By induction, $\widetilde{b}_1$ is a
factor of the right hand side of the above equation. Thus
$\widetilde{b}_1 | b_l$. Let $b_i = s_ib_{1}'$ for $1\le i \le l$,
we have $\frac{c_{1}}{c_l}s_{l} = x^{d_l-d_1}s_1+\sum_{i=1}^{l-1}
f_is_i$ where $\deg(s_i) = d_i-d_1$ and $s_1\in\Z$. Since
$\deg(f_is_i)\le d_l-d_1-1$, we have $\frac{c_1}{c_l} | s_1$ and
$\frac{c_1}{c_l} | b_1$.
For any $1\le i < j < l$, assume $\frac{c_i}{c_l} | b_i$. We will
show that $\frac{c_j}{c_l} | b_j$. Since $S(b_{j-1},b_{j}) =
\frac{c_{j-1}}{c_j}b_{j} -x^{d_j-d_{j-1}}b_{j-1} = \sum_{i=1}^{j-1}
f_ib_i$, we have $\frac{c_{j-1}}{c_l}$ is a factor of the right hand
side of the above equation, for $c_{j-1} | c_{j-2} | \cdots |
c_1$. Then, $\frac{c_{j-1}}{c_l} | \frac{c_{j-1}}{c_j}b_{j}$ and $
\frac{c_{j}}{c_l} | b_{j}$. The claim is proved. \qedd

\begin{exmp}\label{ex-1}
Here are three Gr\"{o}bner bases in $\Z[x]$\hbox{\rm:} $\{ 2, x \}$,
$ \{ 12, 6x+6, 3x^2+3x, x^3+x^2 \}$, $ \{ 9x+3, 3x^2+4x+1 \}$.
\end{exmp}

To give the structure of a reduced Gr\"{o}bner basis similar to that
in Example \ref{ex-1}, we introduce the concept of  generalized
Hermite normal form. Let
{\tiny
\begin{equation}\label{ghf}\C=\left[\begin{array}{lllllllllll}
c_{1,1}   & \ldots & c_{1,l_1}     &c_{1,l_1+1}   &\ldots       &\ldots       &\ldots &\ldots &\ldots            &\ldots  &\ldots   \\
\ldots    & \ldots & \ldots        & \ldots       &\ldots       &\ldots       &\ldots &\ldots &\ldots            &\ldots  &\ldots \\
c_{r_1, 1}& \ldots & c_{r_1, l_1}  &c_{r_1,l_1+1} & \ldots      &\ldots       &\ldots &\ldots & \ldots           &\ldots  &\ldots \\
0         & \ldots & 0             &c_{r_1+1,1}   &    \ldots   &c_{r_1+1,l_2}&\ldots &\ldots &  \ldots          &\ldots  &\ldots  \\
\ldots    & \ldots & \ldots        & \ldots       &\ldots       &\ldots       &\ldots &\ldots &\ldots            &\ldots  &\ldots  \\
0         & \ldots & 0             & c_{r_2,1}    &    \ldots   & c_{r_2,l_2} &\ldots &\ldots & \ldots           &\ldots  &\ldots   \\
\ldots    & \ldots & \ldots        &\ldots        &    \ldots   &\ldots       &\ldots &\ldots & \ldots           &\ldots  &\ldots   \\
0         & \ldots & 0             &0             &    \ldots   &0            &\ldots &0      &c_{r_{t-1}+1,1}   &\ldots  & c_{r_{t-1}+1,l_t} \\
\ldots    & \ldots & \ldots        &\ldots        &    \ldots   &\ldots       &\ldots &\ldots &\ldots            &\ldots  & \ldots \\
0         & \ldots & 0             &0             &    \ldots   &0            &\ldots &0      &c_{r_t,1}         &\ldots  & c_{r_t,l_t} \\
\end{array}\right]_{m\times s}\end{equation}}
whose elements are in $\Zx$.
It is clear that  $m = r_t$ and $s=\sum_{i=1}^t l_i$.
We denote by $\c_{r_i,j}$ to be the column of the matrix $\C$ whose last nonzero
element is
 \begin{equation}\label{eq-cij}
 c_{r_i,j}=c_{i,j,0}x^{d_{ij}}+\cdots+c_{i,j,d_{ij}}.\end{equation}
%
%
%
Then the leading monomial of  $\c_{r_i,j}$ is
$c_{r_i,j,0}x^{d_{r_i,j}}{{\beps}_{r_i}}$.
It is clear that  $\rk(L) = t$.

\begin{defn}\label{def-ghf}
The matrix $\C$ in~\bref{ghf} is called a {\em generalized Hermite
normal form} if it satisfies the following conditions:
\begin{description}
\item[1)] $0\le d_{r_i,1} < d_{r_i,2} < \cdots < d_{r_i,l_i}$ for any $i$.

\item[2)] $ c_{r_i,l_i,0} |\cdots | c_{r_i,2,0}  | c_{r_i,1,0}$.

\item[3)]
$S(\c_{r_i,j_1},\c_{r_i,j_2})=
x^{d_{r_i,j_2}-d_{r_i,j_1}}\c_{r_i,j_1}-\frac{c_{r_i,j_1,0}}{c_{r_i,j_2,0}}\c_{r_i,j_2}$
can be reduced to zero by the column vectors of the matrix for any
$1\le i\le t, 1\le j_1<j_2\le l_i$.

\item[4)]
$\c_{r_i,j}$ is G-reduced w.r.t. the column vectors of the matrix
other than $\c_{r_i,j}$, for any $1\le i\le t, 1\le j\le l_i$.
\end{description}
\end{defn}

It is clear that $\{\c_{r_i,1},\ldots,\c_{r_i,l_i}\}$ is a reduced
Gr\"obner basis in $\Zx$. Then, as a consequence of Theorem
\ref{th-buch} and Lemma \ref{lm-21}, we have
\begin{thm}\label{th-gbh}
$\fb = \{\f_1,\ldots,\f_s \}\subset\Z[x]^n$ is a reduced Gr\"{o}bner
basis such that $\f_1< \f_2< \cdots < \f_s$ if and only if $[\f_1,\ldots,\f_s]$
is a generalized Hermite normal
form.
\end{thm}


\begin{exmp}\label{ex-L1}
The following matrices are generalized Hermite normal forms
 \[M_1=\left [\begin{array}{llll}
x   & 2 & 0   \\
0   & 2 & x    \\
\end{array}\right],\,
 M_2=\left[\begin{array}{llll}
2   & x-1 & 0 & 0  \\
0   & 0 & 2      & x-1      \\
\end{array}\right]
\]
whose columns constitute the reduced  Gr\"{o}bner bases of the $\Zx$-lattices.
%
\end{exmp}

Let $\fb=\{ \f_1,\ldots,\f_s\}$ be a reduced Gr\"{o}bner basis.
Let $S(\f_i,\f_j)=m_{ij}{\f}_i - m_{ji}{\f}_j$ be the S-polynomial of $\f_i,\f_j$
and $\grem(S(\f_i,\f_j),\fb) = \sum_k c_k \f_k$ be the normal
representation of in terms of the Gr\"obner basis $\fb$. Then the
{\em syzygy polynomial} $\widetilde{S}(\f_i, \f_j)$
 $$\widetilde{S}(\f_i, \f_j)=m_{ij}{\beps}_i -m_{ji}{\beps}_j - \sum_k c_k {\beps}_k,$$
is an element in $\Z[x]^s$, where ${\beps}_k$ is the $k$-th standard
basis vector of $\Z[x]^{s}$.
Define an order in $\Z[x]^s$ as follows: $ax^\alpha {\beps}_i \prec
bx^\beta {\beps}_j$ if $\LT (ax^\alpha\f_i)
> \LT(bx^\beta \f_j)$ in $\Z[x]^n$.
By Schreyer's Theorem~\cite[p. 212]{cox-1998}, we have
\begin{thm}\label{lm-free}
Let $F = [\f_1,\ldots,\f_s]_{n\times s}\in\Zx^{n\times s}$ be a generalized
Hermite normal form. Then the syzygy polynomials  $\widetilde{S}(\f_i, \f_j)$
form a Gr\"obner basis of the $\Zx$-lattice $\ker(F) = \{ X\in\Z[x]^s | FX = {\bf 0} \}$
under the newly defined order $\prec$.
\end{thm}

Let $\C$ be defined in \bref{ghf} and $k\in\N$. Introduce the
following notations:
 \begin{eqnarray}
 \C_{-} &=& \cup_{i=1}^t \cup_{k=1}^{l_i-1}    \{\c_{r_i,k}, x\c_{r_i,k}, \ldots,
    x^{\deg(c_{r_i,k+1})-\deg(c_{r_t,k})-1}\c_{r_i,k}\},\nonumber\\
 \C^{+} &=& \cup_{i=1}^t \cup_{k=0}^{\infty}  \{x^k \c_{r_i,l_i}\}.\label{eq-inf} \\
 \C_{\infty} &=& \C_{-}\cup   \C^{+} \nonumber
 \end{eqnarray}

\begin{exmp}\label{ex-ext}
Let $\C =\left [\begin{array}{lllll}
6  & 3x              & 0             &3       & 2x         \\
0   & 0            &    6           & 3x       & x^3+x      \\
\end{array}\right].$

Then $\C_{-} =\left [\begin{array}{lllll}
6        & 0          &3  &3x              \\
0        & 6           & 3x  &3x^2    \\
\end{array}\right]$ and
\[ \C_{\infty} = \left[ \begin{array}{lllllllllll}
6   & 3x    & 3x^2    &  3x^3 &\cdots      & 0      &3      &3x           & 2x   & 2x^2 & \cdots  \\
0    & 0    & 0       &   0   &\cdots      &  6    & 3x    &3x^2        & x^3+x  & x^4+x^2 &  \cdots  \\
\end{array} \right]. \]
\end{exmp}

We need the following properties about $\C_{\infty}$. By
saying the infinite set $\C_{\infty}$ is linear independent over $\Z$,
we mean any finite subset of $\C_{\infty}$ is linear independent over
$\Z$. Otherwise, $\C_{\infty}$ is said to be linear dependent.
\begin{lem}\label{lm-ext1}
The columns of $\C_{\infty}$ in \bref{eq-inf} are linear independent over $\Z$.
\end{lem}
\proof Suppose $\C$ is given in \bref{ghf}. The leading term of
$\c\in\C_{\infty}$ is $\LT(\c) = ax^l{\beps}_{r_i}$ for
$i=1,\ldots,t$ and $l\in\N$. Furthermore, for two different $\c_1$
and $\c_2$ in $\C_S$ such that $\LT(\c_1) = ax^{l_1}{\beps}_{r_i}$
and $\LT(\c_2) = bx^{l_2}{\beps}_{r_i}$, we have $l_1\ne l_2$. Then
$\LT(\C_{\infty})=\{a_{il_i}x^{l_i}{\beps}_{r_i}\,|\, i=1,\ldots,t;
l_i= d_{i1},d_{i1}+1,\ldots;a_{il_i}\in\Z\}$ are linear independent
over $\Z$, where $d_{i1}$ is from \bref{eq-cij}. Then $\C_{\infty}$
are also linear independent over $\Z$.\qedd
%
%

\begin{lem}\label{lm-ext2}
Let $\C$ be a generalized Hermite normal form.
Then any $\g\in (\C)$
can be written uniquely as a linear combination of finitely many
elements of $\C_\infty$ over $\Z$.
\end{lem}
\proof $\g\in (\C)$ can be written as a linear combination of
elements of $\C_\infty$ over $\Z$ by the procedure to compute
$\grem(\g,\C)=0$ \cite{cox-1998}. The uniqueness is a consequence of
Lemma~\ref{lm-ext1}.\qedd


\section{Canonical Representations for Laurent binomial $\sigma$-ideal}
\label{sec-lbi}
In this section, we will give three canonical representations for
a proper Laurent binomial ideal.

\subsection{Laurent binomial $\sigma$-ideal}
\label{sec-lbi1}
In this section, several basic properties of
Laurent binomial $\sigma$-ideals will be proved.

By a {\em Laurent $\sigma$-binomial} in  $\Y$,  we mean a
$\sigma$-polynomial with two terms,  that is,
$a\Y^{\g}+b\Y^{\h}$ where $a, b\in \F^*=\F \setminus\{0\}$
and $\g, \h \in\Z[x]^n$.
A Laurent $\sigma$-binomial of the following form is said to be in
{\em normal form}
 $$p=\Y^{\f}- c_\f$$
where $c_\f\in \F^{\ast}=\F\setminus\{0\}$ and $\f\in\Z[x]^n$ is
normal. The vector $\f$ is called the {\em support} of $p$. For
$p=\Y^{\f}- c_\f$,  we denote $\widehat{p}= -c_\f^{-1}\Y^{-\f} p=
\Y^{-\f}- c_\f^{-1}$ which is called the {\em inverse} of $p$.
%
%
It is clear that any Laurent $\sigma$-binomial $f$ can be written uniquely as
$ f = aM(\Y^{\f}- c_\f)$
where $a\in \F^{\ast}$, $M$ is a Laurent $\sigma$-monomial, and
$\Y^{\f}- c_\f$ is in normal form. Since $aM$ is a unit in
$\F\{\Y^{\pm}\}$, we can use the normal $\sigma$-binomial $\Y^{\f}-
c_\f$ to represent $f$, and when we say a Laurent $\sigma$-binomial
we always use its normal representation.

A Laurent $\sigma$-ideal is called {\em binomial} if it is generated by
Laurent $\sigma$-binomials.

\begin{lem}\label{lm-bi1}
Let $\Y^{\f_{i}}-c_i,i=1, \ldots,s$ be contained in a Laurent
binomial $\sigma$-ideal $\I$ and $\f=a_1\f_1+ \cdots +a_s\f_s$,
where $a_i\in\Zx$. Then $\Y^{\f} - \prod_{i=1}^s c_i^{a_i}$ is in
$\I$.
\end{lem}
\proof It suffices to show that if $p_1=\Y^{{\f}_1}-c_1\in\I$ and
$p_2=\Y^{{\f}_2}-c_2\in\I$, then $\Y^{n{\f}_1}-c_1^n \in\I$ for
$n\in\N$, $\Y^{-{\f}_1}-c_1^{-1}\in\I$,
$\Y^{x{\f}_1}-\sigma(c_1)\in\I$, and
$\Y^{{\f}_1+{\f}_2}-c_1c_2\in\I$, which are indeed true since
$\Y^{n{\f}_1}-c_1^n=(\Y^{{\f}_1})^n-c_1^n$ contains $p_1$ as a
factor, $\Y^{-{\f}_1}-c_1^{-1}=-c_1^{-1}\Y^{-{\f}_1}(\Y^{\f_1}-c_1)
\in\I$, $\Y^{x{\f}_1}-\sigma(c_1)=\sigma(\Y^{{\f}_1}-c_1)\in\I$, and
$\Y^{{\f}_1+{\f}_2}-c_1c_2= \Y^{{\f}_1}(\Y^{{\f}_2}-c_2) +
c_2(\Y^{{\f}_1}-c_1)\in\I$. \qedd

As a direct consequence, we have
\begin{prop}\label{cor-bi1}
Let $\I$ be a proper Laurent binomial $\sigma$-ideal and
\begin{equation}\label{eq-bi1}
\L(\I):=\{\f\in \Z[x]^{n}\, |\, \exists c_\f\in\F^*\,s.t.\,
\Y^{\f}-c_\f\in \I\}.
\end{equation}
Then $\L(\I)$ is a $\Zx$-lattice, which is called the {\em support
lattice} of $\I$.
Furthermore, Let $\L(\I)=(\f_1,\ldots,\f_s)$. Then
$\I=[\Y^{\f_1}-c_{\f_1},\ldots,\Y^{\f_s}-c_{\f_s}]$.
That is, a Laurent binomial $\sigma$-ideal is finitely generated and
$[\f_1,\ldots,\f_s]$ is called a {\em matrix representation} for $\I$
\end{prop}
\proof Let $\I_1=[\Y^{\f_1}-c_{\f_1},\ldots,\Y^{\f_s}-c_{\f_s}]$.
It suffices to show $\I\subset \I_1$.
Since $\I$ is Laurent binomial, it has a set of generators of the form
$f_{\h}=\Y^{\h}-c_{\h}$. Then $\h\in\L(\I)=(\f_1,\ldots,\f_s)$.
By Lemma \ref{lm-bi1}, there exists a $\widetilde{c}_{\h}\in \F$
such that $\widetilde{f}_{\h} = \Y^\h - \widetilde{c}_{\h}\in\I_1$.
Then $f_{\h}- \widetilde{f}_{\h} = \widetilde{c}_{\h}-c_{\h}\in\I$.
Since $\I$ is proper, we have  $f_{\h}- \widetilde{f}_{\h}=0$
or $f_{\h}\in\I_1$ and hence $\I\subset\I_1$.\qedd

Similarly, we can prove
\begin{cor}\label{lm-bi3}
Let $\I=[\Y^{\f_{1}}-c_{1}, \ldots, \Y^{\f_{s}}-c_{s}]$ be a proper
Laurent binomial $\sigma$-ideal and let $\h_1, \ldots, \h_r$ be
another set of generators of
$(\f_1,\ldots,\f_s)$, and
 $\h_i = \sum_{k=1}^s a_{i,k} \f_k, i=1,\ldots,r$, where $a_{i,k}\in\Zx$.
Then $\I = [\Y^{\h_1}-\prod_{i=1}^s c_i^{a_{1, i}} ,\ldots,
 \Y^{\h_r}-\prod_{i=1}^s c_i^{a_{r, i}}]$.
\end{cor}

We now show to check whether a Laurent binomial
$\sigma$-ideal is proper.
\begin{prop}\label{lm-bi2}
Let $\I=[\Y^{\f_{1}}-c_{1}, \ldots, \Y^{\f_{s}}-c_{s}]$ be a Laurent
binomial $\sigma$-ideal and $M=[\f_1, \ldots, \f_s]\in\Zx^{n\times s}$.
Let $\ker(M)=\{\h\in\Zx^s\,|\, M\h=0\}$ be
generated by $\bu_1, \ldots, \bu_t$,  where $\bu_i = (u_{i, 1},
\ldots, u_{is})$.
Then $\I\ne[1]$ if and only if $\prod_{i=1}^s c_i^{u_{l, i}}=1$ for
$l=1, \ldots, t$.
\end{prop}
\proof ``$\Rightarrow$" Let $f_i = \Y^{\f_i}-c_i$. Suppose
$c=\prod_{i=1}^s c_i^{u_{l, i}}\ne1$ for some $l$. Replacing $c_i$
by $\Y^{\f_{i}} - f_i$ in the above equation and noting that
$\bu_l\in\ker(M)$, we have
$c=\prod_{i=1}^s c_i^{u_{l, i}} = \prod_{i=1}^s (y^{\f_{i}} - f_i)
^{u_{l, i}} = \prod_{i=1}^s \Y^{M\cdot \bu_l} + g = 1 + g$ where
$g\in\I$. Then $0\ne c -1 \in\I$ and $\I=[1]$, a contradiction.

``$\Leftarrow$" Suppose the contrary. Then there exist
$g_i\in\F\{\Y^{\pm}\}$ such that
 \begin{equation}\label{eq-lg1} g_1 f_1+\cdots+ g_s f_s=
 1.\end{equation}
Let $l$ be the maximal $c$ such that $y_c^{(k)}$ occurs in some
$f_i$, $o$ the largest $j$ such that $y_l^{(j)}$ occurs in some
$f_k$, and $d=\max_{k=1}^s\deg(f_k,y_l^{(o)})$. Let $f_k =
\Y^{\f_k}-c_k = I_k y_l^{dx^o} - c_k$.
Since \bref{eq-lg1} is an identity about the algebraic variables
$y_i^{x^j}$, we can set $y_l^{dx^o} = c_k/I_k$ in \bref{eq-lg1} to
obtain a new identity. In the new identity, $f_k$ becomes zero and
the left hand side of \bref{eq-lg1} has at most $s-1$ summands.
We will show that this procedure can be continued for the new
identity. Then the left hand side of \bref{eq-lg1} will eventually
becomes zero, and a contradiction is obtained and the lemma is
proved.

If $\ord(f_i,y_l) < o$ or $\ord(f_i,y_l) = o$ and
$\deg(f_i,y_l^{x^o}) < d$ for some $i$, then $f_i$ is not changed in
the above procedure. Let us assume that for some $v$,
$\deg(f_v,y_l^{x^o}) = d$ and $f_v = \Y^{\f_v}-c_v = I_v y_l^{dx^o}
- c_v$. Then after the substitution,
$f_v = c_kI_v/I_k - c_v = c_k \widetilde{f}_v$ where
$\widetilde{f}_v=I_v/I_k - c_v/c_k$.
We claim that either $\widetilde{f}_v=0$ or $I_v/I_k$ is a proper
monomial, and as a consequence, the above substitution can continue.
To prove the claim, it suffices to show that if $I_v=I_k$ then
$c_v=c_k$. If $I_v=I_k$, then $\f_v=\f_k$, that is $\f_v-\f_k=0$ is
a syzygy among $\f_i$ and let ${\beps}_{vk}$ be the corresponding
syzygy vector. Then ${\beps}_{vk}\in\ker(M)$ can be written as a
linear combination of $\bu_1,\ldots,\bu_s$. Let
${\bf{c}}=(c_1,\ldots,c_s)^\tau$. Then
$c_vc_k^{-1}={\bf{c}}^{{\beps}_{vk}}$ can be written as a product of
${\bf{c}}^{\bu_l}=\prod_{i=1}^s c_i^{u_{l, i}}=1$, and thus
$c_vc_k^{-1}=1$. \qedd

\subsection{Characteristic set of Laurent binomial $\sigma$-ideal}
\label{sec-lbi2}
We show how to modify the characteristic set method presented in section
\ref{sec-cs} in the case of Laurent binomial $\sigma$-ideals.
First, assume that all Laurent $\sigma$-binomials are in normal
form, which makes the concepts of order and leading  variables
unique.

Second, when defining the concepts of rank and $q$ to be reduced
w.r.t. $p$, we need to replace $\deg(p, y_j^{(o)})$ by $|\deg(p,
y_j^{(o)})|$. Precisely,  $q$ is said to be {\em reduced} w.r.t. $p$
if  $|\deg(q, y_j^{(k+l)})|<|\deg(p, y_j^{(k)})|$  for all $l\in\N$,
where $\lead(p)=y_j^{(k)}$.
For instance, $y_1^{-2x}y_2-1$ is not reduced w.r.t. $y_1^2-1$. With
these changes, the concepts of $\sigma$-chain and characteristic set
can be defined in the Laurent $\sigma$-binomial case.
For instance, the $\sigma$-chain $\A_2$ in \bref{ex-L11} becomes the
following Laurent normal form:
\begin{equation}\label{eq-T11}
 \begin{array}{llllll}
 \widetilde{\A}_2& =& y_1^2-1, &y_1^{-1}y_1^{x}-1, &y_2^2-1, &y_2^{-1}y_2^{x}-1\\
\end{array}
\end{equation}

Third, the $\sigma$-remainder for two Laurent $\sigma$-binomials
need to be modified as follows.
We first consider how to compute $\prem(f,g)$ in the simple case:
$o=\ord(f,y_l)=\ord(g,y_l)$, where $y_l=\lvar(g)$.
Let $g = I_g (y_{l}^{(o)})^{d} - c_g$, where $d=\deg(g,y_{l}^{(o)})$
and $I_g$ is the initial of $g$. As mentioned above,  $g$ is in
normal form,  that is $d>0$. Let $d_f = \deg(f, y_{l}^{(o)})$ and $f
= I_f (y_{l}^{(o)})^{d_f} - c_f$.
We consider two cases.

In the first case,  let us assume  $d_f\ge0$. If $d_f < d_g$,  then
set $r=\prem_1(f, g)$ to be $f$.
Otherwise,  perform the following basic step
 \begin{eqnarray}
r :=\prem_1(f,g)=
 (f - g \frac{I_f}{I_g}(y_{l}^{(o)})^{d_f-d_g})/c_g
  = \frac{I_f}{I_g} (y_{l}^{(o)})^{d_f-d_g} -\frac{c_f}{c_g}. \label{eq-prem1}
 \end{eqnarray}
Let  $\h_r,\h_f,\f_g$ be the supports of $r,f,g,$ respectively. Then
 \begin{eqnarray}\label{eq-prem2}
 \h_r=\h_f-\h_g.
 \end{eqnarray}
Set $f=r$ and repeat the procedure  $\prem_1$ for $f$ and $g$. Since
$d_f$ decreases strictly after each iteration, the procedure will
end and return $\prem(f,g) = r$ which satisfies
 \begin{eqnarray}
 r &=&  \frac{f}{c_g^k} -hg
     =  \frac{I_f}{I_g^{k}} (y_{l}^{(o)})^{d_f-kd_g} -\frac{c_f}{c_g^k}
 \label{eq-prem3}\\
 \h_r&=&\h_f-k \h_g\label{eq-prem4}
 \end{eqnarray}
where $k=\lfloor\frac{d_f}{d_g}\rfloor$ and $h\in\F\{\Y^{\pm}\}$.

In the second case,  we assume  $d_f<0$. The $\sigma$-remainder can
be computed similar to the first case. Instead of $g$, we consider
$\widehat{g} = (I_g)^{-1} (y_{l}^{(o)})^{-d_g} - c_g^{-1}$.
If $|d_f| < d_g$,  then set $r=\prem_1(f, g)$ to be $f$.
Otherwise,  perform the following basic step
 \begin{eqnarray*}
 r := \prem_1(f, g)=c_g(f - \widehat{g} I_gI_f (y_{l}^{(o)})^{d_f+d_g})
 = I_f I_g (y_{l_g}^{(o)})^{d_f+d_g} -c_fc_g.
 \end{eqnarray*}
In this case, equation \bref{eq-prem2} becomes $\h_r=\h_f+\h_g$. To
compute $\prem(f, g)$, repeat the above basic step for $f=r$ until
$|d_f| < d_g$.

For two general $\sigma$-binomials $f$ and $g$, $\prem(f,g)$ is
defined as follows: if $f$ is reduced w.r.t $g$, set $\prem(f,g)=f$.
Otherwise, let $y_l=\lvar(g)$, $o_f=\ord(f,y_l)$, and
$o_g=\ord(g,y_l)$. Define
 $$\prem(f,g)=\prem(\ldots,\prem(\prem(f,g^{(o_f-o_g)}),g^{(o_f-o_g-1)}), \ldots,g ).$$
Let $\A: A_1,\ldots,A_s$ be a Laurent binomial $\sigma$-chain and
$f$ a $\sigma$-binomial. Then define
 $$ \prem(f,\A)=\prem(\ldots,\prem(\prem(f,A_s),A_{s-1}),
 \ldots,A_1).$$
In summary,  we have
\begin{lem}\label{lm-rem1}
Let $\A= A_1,\ldots,A_s$ be a Laurent binomial $\sigma$-chain, $f$ a
$\sigma$-binomial, and $r=\prem(f,\A)$. Then $r$ is  reduced w.r.t.
$\mathcal {A}$ and   satisfies
   \begin{equation}\label{eq-rem2}
   cf \equiv r,  \mod \,  [\mathcal {A}],
   \end{equation}
where $c\in\F^*$. Furthermore, let the supports of $r$ and $f$ be
$\h_r$ and $\h_f$, respectively. Then $\h_f-\h_r$ is in the $\Z[x]$-lattice generated by the supports of $A_i$.
\end{lem}

Similar to section \ref{sec-cs}, the concepts of coherent and regular
$\sigma$-chains can be extended to the Laurent case.
Since any $\sigma$-monomial is a unit in $\F\{\Y^\pm\}$,
the concept of regular $\sigma$-chain need to be strengthened
as follows.
A $\sigma$-chain $\A$ is called {\em Laurent regular}
if $\A$ is regular and any $\sigma$-monomial is invertible w.r.t $\A$.
Then, following \cite{gao-dcs}, Theorem \ref{th-rp} can be extended to the
following Laurent version straightforwardly.
\begin{thm}\label{th-rpl}
A Laurent $\sigma$-chain $\A$ is a characteristic set of $\sat(A)$ if and
only if $\A$ is coherent and Laurent regular.
\end{thm}

For Laurent binomial $\sigma$-chains, we have
\begin{lem}\label{lm-reg}
Any Laurent binomial $\sigma$-chain $\A$ is Laurent regular.
\end{lem}
\proof
%
Since the initials of $\A$ are $\sigma$-monomials, it suffices
to show that any $\sigma$-monomial is invertible w.r.t $\A$.
By \cite[p.248]{gao-dcs}, a $\sigma$-monomial $M$ is invertible w.r.t $\A$ is
$M$ is invertible  w.r.t an extension $\A_M$ of $\A$ when both
$M$ and $\A_M$ are treated as algebraic polynomials in $y_i^{x^j}$, where $\A_M$
is an algebraic Laurent binomial chain.
By \cite[p.1150]{jsc-csdd}, $M$ is invertible  w.r.t $\A_M$
if the successive Sylvester resultant ${\rm{Resl}}(M,\A_M)$
of $M$ and $\A_M$ is nonzero.
Since $\A$ is Laurent binomial, $B\in\A_M$
has the form $B=I (y_k^{x^o})^m+U$, where $I$ is the initial of $B$
and $U$ a $\sigma$-monomial does not contain $y_k^{x^o}$.
Let $N=J (y_k^{x^o})^n$ be any $\sigma$-monomial with $J$ as the initial.
Then the Sylvester resultant of $M$ and $B$ w.r.t
$y_k^{x^o}$ is $J^mU^n$ which is a nonzero $\sigma$-monomial.
As a consequence, ${\rm{Resl}}(M,\A_M)$ is also a nonzero $\sigma$-monomial
and hence $\A$ is Laurent regular.\qedd
%
%

We now give the first canonical representation for Laurent binomial $\sigma$-ideals.
\begin{thm}\label{th-rp1}
$\I$ is a proper Laurent binomial $\sigma$-ideal if and only
there exists a Laurent coherent $\sigma$-chain $\A$ such that
$\I= \sat(\A)=[\A]$.
%
\end{thm}
\proof Let $\I\ne[1]$ and $\A$ the characteristic set of $\I$.
Then $[\A]\subset\I\subset\sat(\A)$.
From \bref{eq-rem2}, we have $\sat(\A)\subset[\A]$ and then $\I=\sat(\A)=[\A]$.
By Theorem \ref{th-rpl}, $\A$ is coherent.
To prove the other side of the theorem, let $\A$ be a Laurent coherent $\sigma$-chain.
By Lemma \ref{lm-reg}, $\A$ is also Laurent regular.
By Theorem \ref{th-rpl}, $\A$ is a characteristic set of $I=\sat(\A)$.
Then $\I$ is proper.\qedd

\begin{cor}\label{cor-dim}
Let $\I$ be a Laurent reflexive prime  binomial $\sigma$-ideal in
$\F\{\Y^{\pm}\}$. Then $\dim(\I)=n-\rk(\L(\I))$.
\end{cor}
\proof
By Theorem \ref{th-rp1}, $\I=[\A]$, where $\A:\Y^{\c_1}-c_1,\ldots\Y^{\c_s}-c_s$.
Let $\C=[\c_1,\ldots,\c_s]$ is the matrix
representation for $\I$ and in the form of \bref{ghf}. Since $\I$ is reflexive and prime, by Theorem 4.3 of \cite{gao-dcs}, $\dim(\I)=n-t=n-\rk(\L(\I))$.\qedd

\begin{cor}\label{cor-rad}
A Laurent binomial $\sigma$-ideal is radical.
\end{cor}
\proof By Theorem \ref{th-rp1}, $\I=[\A]$, where $\A: \Y^{\h_1}-c_1, \ldots, \Y^{\h_r}-c_r$ is
the characteristic set of $\I$.
Let $A_i=\Y^{\h_i}-c_i$ and $y_{l_i}^{(o_i)} = \lead(A_i)$. $\A$ is
also {\em saturated} in the sense that its separant $\frac{\partial
A_i}{\partial y_{l_i}^{(o_i)}}$ are $\sigma$-monomials and  hence
units in $\F\{\Y^{\pm}\}$. Then similar to the differential case
\cite{bouziane}, it can be shown that $\sat(\A) = [\A]$ is a radical
$\sigma$-ideal.\qed

\vskip5pt
Let $\f_1 < \f_2 < \cdots <\f_s$ be elements in $\Zx^n$, $c_i\in\F^*$, and
\begin{eqnarray}
&&\fb=\{ \f_{1}, \ldots, \f_{s}\}\subset\Zx^n\label{eq-FA}\\
&&\A_{\fb}=\{A_1, \ldots, A_a\}\subset\F\{\Y^\pm\}\hbox{ with }
A_i=\Y^{\f_{i}}-c_{i},i=1,\ldots,s \nonumber
\end{eqnarray}
In the rest of this section, we will establish a connection between
$\fb$ and $\A_{\fb}$.
%
From Definition \ref{def-ghf}, we have
\begin{lem}\label{lm-t2}
For $i< j$,  $A_j$ is reduced w.r.t. $A_i$ if and only if $\f_j$ is G-reduced w.r.t.
$\f_i$.
\end{lem}
%

%
%
%
\begin{lem}\label{lm-t3}
For $\fb$  and  $\A_{\fb}$ in \bref{eq-FA} and a Laurent $\sigma$-binomial
$f=\Y^\f-c$,
if $\prem(f,\A_{\fb})=\Y^{\g}-c_\g$,  then $\g=\grem(\f,\fb)$.
\end{lem}
\proof Let us first consider  $\prem_1$ in \bref{eq-prem1} for $f$
and $A_i=\Y^{\f_i}-c_i=I_i (y_{l_i}^{(o_i)})^{d_i} - c_i$, where
$\lead(A_i)=y_{l_i}$ and $I_i$ is the initial of $A_i$.
From \bref{eq-prem2}, the support of $r=\prem_1(f,A_i)$ is  $\f -
\f_i$.
It is clear that $\LT(\f_i) = d_i x^{o_i} {\beps}_{l_i}$. Let $\f_i
=  d_i x^{o_i} {\beps}_{l_i}+ \overline{\f}_i$. Similarly, write $\f
= d_f x^{o_i}{\beps}_{l_i} +\overline{m}$ where $d_f
x^{o_i}{\beps}_{l_i}$ is the leading term of $\f$ w.r.t.
${\beps}_{l_i}$ and $d_f\ge d_i\ge0$.
Then a basic step to compute $\grem(\f,\f_i)$ is to compute
$\grem_1(\f,\f_i)= \f - \f_i = (d_f-d_i) x^{o_i}{\beps}_{l_i}
+\overline{\f}- \overline{\f}_i$, which is the support of
$\prem_1(f,A_i)$.

Using the basic step $\grem_1$ to compute
$\grem(\f,\fb)$, we have a sequence of elements in $\Z[x]^n$:
$\g_0=\f, \g_1,\ldots,\g_t=\grem(\f,\fb)$.
Correspondingly,  using the basic step $\prem_1$ to compute
$\prem(\f,\A_{\fb})$, we have a sequence of $\sigma$-binomials
$f_0=f,f_1,\ldots,f_t=\prem(f,\A_{\fb})$ such that the support of $f_i$ is
$\g_i$ for $i=1,\ldots,t$.\qedd
%
%

\begin{lem}\label{lm-t4}
If $\fb$ in \bref{eq-FA} is a reduced Gr\"obner basis
and  $[\A_{\fb}]\ne[1]$,  then $\A_{\fb}$
is a coherent $\sigma$-chain.
\end{lem}
\proof By Lemma \ref{lm-t2}, $\A_{\fb}$ is a $\sigma$-chain.
%
%
%
Let $A_i=\Y^{\f_{i}}-c_{i}$ and $A_j=\Y^{\f_{j}}-c_{j}$ ($i<j$) have
the same leading variable $y_l$, and
    $A_i=I_i y_l^{d_i x^{o_i}}-c_i$,
    $A_j=I_j y_l^{d_j x^{o_j}}-c_j$.
From Definition \ref{def-ghf},  we have $o_i < o_j$ and $d_j| d_i$.
Let $d_i = t d_j$ where $t\in\N$.
According to \bref{eq-prem3}, we have
 \begin{equation}\label{eq-delta1}
 \Delta(A_i,A_j)
=\prem((A_i)^{x^{o_j-o_i}},A_j) =
       \frac{(I_i)^{x^{o_j-o_i}}}{I_j^t} -
         \frac{(c_i)^{x^{o_j-o_i}}}{c_j^t}.
 \end{equation}
Then the support of $\Delta(A_i,A_j)$ is $x^{o_j-o_i} \f_i -
\frac{d_i}{d_j} \f_j$.

Since $\LT(A_i)=d_i x^{o_i} {\beps}_l$ and  $\LT(A_j)=d_j x^{o_j}
{\beps}_l$,  we have $N=\lcm(d_i x^{o_i}, d_j x^{o_j}) = d_i
x^{o_j}$.
According to Definition \ref{def-sv}, the S-vector of $\f_i$ and
$\f_j$ is
 $$S(\f_i,\f_j) = x^{o_j-o_i} \f_i - \frac{d_i}{d_j}  \f_j.$$
Since $\fb$ is a Gr\"obner basis, we have
$\g=\grem(S(\f_i,\f_j),\fb)=0$. Since the support of
$\Delta(A_i,A_j)$ is $S(\f_i,\f_j)$, by Lemma \ref{lm-t3},
$R=\prem(\Delta(A_i,A_j),\A_{\fb})=\Y^{\g}-c=1-c$ for some $c\in\F$.
Since $[\A_\fb]$ is proper and $R=1-c\in[\A_\fb]$,
we have $R=0$ and hence $\A_{\fb}$ is coherent.\qedd

We now give the main result of this section.
\begin{thm}\label{th-t4}
For $\fb$ and $\A_{\fb}$ defined in \bref{eq-FA},
$\A_{\fb}$ is a coherent $\sigma$-chain if and only if $\fb$
is a reduced Gr\"obner basis and $[\A_{\fb}]\ne[1]$.
%
\end{thm}
\proof Lemma \ref{lm-t4} proves one side of the theorem.
For the other direction, let $\A_{\fb}$ be a coherent
$\sigma$-chain. By Lemma \ref{lm-t2}, $\f_i$ is
G-reduced to $\f_j$ for $i\ne j$.
By Theorem \ref{th-rp1}, $[\A_{\fb}]$ is proper.
Use the notations introduced in the proof of Lemma \ref{lm-t4}.
Since $S(\f_i,\f_j)$ is the support of $\Delta(A_i,A_j)$,
by Lemma \ref{lm-t3}, $\f_{ij}=\grem(S(\f_i,\f_j),\fb)$ is the
support of $\prem(\Delta(A_i,A_j),\A_{\fb})$.
Since $\A_{\fb}$ is coherent, $\prem(\Delta(A_i,A_j),\A_{\fb})=\Y^{\f_{ij}}-c=0$
for any $i$ and $j$, and this is possible only
when $\f_{ij}=\grem(S(\f_i,\f_j),\fb)=0$ and $c=1$ due to the fact
$[\A_{\fb}]\ne[1]$.
Hence $\fb$ is a reduced
Gr\"obner basis.\qedd
%
%


\subsection{Partial character and Laurent binomial $\sigma$-ideal}
\label{sec-lbi3} In this section,  we will show that proper Laurent
binomial $\sigma$-ideals can be described uniquely with their
partial characters.

\begin{defn}
A {\em partial character} $\rho$ on $\Z[x]^{n}$ is a homomorphism
from a $\Z[x]$-lattice $L_{\rho}$ to the multiplicative group
$\F^{\ast}$ satisfying $\rho(x\f)=(\rho(\f))^x=\sigma(\rho(\f))$ for $\f\in
L_\rho$.
\end{defn}

Let $\rho$ be a partial character over $\Zxn$ and $L_\rho=(\f_{1},
\ldots, \f_{s})$, where $\fb=\{\f_{1}, \ldots, \f_{s}\}$ is a
reduced Gr\"obner basis. Define
\begin{eqnarray}
 \I(\rho)&:=&[\Y^{\f}-\rho(\f)\,|\, \f\in L_{\rho}].\label{eq-I}\\
 \A(\rho)&:=& \Y^{\f_{1}}-\rho(\f_{1}), \ldots, y^{\f_{s}}-\rho(\f_{s}).\label{eq-Arho}
\end{eqnarray}
The Laurent binomial $\sigma$-ideal $\I(\rho)$ has the following
properties.


\begin{lem}\label{lm-l1}
$\I(\rho)=[\A(\rho)]\ne [1]$ and $\A(\rho)$ is a characteristic set
of $\I(\rho)$.
\end{lem}
\proof
By Lemma \ref{lm-bi1} and the property of partial character, $\I(\rho)=[\A(\rho)]$.
%
%
By Proposition \ref{lm-bi2}, in order to prove $\I(\rho)\ne [1]$, it
suffices to show that for any syzygy $\sum_i a_i \f_i = 0$ among
$\f_i$, we have $\prod_i \rho(\f_i)^{a_i}=1$. Indeed,  $\rho(\sum_i
a_i \f_i)=\prod_i\rho(\f_i)^{a_i}=1$, since $\rho$ is a homomorphism
from the $\Zx$-module $L_\rho$ to $\F^{*}$.
%
Since $\fb$ is a reduced Gr\"ober basis, by Theorem \ref{th-t4},
$\A$ is a characteristic set of $\I(\rho)$.\qedd

\begin{lem}\label{lm-l3}
A Laurent $\sigma$-binomial $\Y^{\f}-c_\f$ is in $\I(\rho)$ if and
only if $\f\in L_\rho$ and $c_\f=\rho(\f)$.
\end{lem}
\proof By Lemma \ref{lm-l1}, $\A(\rho)$ is a characteristic set of
$\I(\rho)$.
Since $f=y^{\f}-c_{\f}$ is a $\sigma$-binomial in $\I(\rho)$, we
have $r=\prem(f,\A)=0$. By Lemma \ref{lm-rem1}, $\f$ is in the
$\Z[x]$-module $L_\rho$. The other side is obviously true and the
lemma is proved.\qedd

We now show that all Laurent binomial $\sigma$-ideals are defined by
partial characters.
\begin{thm}\label{th-l1}
The map $\rho\Rightarrow\I(\rho)$  gives a one to one correspondence
between the set of proper Laurent binomial $\sigma$-ideals and
partial characters on $\Z[X]^n$.
\end{thm}
\proof By Lemma \ref{lm-l1}, a partial character defined a proper
Laurent binomial $\sigma$-ideal.
For the other side, let $\I\subseteq \F\{\Y^{\pm}\}$ be a proper
Laurent binomial $\sigma$-ideal.
$\I$ is generated by its members of the form $y^{\f}-c_{\f}$ for
$\f\in \Z[x]^{n}$ and $c_\f\in \F^{\ast}$. Let $L_\rho = \L(\I)$
which is defined in  \bref{eq-bi1} and  $\rho(\f)=c_\f$. Since $\I$
is proper, $c_{\f}$ is uniquely determined by $\f$. By Lemma
\ref{lm-bi1} and Proposition \ref{cor-bi1}, $\rho$ is a partial
character which is uniquely determined by $\I$. It is clear
$\I(\rho)=\I$.
To show the correspondence is one to one, it suffices to show
$\rho(\I(\rho)) = \rho$ which is a consequence of Lemma \ref{lm-l3}.
The theorem is proved.\qedd


As a summary of this section,  we have the following canonical representations for a proper Laurent binomial $\sigma$-ideal, which follows directly from Theorems \ref{th-rp1}, \ref{th-t4}, \ref{th-l1}.
\begin{thm}\label{cor-bic}
$\I$ is a proper Laurent binomial $\sigma$-ideal if and only if
\begin{description}
\item[(1)] $\I=[\A]$, where $\A$ is a  coherent
Laurent binomial $\sigma$-chain.
\item[(2)]$\I=[\A]$, where $\A=\: \Y^{\f_{1}}-c_1, \ldots, \Y^{\f_{s}}-c_s$,
$\f_i\in\Zxn$, $c_i\in\F^*$, $\fb=\{\f_{1}, \ldots, \f_{s}\}$ is a
reduced Gr\"obner basis of a $\Zx$-lattice, and $[\A]\ne[1]$.
\item[(3)]$\I=\I(\rho)=[\A]$, where $\rho$ is a partial character on $\Zx^n$  and $\A=\A(\rho)$.
\end{description}
Furthermore,  $\A$ is a characteristic set of $\I$ and $(\fb)$ is the support lattice of $\I$.
\end{thm}

\section{Criteria for prime, reflexive, and perfect Laurent binomial $\sigma$-ideals}
\label{sec-lbicr}
In this section, we  give criteria for a
Laurent binomial $\sigma$-ideal to be prime, reflexive,
well-mixed, and perfect in terms of  its support lattice.

\subsection{Reflexive and prime  Laurent binomial $\sigma$-ideals}
\label{sec-pr1}

In this section, we first give criteria for reflexive and prime
Laurent binomial $\sigma$-ideals and then give a decomposition
theorem for perfect Laurent binomial $\sigma$-ideals.
%

For the $\sigma$-indeterminates $\Y=\{y_1,\ldots,y_n\}$ and
$t\in\N$, we will treat the elements of $\Y^{[t]}$ as algebraic
indeterminates, and $\F[\Y^{[\pm t]}]$ is the Laurent polynomial
ring in $\Y^{[t]}$.
Let $\I$ be a Laurent binomial $\sigma$-ideal in $\F\{\Y^{\pm}\}$.
Then it is easy to check that
$$\I_t =\I\cap \F[\Y^{[\pm t]}]$$
is a Laurent binomial ideal in $\F[\Y^{[\pm t]}]$.

Denote $\Zx_t$ to be the set of elements in $\Zx$ with degree $\le
t$. Then $\Zxn_t$ is the $\Z$-module generated by $x^i{\beps}_l$ for
$i=0,\ldots,t,l=1,\ldots,n$. It is clear that $\Zxn_t$ is isomorphic
to $\Z^{n(t+1)}$ as $\Z$-modules by mapping $x^i{\beps}_l$ to the
$((l-1)(t+1) + i+1)$-th standard basis vector in $\Z^{n(t+1)}$. Hence,
we treat them as the same in this section.
Let $L$ be a $\Zx$-lattice and $t\in\N$. Then
 $$L_t =L\cap \Zxn_t =L\cap \Z^{n(t+1)}$$
is a $\Z$-module in $\Z^{n(t+1)}$.
Similarly, it can be shown that when restricted to $\Zxn_t$, a
partial character $\rho$ on $\Zxn$  becomes a partial character
$\rho_t$ on $\Z^{n(t+1)}$.

\begin{lem}\label{lm-pr1}
With the notations introduced above,  we have $\I_t =\I\cap
\F[\Y^{[\pm t]}] =\I(\rho_t)$.
\end{lem}
\proof It suffices to show that the support lattice of $\I_t$ is
$L_{\rho_t} = L_t$. By Lemma \ref{lm-l3}, $\Y^\f-c_m\in\I_t$ if and
only if $\f\in L\cap\Zxn_t$, or equivalently, $\max_{m\in\f}
\deg(m,x)\le t$, which is equivalent to $\f\in L_t$.\qedd
%
%

\begin{defn} Let $L$ be a $\Z[x]$-module in $\Z[x]^n$.
\begin{itemize}
\item $L$ is called {\em $\Z$-saturated} if,  for any
$a\in\Z$ and $\f\in\Z[x]^n$,  $a\f\in L$ implies $\f\in L$.

\item
$L$ is called {\em $x$-saturated} if,  for any $\f\in\Z[x]^n$,
$x\f\in L$ implies $\f\in L$.

\item
$L$ is called {\em saturated} if it is both $\Z$- and $x$-
saturated.
\end{itemize}
\end{defn}

\begin{thm}\label{th-pr1}
Let $\rho$ be a partial character over $\Zxn$. If $\F$ is
algebraically closed and inversive, then
\begin{description}
\item[(a)] $L_\rho$ is $\Z$-saturated if and only if $\I(\rho)$ is prime;
\item[(b)] $L_\rho$ is $x$-saturated if and only if $\I(\rho)$ is reflexive;
\item[(c)] $L_\rho$ is saturated if and only if $\I(\rho)$ is reflexive prime.
\end{description}
\end{thm}
\proof  It is clear that (c) comes from (a) and (b). Let
$\I=\I(\rho)$ and $L=L_\rho$.

$(a)$: $\I$ is a Laurent prime $\sigma$-ideal if and only if $\I_t$
is a Laurent prime ideal for all  $t$.
From Lemma \ref{lm-pr1}, the support of $\I_t$ is $L_t$. Then by
\cite[Thm 2.1]{es-bi}, $\I_t$ is a Laurent prime ideal if and only
if $L_t$ is a $\Z$-saturated $\Z$-module.
Furthermore, a $\Zx$-lattice $L$ is $\Z$-saturated if and only if
$L_t$ is a $\Z$-saturated $\Z$-module for all $t$. Thus, (a) is
valid.

(b): Suppose $\I$ is reflexive. For $x\f\in L$, by Lemma
\ref{lm-l3}, there is a $\Y^{x\f}-c \in \I$. Since $\F$ is
reflexive, $c = d^x$ for $d\in\F$. Then $\sigma(\Y^{\f}-d) \in \I$
and hence $\Y^{\f} - d\in\I$ since $\I$ is reflexive. By Lemma
\ref{lm-l3} again, $\f\in L$ and $L$ is $x$-saturated.
To prove the other direction, assume $L$ is $x$-saturated.
For $f^{x} \in \I$, we have an expression
\begin{equation}\label{eq-pr2}
f^{x}=\sum_{i=1}^{s}f_{i}(\Y^{\f_{i}}-c_i)
\end{equation}
where $\Y^{\f_{i}}-c_i\in\I$ and $f_{i}\in \F\{\Y^{\pm}\}$.
Let $d=\max_{i=1}^s\deg(\Y^{\f_{i}}-c_i,y_1)$ and assume
$\Y^{\f_{1}} =M_1 y_1^d$.
Replace $y_1^d$ by $c_1/M_1$ in \bref{eq-pr2}. Since \bref{eq-pr2}
is an identity for the variables $y_i^{(j)}$, this replacement is
meaningful and we obtain a new identity. $\Y^{\f_{1}}-c_1$ becomes
zero after the replacement. Due to the way to chose $d$, if another
summand, say $\Y^{\f_{2}}-c_2$, is affected by the replacement, then
$\Y^{\f_{2}} = M_2 y_1^d$. After the replacement, $\Y^{\f_{2}}-c_2$
becomes $c_1(M_2/M_1-c_2/c_1)$ which is also in $\I$ by Lemma
\ref{lm-l3}.
In summary, after the replacement, the right hand side of
\bref{eq-pr2} has less than $s$ summands and the left hand side of
\bref{eq-pr2} does not changed. Repeat the above procedure, we will
eventually obtain a new indenity
\begin{equation}\label{eq-pr21}
f^{x}=\sum_{i=1}^{\bar{s}} \bar{f}_{i}(\Y^{x\g_{i}}-\bar{c}_i)
\end{equation}
where $\Y^{x\g_{i}}-\bar{c}_i\in\I$ and $\bar{f}_{i}\in
\F\{\Y^{\pm}\}$.
We may assume that any $y_i$ does not appear in $\bar{f}_{i}$.
Otherwise, by setting $y_i$ to be $1$, the left hand side of
\bref{eq-pr21} is not changes and a new identity is obtained.
Since $\F$ is inversive, $\bar{c}_i = e_i^x$ and $\bar{f}_{i} =
g_i^x$ for $e_i\in\F$ and $g_i\in\F\{\Y^{\pm}\}$.
By Lemma \ref{lm-l3}, $\Y^{x\g_{i}}-e_i^x \in I$ implies $x\g_i\in
L$. Since $L$ is $x$-saturated, $x\g_i\in L$ implies $\g_i\in L$ and
hence $\Y^{\g_{i}}-e_i \in I$ by Lemma \ref{lm-l3} again.
From \bref{eq-pr21}, $\sigma(f-\sum_{i=1}^{\bar{s}}
g_{i}(\Y^{\g_{i}}-e_i))=0$ and hence $f=\sum_{i=1}^{\bar{s}}
g_{i}(\Y^{\g_{i}}-e_i)\in \I$. (b) is proved.\qedd

\begin{defn}
Let $L\subset\Z[x]^{n}$ be a $\Z[x]$-lattice.
The {\em $\Z$-saturation} of $L$ is $\sat_{\Z}(L)=\{\f\in
\Z[x]^{n}\, |\, \exists a \in \Z \st a\f\in L\}$.
The {\em $x$-saturation} of $L$ is $\sat_{x}(L)=\{\f\in \Z[x]^{n}\,
|\, x \f\in L\}$.
The {\em saturation} of $L$ is $\sat(L)=\{\f\in \Z[x]^{n}\, |\,
\exists a\in\Z, \exists k \in \N \st ax^k \f\in L\}$.
\end{defn}
It is clear that the $\Z$-saturation ($x$-saturation) of $L$ is
$\Z$-saturated ($x$-saturated) and
 $$\sat(L) =
\sat_{\Z}(\sat_{x}(L))=\sat_{x}(\sat_{\Z}(L)).$$
%

\begin{thm}\label{lm-ld0}
Let $\I$ be a Laurent binomial $\sigma$-ideal and $L$ the support
lattice of $\I$. If $\F$ is inversive, then the reflexive closure of
$\I$ is also a Laurent binomial $\sigma$-ideal whose support lattice
is the $x$-saturation of $L$.
\end{thm}
\proof Let $\I_x$ be the reflexive closure of $\I$ and
$L_x=\sat_x(L)$. Suppose $\I=[f_1,\ldots,f_r]$, where $f_i =
\Y^{\f_i}-c_i$. Then $L=(\f_1,\ldots,\f_r)$.
If $L$ is $x$-saturated, by Theorem \ref{th-pr1}, $\I$ is reflexive.
Otherwise,  there exist $k_1\in\N$, $b_{i}\in\Zx$, and $\h_1\in\Zxn$
such that $\h_1\not\in L$ and
 \begin{equation}\label{eq-ldec1}
  x^{k_1} \h_1 = \sum_{i=1}^r b_{i} \f_i\in L.\end{equation}
By Lemma \ref{lm-bi1}, $\Y^{x^{k_1} \h_1} - \widetilde{a}$ is in
$\I$, where $\widetilde{a} = \prod_{i=1}^r c_i^{b_{i}}$. Since $\F$
is inversive, $\overline{a}=\sigma^{-k_1}(\widetilde{a})\in\F$.
Then, $\sigma^{k_1} (\Y^{\h_1} - \overline{a})\in\I$, and hence
$\Y^{\h_1} - \overline{a}\in \I_x$.
Let $\I_1=[f_1,\ldots,f_r,\Y^{\h_1} - \overline{a}]$. It is clear
that $L_1=(\f_1,\ldots,\f_r,\h_1)$ is the support lattice of $\I_1$.
Then $\I\varsubsetneq\I_1\subset \I_x$ and $L\varsubsetneq
L_1\subset L_x$.
Repeating the above procedure for $\I_1$ and $L_1$, we obtain $\I_2$
and $L_2=(\f_1,\ldots,\f_r,\h_1,\h_2)$ such that $\h_2\not\in L_1$
and $x^{k_2}\h_2\in L_1$.
We claim that $L_2\subset L_x$. Indeed, let $x^{k_2}\h_2 =
\sum_{i=1}^r e_i \f_i + e_0 \h_1$. Then by \bref{eq-ldec1},
$x^{k_1+k_2}\h_2=x^{k_1} (x^{k_2}\h_2)
 =x^{k_1}\sum_{i=1}^r e_i \f_i + e_0 (x^{k_1}\h_1)
 =x^{k_1}\sum_{i=1}^r e_i \f_i + e_0 \sum_{i=1}^r b_i \f_i\in L$ and
 the claim is proved.
As a consequence, $\I_2\subset \I_x$.

Continuing the process, we have
 $\I\varsubsetneq\I_1\varsubsetneq\cdots\varsubsetneq\I_t\subset \I_x$ and $L\varsubsetneq
L_1\varsubsetneq\cdots\varsubsetneq L_t\subset L_x$ such that $L_i$
is the support lattice of $\I_i$.
The process will terminate, since $\Zxn$ is Northerian. The final
$\Zx$-lattice $L_t$ is $x$-saturated and hence $\I_t$ is reflexive
by Theorem \ref{th-pr1}. Since $L_x$ is the smallest $x$-saturated
$\Zx$-lattice containing $L$ and $L\subset L_t\subset L_x$, we have
$L_t=L_x$ and $\I_t=\I_x$.\qedd

\begin{cor}\label{cor-ld0}
Let $L\subset\Z[x]^{n}$ be a $\Z[x]$-lattice.
Then $\rk(L)=\rk(\sat_x(L))$ and $\rk(L)=\rk(\sat_\Z(L))$.
%
\end{cor}
\proof From the proof of Theorem \ref{lm-ld0}, $\sat_x(L)=
(L,\h_1,\ldots,\h_t)$ and  for each $\h_i$, there is a positive
integer $n_i$ such that $x^{n_i}\h_i\in L$.
Let $A$ be a representation matrix of $L$. Then a representation
matrix $B$ of $L_x$ can be obtained by adding to $A$ a finite number
of new columns which are linear combinations of columns of $A$
divided by some $x^d$. Therefore, $\rk(A)=\rk(B)$.
We can prove $\rk(L)=\rk(\sat_\Z(L))$ similarly. \qedd

%
%
%

We now give a decomposition theorem for perfect $\sigma$-ideals.
\begin{thm}\label{th-l2}
Let $\I$ be a Laurent binomial $\sigma$-ideal,  $L$ the support
lattice of $\I$, and $L_S$ the saturation of $L$. If $\F$ is
algebraically closed and inversive,  then $\{\I\}$ is either $[1]$
or can be written as the intersection of Laurent reflexive prime
binomial $\sigma$-ideals whose support lattice is $L_S$.
\end{thm}
\proof  Let $\I_x$ be the reflexive closure of $\I$ and
$L_x=\sat_x(L)$. By Theorem \ref{lm-ld0}, $L_x$ is the support lattice
of $\I_x$.
Suppose $\I_x=[f_1,\ldots,f_r]$, $f_i = \Y^{\f_i} -
c_i,i=1,\ldots,r$, and $L_x=(\f_1,\ldots,\f_r)$. If $L_x$ is
$\Z$-saturated, then by Theorem \ref{th-pr1}, $\I_x$ is reflexive
prime. Otherwise, there exist $k_1\in \N$, $a_i\in\Z[x]$, and
$\h_1\in \Z[x]^n$ such that\ $\h_1\not\in L_x$ and
 \begin{equation}\label{eq-ldec2}
  k_1\h_1 = a_1\f_1 + \cdots + a_r \f_r \in L_x.
 \end{equation}
By Lemma \ref{lm-bi1}, $\Y^{k_1\h_1}- \widetilde{a}\in\I$, where
$\widetilde{a}=\prod_{i=1}^r c_i^{a_i}$. Since $\F$ is algebraically
closed,
$$\Y^{k_1\h_1}- \widetilde{a}= \prod_{l=1}^{k_1} (\Y^{\h_1}-\widetilde{a}_{l})\in\I_x$$
where $\widetilde{a}_{l}, l=1, \ldots, k_1$ are the $k_1$ roots of
$\widetilde{a}$. By the difference Nullstellensatz
\cite[p.87]{cohn}, we have the following decomposition
$$\{\I\} = \cap_{l_1=1}^{k_1} \{\I_{1l_1}\}$$
where $\I_{1l} = [f_1,\ldots,f_r, \Y^{\h_1}-\widetilde{a}_{l}]$.
Check whether $\I_{1l_1}=[1]$ with Proposition \ref{lm-bi2} and discard
those trivial ones. Then the support lattice for any of $\I_{1l}$ is
$L_1 =(\f_1,\ldots,\f_r,\h_1)$. Similar to the proof of Theorem
\ref{lm-ld0}, we can show that $\I_x\varsubsetneq\I_{1l}$ and
$L_x\varsubsetneq L_1 \subset  L_S$.

Repeating the process, we have
 $\I_x\varsubsetneq\I_{1l_1}\varsubsetneq\cdots\varsubsetneq\I_{tl_t}$ for $l_i=1,\ldots,k_i$
and $L_x\varsubsetneq L_1\varsubsetneq
L_2\varsubsetneq\cdots\varsubsetneq L_t\subset L_S$ such that $L_i$
is the support lattice of $\I_{il_i}$ for $l_i=1,\ldots,k_i$ and
 $$\{\I\}=\cap_{l_i=1}^{k_i}\{\I_{il_i}\}, i=1,\ldots,t.$$
The process will terminate, since $\Zxn$ is Northerian. Since $L_S$
is the smallest $\Z$-saturated $\Zx$-lattice containing $L_x$ and
$L_x\subset L_t\subset L_S$, we have $L_t=\sat_{\Z}(L_x)
=\sat_{\Z}(\sat_{x}(L)) = L_S$. Then $\I_{tl_t}$ is reflexive prime
and the theorem is proved. \qedd

Since the reflexive prime components of $\I$ have the same support
lattice, by Corollary \ref{cor-dim}, they also have the same dimension.
\begin{cor}\label{cor-l2}
Any Laurent binomial $\sigma$-ideal $\I$ is dimensionally unmixed.
\end{cor}
%


\subsection{Well-mixed and perfect Laurent binomial $\sigma$-ideals}
\label{sec-pr3}

In this section, we give criteria for a Laurent binomial
$\sigma$-ideal to be well-mixed and perfect in terms of its support
lattice and show that the well-mixed and perfect closures of a
Laurent binomial $\sigma$-ideal are still binomial.

For $S\subset\F\{\Y^{\pm}\}$, let $S' =
\{ fg^x | fg\in S \}$. We define inductively: $S_0 = S, S_n =
[S_{n-1}]', n=1, 2, \ldots$. The union of the $S_n$ is clearly a
well-mixed $\sigma$-ideal and is contained in every well-mixed
$\sigma$-ideal containing $S$. Hence this union is $\langle
S\rangle$. If $\I\subset\F\{\Y^{\pm}\}$ is a Laurent $\sigma$-ideal,
then $\langle\I\rangle$ ia called the {\em well-mixed closure} of $\I$.
We first prove some basic properties of well-mixed $\sigma$-ideals. Note that these properties are also valid in $\F\{\Y\}$.

\begin{lem}\label{lm-wm}
Let $\I_1, \ldots,\I_s$ be prime $\sigma$-ideals.
Then $\I = \cap_{i=1}^s \I_i$ is a well-mixed $\sigma$-ideal.
\end{lem}
\proof It is obvious.\qedd

\begin{lem}\label{lm-wm1}
Let $S_1, S_2$ be two subsets of $\F\{\Y^{\pm}\}$ which satisfy $a\in S_i$ implies $\sigma(a)\in S_i, i=1,2$. Then $[S_1]_n [S_2]_n \subset [S_1S_2]_{n}$.
\end{lem}
\proof Let $s\in [S_1]_1$ and $t\in [S_2]_1$. Then $s= f_1g_1^x$ and
$t=f_2g_2^x$ where $f_1g_1\in[S_1], f_2g_2\in[S_2]$. Then, $f_1g_1f_2g_2\in
[S_1S_2]$, and $st=f_1f_2(g_1g_2)^x\in [S_1S_2]_1$. Hence,
$[S_1]_1[S_2]_1 \subset [S_1S_2]_1$. By induction, $[S_1]_n [S_2]_n
\subset [S_1S_2]_{n}$. \qedd

\begin{lem}\label{lm-wm2}
Let $S_1, S_2$ be two subsets of $\F\{\Y^{\pm}\}$ which satisfy
$a\in S_i$ implies $\sigma(a)\in S_i, i=1,2$. Then
$\sqrt{[S_1S_2]_n} = \sqrt{[S_1]_n\cap [S_2]_n}$ for $n\ge 1$, and
$\sqrt{\langle S_1 \rangle} \cap \sqrt{\langle S_2 \rangle} =
\sqrt{\langle S_1S_2 \rangle}$.
\end{lem}
\proof  The last statement is an immediate consequence of the first one. Since $[S_1S_2] \subset [S_i]$, we have $[S_1S_2]_n \subset [S_i]_n$ for $i=1,2$, and $[S_1S_2]_n \subset [S_1]_n\cap [S_2]_n$ follows. Hence, $\sqrt{[S_1S_2]_n} \subset \sqrt{[S_1]_n\cap [S_2]_n}$. Let $a\in [S_1]_n\cap [S_2]_n$ we have
$a^2 \in [S_1]_n[S_2]_n$. By Lemma~\ref{lm-wm1}, $a^2 \in [S_1S_2]_{n} $. Hence $a\in \sqrt{[S_1S_2]_n}$, and $ \sqrt{[S_1]_n\cap [S_2]_n} \subset \sqrt{[S_1S_2]_n}$ follows.
\qedd

\begin{lem}\label{lm-wm3}
Let $\I_1,\ldots,\I_m$ be Laurent $\sigma$-ideals.\newline
 Then
$\sqrt{\langle \cap_{i=1}^m \I_i \rangle} = \cap_{i=1}^m
\sqrt{\langle \I_i \rangle} $.
\end{lem}
\proof  Let $\I = \cap_{i=1}^m \I_i$. Then $\sqrt{\I} = \sqrt{[\prod_{i=1}^m\I_i]}$.
By Lemma~\ref{lm-wm2}, we have $\sqrt{\langle \prod_{i=1}^m\I_i \rangle } = \sqrt{\prod_{i=1}^{n-1} \langle \I_i \rangle} \cap \sqrt{\langle \I_n \rangle} = \ldots = \cap_{i=1}^m \sqrt{\langle \I_i \rangle}$. Now we show that $\sqrt{ \langle \I \rangle } = \sqrt{\langle \prod_{i=1}^m\I_i \rangle  }$. Since $\prod_{i=1}^m\I_i \subset \I$, we have $\sqrt{\langle \prod_{i=1}^m\I_i \rangle} \subset \sqrt{ \langle \I \rangle }$. By Lemma~\ref{lm-wm2},  $\sqrt{\langle I \rangle} = \sqrt{\langle I \rangle} \cap \cdots\cap \sqrt{\langle I \rangle} =  \sqrt{\langle I^m \rangle} \subset \sqrt{ \langle \prod_{i=1}^m\I_i \rangle }$, and hence $\sqrt{ \langle \I \rangle } = \sqrt{\langle \prod_{i=1}^m\I_i \rangle  }$. Then, $\sqrt{ \langle \I \rangle } =  \cap_{i=1}^m \sqrt{\langle \I_i \rangle}$.
\qedd

Now, we prove a basic property for a $\sigma$-field $\F$.
\begin{lem}\label{lm-per1}
Let $\zeta_m = e^{\frac{2\pi \ix}{m}}$ be the primitive $m$-th root
of unity, where $\ix=\sqrt{-1}$ and $m\in\Z_{\ge2}$. If $\F$ is
algebraically closed, then there exists an
$o_m\in[0,m-1]$ such that $\gcd(o_m,m)=1$ and
$\sigma(\zeta_m)=\zeta_m^{o_m}$. Furthermore, the perfect
$\sigma$-ideal $\{y^m-1\}$ in $\F\{y\}$ is
 \begin{equation}\label{eq-per1}
 \{y^m-1\}=[y^m-1,y^{x}-y^{o_m}]\end{equation}
where $y$ is a $\sigma$-indeterminate.
\end{lem}
\proof Since  $\F$ is algebraically closed, $\zeta_m$ is in $\F$.
From $y^m-1=\prod_{j=0}^{m-1} (y-\zeta_m^j)=0$, we have
$\sigma(y)^m -1=\prod_{j=0}^{m-1} (\sigma(y)-\zeta_m^j)=0$.
Then, there exists an $o_m$ such that $0\le o_m\le m-1$ and
$\sigma(\zeta_m) = \zeta_m^{o_m}$.
Suppose $\gcd(o_m,m)=d>1$ and let $o_m=dk, m=ds$, where
$s\in[1,m-1]$. Then
$\sigma(\zeta_m^s)=\zeta_m^{o_ms}=\zeta_m^{dks}=\zeta_m^{km}=1$,
which implies $\zeta_m^s=1$, a contradiction.

By the difference Nullstellensatz \cite[p.87]{cohn}, we have
$\{y^m-1\}=\cap_{j=0}^{m-1} [y-\zeta_m^j]$. In order to show
\bref{eq-per1}, it suffices to show
 $\cap_{j=0}^{m-1} [y-\zeta_m^j]=[y^m-1,y^x-y^{o_m}]$.
 Since $y^{x}-y^{o_m} = (y- \zeta_m^j)^x +
\zeta_m^{xj} - y^{o_m} = (y- \zeta_m^j)^x + \zeta_m^{jo_m} - y^{o_m}
\in[y -\zeta_m^j]$ for any $0\le j\le m-1$, we have $y^{x}-y^{o_m}
\in \cap_{j=0}^{m-1} [y- \zeta_m^j]$ and hence
$[y^m-1,y^x-y^{o_m}]\subset \cap_{j=0}^{m-1} [y-\zeta_m^j]$.
Let $f\in\cap_{j=0}^{m-1} [y-\zeta_m^j]$. Since
$y^{x}-y^{o_m}\in[y-\zeta_m^j]$,  for $j=0,\ldots,m-1$, from
$f\in[y-\zeta_m^j]$, we have $f= g_j(y-\zeta_m^j)+\sum_k
h_{jk}(y^{x}-y^{o_m})^{x^k}$, where $g_j,h_{jk}\in\Q\{y\}$.
Then $f^m = \prod_{j=0}^{m-1} (g_j(y-\zeta_m^j)+\sum_k
h_{jk}(y^{x}-y^{o_m})^{kx}) = \prod_{j=0}^{m-1}g_j(y^m-1)+p$, where
$p\in[y^{x}-y^{o_m}]$. Hence, $f\in[y^m-1,y^x-y^{o_m}]$ and
$\cap_{j=0}^{m-1} [y-\zeta_m^j]\subset[y^m-1,y^x-y^{o_m}]$. The
lemma is proved.\qedd

The number $o_m$ introduced in Lemma \ref{lm-per1} depends on $\F$
only and is called the {\em $m$-th transforming degree} of unity. In
the following corollaries, $\F$ is assumed to be algebraically
closed and hence $o_m$ is fixed for any $m\in\N$. From the proof of
Lemma \ref{lm-per1}, we have

\begin{cor}\label{cor-per1}
$y^{x}-y^{o_m}\in \cap_{j=0}^{m-1} [y-\zeta_m^j]$.
\end{cor}

\begin{cor}\label{cor-per2}
For $n,m,k$ in $\N$, if $n=km$ then $o_n = o_m\,\mod\, m$.
\end{cor}
\proof By definition, $\zeta_n^{k} =\zeta_m$. Then,
$\sigma(\zeta_n^{k})=\zeta_n^{ko_n}= \zeta_m^{o_n}$. From,
$\sigma(\zeta_n^{k})=\sigma(\zeta_m)=\zeta_m^{o_m}$, we have
$\zeta_m^{o_n}=\zeta_m^{o_m}$. Then $o_n=o_m\,\mod\, m$.\qedd

\begin{lem}\label{lm-per0}
$\langle y^m-1\rangle=\{y^m-1\}=[y^m-1,y^{x}-y^{o_m}]$.
\end{lem}
\proof By Lemma~\ref{lm-per1}, it suffices to show
$y^{x}-y^{o_m} \in \langle y^m-1\rangle$. Since
$y^m-1 = \prod_{j=0}^{m-1} (y-\zeta_m^j)$ and $(y-\zeta_m^{i})^x=(y^x-\zeta_m^{o_m i})$, we have
$f_i = (y^x-\zeta_m^{o_m i})\prod_{0\le j \le m-1, j\ne i} (y-\zeta_m^j) \in \langle y^m-1\rangle$
for $i=0, \ldots, m-1$.
We will show that $y^{x}-y^{o_m} \in (f_0, \ldots, f_{m-1})$.
To show this, we need the formula  $\frac{1}{y^m-1} = \sum_{i=0}^{m-1} \frac{1}{m(\zeta_m^i)^{m-1}(y-\zeta_m^i)}  = \frac{1}{m}\sum_{i=0}^{m-1}\frac{\zeta_m^{i}}{y-\zeta_m^i}$ from \cite[p. 494]{GCG1992}.
We have \begin{eqnarray*}
\frac{1}{m}\sum_{i=0}^{m-1} \zeta_m^{i}f_i  & =& \frac{1}{m}\sum_{i=0}^{m-1} \zeta_m^{i}\frac{y^m-1}{y-\zeta_m^i} (y^x-\zeta_m^{o_m i}) \\
& =& \frac{1}{m}\sum_{i=0}^{m-1} \zeta_m^{i}\frac{y^m-1}{y-\zeta_m^i} y^x - \frac{1}{m}\sum_{i=0}^{m-1} \zeta_m^{i}\frac{y^m-1}{y-\zeta_m^i}\zeta_m^{o_m i} \\
& =& y^x - \frac{1}{m}\sum_{i=0}^{m-1} \frac{y^m-1}{y-\zeta_m^i}\zeta_m^{(o_m+1)i}.
\end{eqnarray*}
Let $g(y) = \frac{1}{m}\sum_{i=0}^{m-1} \frac{y^m-1}{y-\zeta_m^i}\zeta_m^{(o_m+1)i}$.
Then, $g(\zeta_m^j) = \frac{1}{m}\zeta_m^{(o_m+1)j} \frac{y^m-1}{y-\zeta_m^j}|_{y=\zeta_m^j} $ $=  \frac{1}{m}\zeta_m^{(o_m+1)j} $\,\,\,\, $\prod_{0\le i\le m-1, i\ne j}$ $(\zeta_m^i-\zeta_m^j) = \frac{1}{m}\zeta_m^{(o_m+1)j}\zeta_m^{j(m-1)}\prod_{1\le i\le m-1} (\zeta_m^i -1) = \frac{1}{m}\zeta_m^{o_m j}m =(\zeta_m^j)^{o_m}$. Since $\deg(g(y)) \le m-1$ and $g(\zeta_m^j) = (\zeta_m^j)^{o_m}$ for $j=0,\ldots,m-1$, we have $g(y) = y^{o_m}$.
Hence $y^x-y^{o_m} \in (f_0, \ldots, f_{m-1}) \subset \langle y^m-1\rangle$.
\qedd

\begin{cor}\label{cor-per3}
For $m\in\N$, $a\in\F^*$, and $\f\in\Zxn$, we have
$\Y^{(x-o_m)\f}-a^{x-o_m} \in \langle\Y^{m\f} - a^m \rangle$.
\end{cor}
\proof Let $z = \frac{\Y^{\f}}{a}$ and $\I=[\Y^{m\f} - a^m]$. Then
$z^{m} - 1 \in \I$. By Lemma \ref{lm-per0},  $z^{x-o_m}-1
\in\langle z^m-1\rangle\subset \langle \I \rangle$.
Then $(\frac{\Y^\f}{a})^{x-o_m}-1 \in \langle\I\rangle$ or
$\Y^{(x-o_m)\f}-a^{x-o_m} \in \langle \I \rangle$.\qedd

The following example shows that the generators of a well-mixed or
perfect ideal depend on the difference field $\F$.
\begin{exmp}\label{ex-per1}
%
Let $\F=\Q(\sqrt{-3})$ and $p=y_1^{3}-1$. Following Lemma
\ref{lm-per1}, if $\sigma(\sqrt{-3})=\sqrt{-3}$, we have $o_3=1$ and
$\langle p\rangle =\{p\}=[p,y_1^x - y_1]$.
If $\sigma(\sqrt{-3})=-\sqrt{-3}$, we have $o_3=2$ and
$\langle p\rangle =\{p\}=[p,y_1^x - y_1^2]$.
%
%
%
%
%
%
%
\end{exmp}

Motivated by Corollary \ref{cor-per3}, we have the following
definition.
\begin{defn}
If a $\Zx$-lattice $L$ satisfies
  \begin{equation}\label{eq-wmc}
  m\f \in L \Rightarrow (x-o_m)\f\in  L\end{equation}
where $m\in\N$, $\f\in\Zxn$, and  $o_m$ is defined in Lemma
\ref{lm-per1}, then it is called {\em {\rm{M}}-saturated}. For any
$\Zx$-lattice $L$, the smallest {\rm{M}}-saturated $\Zx$-lattice
containing $L$ is called the {\rm{M}}-saturation of $L$ and is
denoted by $\sat_M(L)$.
\end{defn}

The following result gives an effective version for condition
\bref{eq-wmc}.
\begin{lem}\label{lm-wmc}
A $\Zxn$-lattice $L$ is M-saturated if and only
if the following condition is true:
Let $L_1 = \sat_\Z (L) = (\g_1,\ldots,\g_s)$  such that
$m_i\g_i\in L$ for $ m_i\in\N$. Then $(x-o_{m_i})\g_i\in
L$.
\end{lem}
\proof We need only to show $(x-o_{m_i})\g_i\in L$ implies
\bref{eq-wmc}.
For any $ m\f \in L$, we have $\f\in L_1$ and hence $\f =
\sum_{i=1}^r q_i\g_i$, where $q_i\in\Zx$. Let
$t=\lcm(m,m_1,\ldots,m_s)$. By Corollary \ref{cor-per2}, we have
$o_t=o_{m_i} +c_i m_i$, where $c_i\in\Z$.
Then $(x-o_t)\f = \sum_{i=1}^r q_i(x-o_t)\g_i = \sum_{i=1}^r
q_i(x-o_{m_i})\g_i - \sum_{i=1}^r q_ic_i m_i\g_i \in L$.
By Corollary \ref{cor-per2}, $o_t = o_m + c m$, where $c\in\Z$. Then
$(x-o_m)\f = (x-o_t)\f +c m\f\in L$.\qedd

We now give a criterion for a Laurent binomial $\sigma$-ideal to be
well-mixed.
\begin{thm}\label{th-wm}
Let $\rho$ be a partial character and  $\F$ an algebraically closed
and inversive $\sigma$-field. If $\I(\rho)$ is well-mixed, then
$L_\rho$ is {\rm{M}}-saturated. Conversely, if $L_\rho$ is
{\rm{M}}-saturated, then either $\langle \I(\rho)\rangle=[1]$ or $\I(\rho)$ is
well-mixed.
\end{thm}
\proof Suppose that $\I(\rho)$ is well-mixed.
If there exists an $m\in\N$ such that $m\f \in
L_\rho$, then by Lemma \ref{lm-l3}, there exists a $c\in\F^*$ such
that $\Y^{m\f} - c\in\I(\rho)$. Since $\F$ is algebraically closed,
there exists an $a\in\F^*$ such that $c = a^m$. Then, $\Y^{m\f} -
a^m\in\I(\rho)$.
Since $\I(\rho)$ is well-mixed, by Corollary \ref{cor-per3},
$\Y^{(x-o_m)\f}-a^{x-o_m} \in \I(\rho)$, and by Lemma \ref{lm-l3}
again, $(x-o_m)\f\in L_\rho$ follows and $L_\rho$ is {\rm{M}}-saturated.

Conversely, let $L_\rho$ be {\rm{M}}-saturated. If $L_\rho$ is
$\Z$-saturated, then by Theorem \ref{th-pr1}, $\I(\rho)$ is
prime and hence well-mixed by Lemma~\ref{lm-wm}.
Otherwise, there exists an
$m_1\in \N$, and $\f\in \Z[x]^n$ such that $\f\not\in L_\rho$ and
$m_1\f\in L_\rho$.
By Lemma \ref{lm-l3}, there exists an $a\in\F^*$ such that
$\Y^{m_1\f} - a^{m_1}\in\I(\rho)$.  We claim that either
$\langle \I(\rho) \rangle=[1]$ or
 \begin{equation}\label{eq-wmt1}\I(\rho) = \cap_{l_1=0}^{m_1-1} \I_{l_1}\end{equation}
where $\I_{l_1} = [\I(\rho), \Y^{\f}-a\zeta_{m_1}^{l_1}]$ and
$\zeta_{m_1} = e^{\frac{2\pi \ix}{m_1}}$.
By \bref{eq-wmc}, $(x-o_{m_1})\f\in L_\rho$. By Lemma \ref{lm-l3},
there exists a $b\in\F^*$ such that $\Y^{(x-o_{m_1})\f} -
b\in\I(\rho)$.
Since $\Y^{m_1\f} - a^{m_1}\in\I(\rho)$, by Corollary
\ref{cor-per1}, we have $\Y^{(x-o_{m_1})\f} -
a^{x-o_{m_1}}\in[\Y^{\f}-a\zeta_{m_1}^{l_1}]$ for any $l_1$. Then
$b-a^{x-o_{m_1}}= \Y^{(x-o_{m_1})\f} - a^{x-o_{m_1}} -
(\Y^{(x-o_{m_1})\f} - b) \in\I_{l_1}$ for any $l_1$.
If $b\ne a^{x-o_{m_1}}$, $\I_{l_1}=[1]$ for all $l_1$, and hence $1\in
\cap_{l_1=0}^{m_1-1} \I_{l_1} \subset \langle \I(\rho) \rangle $ by Lemma~\ref{lm-per0}
and $\langle \I(\rho) \rangle=[1]$ follows.
Now suppose $b=a^{x-o_{m_1}}$ or $a^{x}=ba^{o_{m_1}}$. To prove \bref{eq-wmt1},
it suffices to show  $\cap_{l_1=0}^{m_1-1} \I_{l_1} \subset \I(\rho)$.
Let $f\in \cap_{l_1=0}^{{m_1}-1} \I_{l_1}$. From $f\in\I_{l_1}$, we
have $f = f_{l_1} + \sum_{j=0}^s
p_{j}\sigma^j(\Y^{\f}-a\zeta_{m_1}^{l_1})$, where $f_{l_1}\in\I(\rho)$.
By Lemma \ref{lm-per1}, $\sigma(\zeta_{m_1})=\zeta_{m_1}^{o_{m_1}}$.
We thus have
\begin{eqnarray*}
&&\sigma(\Y^{\f}-a\zeta_{m_1}^{l_1})
 = \Y^{x\f} - b\Y^{o_{m_1}\f} + b\Y^{o_{m_1}\f}- \sigma(a\zeta_{m_1}^{l_1})\\
 &=& \Y^{o_{m_1}\f}(\Y^{(x-o_{m_1})\f}-b) + b(\Y^{o_{m_1}\f}- a^{o_{m_1}} \zeta_{m_1}^{l_1o_{m_1}})
 + (ba^{o_{m_1}}  -\sigma(a))\zeta_{m_1}^{l_1o_{m_1}}.
\end{eqnarray*}
Since  $\Y^{(x-o_{m_1})\f}-b\in\I(\rho)$ and
 $ba^{o_{m_1}}-\sigma(a)=ba^{o_{m_1}}-a^x=0$, we have $\sigma(\Y^{\f}-a\zeta_{m_1}^{l_1}) =
g_{l_1} + q_{l_1}(\Y^{\f}- a\zeta_{m_1}^{l_1})$, where
$g_{l_1}\in\I(\rho)$. Using the above equation repeatedly, we have
$f = h_{l_1} + p_{l_1}(\Y^{\f}- a\zeta_{m_1}^{l_1})$, where
$h_{l_1}\in\I(\rho)$.
Then, $f^{m_1}= \prod_{l_1=0}^{{m_1}-1}(h_{l_1}+p_{l_1}(\Y^{\f}-
a\zeta_{m_1}^{l_1})) = s +
\prod_{l_1=0}^{{m_1}-1}p_{l_1}(\Y^{\f}-a\zeta_{m_1}^{l_1}) = s +
(\Y^{{m_1}\f}-a^{m_1})\prod_{l_1=0}^{{m_1}-1}p_{l_1}\in \I(\rho)$,
where $s$ is in $\I(\rho)$. By Corollary~\ref{cor-rad}, we have
$f\in \I(\rho)$. The claim is proved.

The support lattice for any of $[\I_{l_1}]$ is $L_1 =(L_\rho,\f)$.
Similar to the proof of Theorem \ref{lm-ld0}, we can show that
$\I(\rho)\varsubsetneq\I_{l_1}$ and $L_\rho\varsubsetneq L_1$.
If $L_1$ is not $\Z$-saturated, there exists a $k>1$ and $\g\in\Zxn$
such that $\g\not\in\L_1$ and $k\g\in L_1$. Let $m_2=k{m_1}$. We
have $m_2\g=k{m_1}\g\in L_\rho$ and there exists a $c\in\F^*$
 such that $\Y^{m_2\g}-c^{m_2}\in\I(\rho)$.
Hence, $(x-o_{m_2})\g \in L_\rho\subset L_1$ and there exists a
$d\in\F^*$, such that $\Y^{(x-o_{m_2})\g}-d\in I(\rho)$.
Let $L_2 = (L_1, \g)$ and
$\I_{l_1,l_2}=[\I_{l_1},\Y^{\g}-c\zeta_{m_2}^{l_2}]$,
$l_2=0,\ldots,m_2-1$. Then $L_1\varsubsetneq L_2$ and $L_2$ is the
support lattice for all $\I_{l_1,l_2}$ provided
$\I_{l_1,l_2}\ne[1]$. Similar to the above, it can be shown that
$d-c^{x-o_{m_2}} \in \I_{l_1,l_2}$ for any $l_1,l_2$.
If $d-c^{x-o_{m_2}}\ne0$, then $\I_{l_1,l_2}=[1]$ for any $l_1,l_2$
and $\langle \I_{l_1}\rangle =[1]$ by Lemma~\ref{lm-per0}.
Since Laurent binomial $\sigma$-ideals are radical, $\langle
\I(\rho)\rangle = \cap_{l_1=0}^{m_1-1} \langle \I_{l_1}\rangle =[1]$
by Lemma~\ref{lm-wm3} and \bref{eq-wmt1}.
If $d-c^{x-o_{m_2}}=0$, it can be similarly proved that
 $\I_{l_1} = \cap_{l_2=0}^{m_2-1}\I_{l_1,l_2}$ for any $l_1$.
%
%
As a consequence, we have either $\langle I(\rho)\rangle =[1]$ or
$\I(\rho)=\cap_{l_1=0}^{m_1-1} \I_{l_1} =
\cap_{l_1=0}^{m_1-1}\cap_{l_2=0}^{m_2-1} \I_{l_1,l_2}.$

Repeating the process, we have either $\langle I(\rho)\rangle =[1]$ or
 $$\I(\rho)=\cap_{l_1=0}^{m_1-1} \I_{l_1}= \cdots = \cap_{l_1=0}^{m_1-1}\cdots\cap_{l_t=0}^{m_t-1} \I_{l_1,\ldots,l_t}$$
where $L_\rho\varsubsetneq L_1\varsubsetneq\cdots\varsubsetneq L_t \subset sat_\Z(L_\rho)$.
 Since $\Zxn$ is Notherian, the procedure will terminate and
$L_t$ is  $\Z$-saturated. Since each $\I_{l_1,\ldots,l_t}$ is
either $[1]$ or a prime $\sigma$-ideal, and hence
either $\langle \I(\rho)\rangle=[1]$ or $ \I(\rho)$ is well-mixed by Lemma~\ref{lm-wm}. \qedd

The following example shows that $\langle\I(\rho)\rangle=[1]$ can
indeed happen in Theorem \ref{th-wm}.
\begin{exmp}\label{ex-311}
Let $\I=[\A]$,  where $\A = \{y_1^2+1, y_1^x-y_1, y_2^2+1,
y_2^x+y_2\}$ is a $\sigma$-chain. The support lattice of
$\I$ is M-saturated.  We have $y_2^2-y_1^2=y_2^2+1-(y_1^2+1)\in\I$.
Then by Corollary \ref{cor-per3},
$y_1y_2^x-y_1^xy_2\in\langle\I\rangle$. From $y_1^x-y_1,
y_2^x+y_2\in\I$, we have $y_1y_2\in\langle\I\rangle$ and hence $1\in
\langle\I\rangle$. This also shows that a binomial $\sigma$-ideal is
generally not well-mixed.
\end{exmp}


\begin{thm}\label{th-wm2}
Let $\F$ be an algebraically closed and inversive $\sigma$-field and
$\I=\I(\rho)$ a Laurent binomial $\sigma$-ideal. Then the well-mixed
closure of $\I$ is either $[1]$ or a Laurent binomial $\sigma$-ideal
whose support lattice is $\sat_M(L_\rho)$.
\end{thm}
\proof
Suppose that $\langle\I(\rho)\rangle\ne[1]$.
If $L$ is not M-saturated, then there exists an $m\in\N$ and
$\f\in\Zxn$ such that $\f\not\in L$, $m\f\in L$, and
$(x-o_m)\f\not\in L$.
By Lemma \ref{lm-l3}, there exists a $c\in\F^*$ such that $\Y^{m\f}
- c^m\in\I(\rho)$.
Let $\I_1 =[\I, \Y^{(x-o_m)\f}-c^{x-o_m}]$ and $L_1 = (L,(x-o_m)\f)$. By
Corollary \ref{cor-per3}, $\Y^{(x-o_m)\f}-c^{x-o_m}\in\langle \I(\rho)\rangle$.
Let $L_M=\sat_M(L)$. Then $\I\varsubsetneq \I_1\subset\langle\I\rangle$
and $L\varsubsetneq L_1\subset L_M$.
Repeat the procedure to construct $I_i$ and $L_i$ for $i=2,\ldots,t$
such that $\I\varsubsetneq
\I_1\varsubsetneq\cdots\varsubsetneq\I_t \subset\{\I\}$ and
$L\varsubsetneq L_1\varsubsetneq\cdots\varsubsetneq L_t\subset
L_M$. Since $\Zxn$ is Notherian, the procedure will terminate at,
say $t$. Then $L_t=L_M$ is M-saturated. By Lemma \ref{lm-pxsat},
$L_t$ is also $x$-saturated. By Theorem \ref{th-wm},
$\I_t\subset\langle\I\rangle$ is well-mixed and hence $\I_t=\langle\I\rangle$. \qedd

By the proof of Theorem \ref{th-wm2},
we have
\begin{cor}
A $\Zx$-lattice and its M-saturation have the same rank.
%
\end{cor}

\begin{exmp}\label{ex-perc1}
 Let $p=y_2^{2}-y_1^{2}$. Following the proof of Theorem \ref{th-wm2},
 it can be shown that
 $\langle p\rangle=\{p\}=[y_1^{-2}y_2^{2}-1,y_1^{1-x}y_2^{x-1}-1]=[y_2^{2}-y_1^{2},y_1y_2^{x}-y_1^{x}y_2]$.
\end{exmp}

\vskip10pt
In the rest of this section, we prove similar results for
the perfect closure of Laurent binomial $\sigma$-ideals.
We first give a definition.

\begin{defn}
If a $\Zx$-lattice is both $x$-saturated and M-saturated, then it is
called {\em {\rm{P}}-saturated}. For any  $\Zx$-lattice $L$, the
smallest {\rm{P}}-saturated $\Zx$-lattice containing $L$ is called
the {\rm{P}}-saturation of $L$ and is denoted by $\sat_P(L)$.
\end{defn}

\begin{lem}\label{lm-pxsat}
For any  $\Zx$-lattice $L$, $\sat_P(L)=\sat_x(\sat_M(L))=\sat_M(\sat_x(L))$.
\end{lem}
\proof Let $L_1=\sat_x(\sat_M(L))$ and $L_2 = \sat_M(\sat_x(L))$.
It suffices to show $L_1=L_2$.
We claim that $L_1$ is P-saturated. Let
$m\f\in L_1$ for $m\in\N$. Then $mx^a\f\in\sat_M(L)$ for some
$a\in\N$, which implies $(x-o_m)x^a\f\in L\subset
\sat_x(\sat_M(L))=L_1$. Since $L_1$ is $x$-saturated, $(x-o_m)\f\in
L_1$ and the claim is proved.
Since $L\subset \sat_M(L)$, $\sat_x(L)\subset\sat_x(\sat_M(L))=L_1$.
From the claim, $L_1$ is P-saturated and hence
$L_2\subset\sat_M(L_1)=L_1$.

For the other direction, we claim that  $L_2$ is
$x$-saturated. Let $x\f\in
\sat_M(\sat_x(L))\subset\sat_\Z(\sat_x(L))$. Then there exists an
$m\in\N$, such that $m\f\in \sat_x(L)$ which implies $(x-o_m)\f\in
\sat_M(\sat_x(L))$ and hence $o_m\f=x\f-(x-o_m)\f\in
\sat_M(\sat_x(L))$ follows. By Lemma \ref{lm-per1}, $\gcd(o_m,m)=1$.
Then $\f \in \sat_M(\sat_x(L))$, and the claim is true. Since
$\sat_M(L)\subset \sat_M(\sat_x(L)) = L_2 = \sat_x(\sat_M(\sat_x(L))
)$, we have $L_1\subset L_2$. \qedd

It is easy to check that a $\sigma$-ideal $\I$ is perfect if and only if $\I$ is reflexive,
radical, and well-mixed. Since a Laurent binomial $\sigma$-ideal
$\I$ is always radical, $\I$ is perfect if and only if $\I$ is
reflexive and well-mixed. From this observation, we can
deduce the following result about perfect Laurent binomial
$\sigma$-ideal ideals.

\begin{thm}\label{th-per1}
Let $\rho$ be a partial character and $\F$ an algebraically closed
and inversive $\sigma$-field. If $\I(\rho)$ is perfect, then
$L_\rho$ is {\rm{P}}-saturated. Conversely, if $L_\rho$ is
{\rm{P}}-saturated, then either $\{\I(\rho)\}=[1]$ or $\I(\rho)$ is
perfect.
Furthermore, the perfect closure of $\I(\rho)$ is either $[1]$ or a Laurent binomial $\sigma$-ideal
whose support lattice is $\sat_P(L_\rho)$.
\end{thm}
\proof If $\I(\rho)$ is perfect, then it is well-mixed and
reflexive. By Theorems~\ref{th-wm} and Theorem~\ref{th-pr1},
$L_\rho$ is {\rm{M}}-saturated and $x$-saturated, and hence
{\rm{P}}-saturated.
Conversely, if $L_\rho$ is {\rm{P}}-saturated, it is
{\rm{M}}-saturated and $x$-saturated. By Theorem~\ref{th-wm},
 either $\langle \I(\rho)\rangle=[1]$ or $\I(\rho)$ is well-mixed.
If $\langle \I(\rho)\rangle=[1]$, $\{\I(\rho)\}=[1]$. Otherwise, by
Theorem~\ref{th-pr1}, $\I(\rho)$ is reflexive. By
Corollary~\ref{cor-rad}, $\I(\rho)$ is radical. Then $\I(\rho)$ is
perfect.

By Lemma \ref{lm-pxsat}, $L_p=\sat_P(L_\rho)=\sat_M(\sat_x(L_\rho))$.
Then the perfect closure of $\I(\rho)$ is the well-mixed closure
of the reflexive closure of $\I(\rho)$, and then is either $[1]$ or a Laurent binomial $\sigma$-ideal whose support lattice is $L_P$ by Theorems \ref{lm-ld0} and \ref{th-wm2}.\qedd

%

\section{Binomial $\sigma$-ideal}\label{sec-bi}
%

\subsection{Basic properties of binomial $\sigma$-ideal}
In this section, it is shown that certain results from \cite{es-bi}
can be extended to the difference case using the theory of infinite Gr\"obner basis.

A {\em $\sigma$-binomial}   in  $\Y$ is  a $\sigma$-polynomial with
at most two terms,  that is, $a\Y^{\a}+b\Y^{\b}$ where $a, b\in \F$
and $\a, \b \in\N[x]^n$.
For $\f\in\Z[x]^n$,  let $\f^{+}, \f^{-}\in \N^{n}[x]$ denote the
positive part and the negative part of $\f$ such that
$\f=\f^{+}-\f^{-}$.
Consider a $\sigma$-binomial $f=a\Y^{\a}+b\Y^{\b}$, where
$a,b\in\F^*$. Without loss of generality, assume $\a > \b$ according
to the order defined in Section \ref{sec-zxm}. Then $f$ has the
following canonical representation
 \begin{equation}\label{eq-nfb}
 f=a\Y^{\a}+b\Y^{\b} = a\Y^\g(\Y^{\f+}-c\Y^{\f^-})
 \end{equation}
where $c=\frac{-b}{a}$, $\f=\a-\b\in\Zxn$  is a normal vector, and
$\g=\a-\f^+\in\N[x]$. The normal vector $\f$ is called the {\em
support} of $f$. Note that $\gcd(\Y^{\f+},\Y^{\f^-})=1$.

A $\sigma$-ideal in $\F\{\Y\}$ is called {\em binomial} if it is
generated by, possibly infinitely many, $\sigma$-binomials.

In this section, $\F\{\Y\}$ is considered as a polynomial ring in
infinitely many algebraic variables $\Theta(\Y)=\{y_i^{x^j}, i=1,\ldots,n;
j\ge 0\}$ and denoted by $S=\F[\Theta(\Y)]$. A theory of Gr\"obner
basis in the case of infinitely many variables is developed in
\cite{Kei} and will be used in this section.
For any $m\in\N$, denote $\Theta^{\lr{m}}(\Y)=\{y_i^{x^j},
i=1,\ldots,n; j=0,1,\ldots,m\}$ and
$S^{\lr{m}}=\F[\Theta^{\lr{m}}(\Y)]$ is a polynomial ring in finitely many variables.

A monomial order in  $S$ is called {\em compatible} with the
difference structure, if $y_i^{x^{k_1}}<y_i^{x^{k_2}}$ for $k_1 <
k_2$.
%
%
Only compatible monomial orders are considered in this section.

Let $\I$ be a  $\sigma$-ideal in $\F\{\Y\}$. Then $\I$ is an
algebraic ideal in $S$. By \cite{Kei}, we have
\begin{lem}\label{lm-gbbi1}
Let $\I$ be a binomial $\sigma$-ideal in $\F\{\Y\}$. Then for a
compatible monomial order, the reduced Gr\"{o}bner basis $\G$ of
$\I$ exists and satisfies
 \begin{equation}\label{eq-gbbi1} \G = \cup_{m=0}^{\infty} \G^{\lr{m}}\end{equation}
where $\G^{\lr{m}}=\G\cap S^{\lr{m}}$ is the reduced Gr\"{o}bner
basis of $\I^{\lr{m}} = \I \cap S^{\lr{m}}$ in
$S^{\lr{m}}$.
\end{lem}

Contrary to the Laurent case, a binomial $\sigma$-ideal may be
infinitely generated, as shown by the following example.
\begin{exmp}\label{ex-id}
Let  $\I=[y_1^{x^{i}}y_2^{x^{j}}-y_1^{x^{j}}y_2^{x^{i}} : 0 \leq i <
j \in \N]$. It is clear that $\I$ does not have a finite set of
generators and hence a finite Gr\"obner basis.
%
%
The Gr\"obner basis of
 $$\I^{\lr{m}}=\I\cap \Q[y_1,
y_2;y^{x}_1,y^{x}_2;\ldots;y^{x^{m}}_1,y^{x^{m}}_2]$$ is
$\{y_1^{x^{i}}y_2^{x^{j}}-y_1^{x^{i}}y_2^{x^{i}} : 0 \leq i < j \leq
m\}$ with a monomial order satisfying $y_1<
y_2<y^{x}_1<y^{x}_2<\cdots<y^{x^m}_1<y^{x^m}_2$. Then
 $\{y_1^{x^{i}}y_2^{x^{j}}-y_1^{x^{j}}y_2^{x^{i}}
: 0 \leq i < j \in \N\}$ is an infinite reduced Gr\"obner basis for
$\I$ in the sense of \cite{Kei} when $y^{x^m}_1$ and $y^{x^m}_2$ are
treated as independent algebraic variables.
\end{exmp}
\begin{rem}\label{rem-bi11}
The above concept of Gr\"obner basis does not consider the
difference structure.
The concept may be refined by introducing the reduced {\em
$\sigma$-Gr\"obner basis} \cite{dd-gb}. A $\sigma$-monomial $M_1$ is
called {\em reduced} w.r.t. another $\sigma$-monomial $M_2$ if there
do not exist a $\sigma$-monomial $M_0$ and a $k\in\N$ such that
$M_1=M_0 M_2^{x^k}$.
Then the reduced $\sigma$-Gr\"obner basis of $\I$ in Example
\ref{ex-id} is $\{y_1y_2^{x^{i}}-y_1^{x^{i}}y_2: i \in \Z_{\ge1}\}$
which is still infinite.
Since the purpose of Gr\"obner basis in this paper is theoretic and
not computational, we will use the version of infinite Gr\"obner
basis in the sense of \cite{Kei}.
\end{rem}

With Lemma \ref{lm-gbbi1}, a large portion of the properties for
algebraic binomial ideals proved by Eisenbud and Sturmfels in
\cite{es-bi} can be extended to the difference case.
The proofs follow the same pattern: to prove a property for $\I$, we
first show that the property is valid for $\I$ if and only if it is
valid for all $\I^{\lr{m}}$, and then the corresponding statement
from \cite{es-bi} will be used to show that the property is indeed
valid for $\I^{\lr{m}}$. We will illustrate the procedure in the
following corollary. For other results, we omit the proofs.

\begin{cor}\label{term}
Let $\I \subset \F\{\Y\}$ be a binomial $\sigma$-ideal. Then the
Gr\"{o}bner basis $\G$ of $\I$ consists of $\sigma$-binomials and
the normal form of any $\sigma$-term modulo $\G$ is again a
$\sigma$-term.
\end{cor}
\proof By a $\sigma$-term, we mean the multiplication of an element
from $\F^*$ and a $\sigma$-monomial. By \bref{eq-gbbi1}, it suffices
to show that corollary is valid for all $\G^{\lr{m}}$, that is, the
Gr\"{o}bner basis $\G^{\lr{m}}$ of $\I^{\lr{m}}$ consists of
binomials and the normal form of any term modulo $\G^{\lr{m}}$ is
again a term.
Since $\G^{\lr{m}}$ is the Gr\"{o}bner basis of $\I^{\lr{m}} = \I
\cap S^{\lr{m}}$ and $\I^{\lr{m}}$ is a binomial ideal in a
polynomial ring with finitely many variables, the corollary follows
from Proposition 1.1 in \cite{es-bi}.\qedd
%
%

\begin{cor}
A $\sigma$-ideal $\I$ is binomial if and only if the reduced
Gr\"{o}bner basis for $\I$ consists of $\sigma$-binomials.
\end{cor}

%
\begin{cor}\label{eli}
If $\I$ is a  binomial $\sigma$-ideal, then the elimination ideal
$\I \cap \F\{y_1,y_2,\ldots,y_r\}$ is binomial for every $r\leq n$.
\end{cor}
%
%

 The following lemma
can be proved similar to its algebraic counterpart.
\begin{lem}\label{lm-di11}
If $\I$ and $\J$ are binomial $\sigma$-ideals in $\F\{\Y\}$ then we
have $\I \cap\J=[t\I+(1-t)\J]\cap \F\{\Y\}$ where $t$ is a new
$\sigma$-indeterminate.
\end{lem}

The intersection of binomial $\sigma$-ideals is not binomial in
general, but from Lemma \ref{lm-di11} and \cite{es-bi} we have
\begin{cor}\label{cor-gbi8}
If  $\I$ and $\I^{'}$ are binomial $\sigma$-ideals and
$\J_{1},\ldots,\J_{s}$ are $\sigma$-ideals generated by
$\sigma$-monomials, then $[\I+\I^{'}]\cap[ \I+\J_{1}]\cap\ldots
\cap[\I+\J_{s}]$ is binomial.
\end{cor}

\begin{cor}\label{cor-tttt1}
Let $\I$ be a binomial $\sigma$-ideal  and let
$\J_{1},\ldots,\J_{s}$ be monomial $\sigma$-ideals.

(a) The intersection $[ \I+\J_{1}]\cap\cdots \cap[\I+\J_{s}]$ is
generated by $\sigma$-monomials modulo $\I$.

(b) Any $\sigma$-monomial in the sum $\I+\J_{1}+\cdots+\J_{s}$ lies
in one of the $\sigma$-ideals $\I+\J_{i}$.

\end{cor}

\begin{cor}\label{diff-quotient}
If $\I$ is a binomial $\sigma$-ideal, then for any $\sigma$-monomial
$M$, the $\sigma$-ideal quotients $[\I:M]$ and $[\I:M^{\infty}]$ are
binomial.
\end{cor}

\begin{cor}
Let $\I$ be a binomial $\sigma$-ideal and $\J$ a monomial
$\sigma$-ideal. If $f\in \I+\J$ and $g$ is the sum of those terms of
$f$ that are not individually contained in $\I+\J$, then $g\in \J$.
\end{cor}

From \cite[Theorem 3.1]{es-bi}, we have
\begin{thm}\label{th-bi11}
If $\I$ is a  binomial $\sigma$-ideal, then the radical of $\I$ is
binomial.
\end{thm}

Finally, we consider the reflexive closure of binomial $\sigma$-ideals.
\begin{lem}\label{lm-biref1}
A binomial $\sigma$-ideal $\I$ is reflexive if and only if  $b^x\in\I\imply
b\in\I$ for any $\sigma$-binomial $b\in\F\{\Y\}$.
\end{lem}
\proof It suffices to prove one side of the statement, that is, if
$b^x\in\I\imply b\in\I$ for any $\sigma$-binomial $b$ then $\I$ is
reflexive.
Let $p$ be a $\sigma$-polynomial such that $p^x\in\I$. Then, there
exists an $m\in\N$ such that $p^x\in\I^{\lr{m}} = \I \cap
S^{\lr{m}}$. Let $\G$ be the (finite) reduced Gr\"obner basis of
$\I^{\lr{m}}$ in $S^{\lr{m}}$ under the variable order $y_i^{x^j}<
y_k$ for any $i,k,j>0$. By Proposition 1.1 in \cite{es-bi}, $\G$
consists of binomials. $p^x$ can be reduced to zero by $\G$. Due to
the chosen variable order, we have $p^x =\sum_i e_i^x g_i^x$, where
$e_i^x\in S^{\lr{m}}$ and $g_i^x$ is a $\sigma$-binomial in
$S^{\lr{m}}$.
Since $g_i^x$ are $\sigma$-binomials in $\I$, we have $g_i\in\I$.
Then, $p =\sum_i e_i g_i\in\I$ and $\I$ is reflexive.\qedd

\begin{thm}\label{th-bi12}
If $\I$ is a  binomial $\sigma$-ideal, then the reflexive closure of
$\I$ is binomial.
\end{thm}
\proof Let $\I_1$ be the $\sigma$-ideal generated by the
$\sigma$-binomials $p$ such that $p^{x^{k}} \in\I$ for a
$k\in\N$. We claim that $\I_1$ is the reflexive closure of $\I$ and
it suffices to show that $\I_1$ is reflexive. Let $p$ be a
$\sigma$-binomial such that $p^x\in\I_1$. Then for some $s\in\N$,
$(p^x)^{x^s} = p^{x^{s+1}}\in\I$. Thus $p\in\I_1$ and $\I_1$ is
reflexive by Lemma \ref{lm-biref1}.\qedd

\subsection{Normal binomial $\sigma$-ideal}
In this section, most of the results about Laurent binomial
$\sigma$-ideals proved in Sections \ref{sec-lbi} and \ref{sec-lbicr} will be extended to
normal binomial $\sigma$-ideals.

Let $\mb$ be the multiplicative set generated by $y_i^{x^j}$ for
$i=1, \ldots, n,  j\in\N$.
A $\sigma$-ideal $\I$ is called {\em normal} if for $M\in\mb$
and $p\in\F\{\Y\}$,  $Mp\in\I$ implies $p\in\I$. For any
$\sigma$-ideal $\I$,
$$\I:\mb=\{f\in \F\{\Y\}\,|\,\exists M\in \mb \st M f\in \I\}$$
is a normal $\sigma$-ideal.
For any $\sigma$-ideal $\I$ in $\F\{\Y\}$, it is easy to check that
 \begin{equation}\label{eq-nbi1}
 \F\{\Y^{\pm}\}\I\cap\F\{\Y\}=\I:\mb.\end{equation}
We first prove a property for general normal $\sigma$-ideals.
\begin{lem}\label{lm-gni1}
A normal $\sigma$-ideal $\I$  in $\F\{\Y\}$ is
reflexive (radical, well-mixed, perfect, prime) if and only if
$\F\{\Y^{\pm}\}\I$ is reflexive (radical, well-mixed, perfect, prime) in
$\F\{\Y^{\pm}\}$.
\end{lem}
\proof Let $\overline{\I}=\F\{\Y^{\pm}\}\I$ be a Laurent
$\sigma$-ideal. Since $\I$ is normal, from \bref{eq-nbi1} we have
$\overline{\I}\cap\F\{\Y\}=\I$.
If $\overline{\I}$ is reflexive, it is clear that $\I$ is reflexive.
For the other direction, if $f^{x}\in \overline{\I}$, then by
clearing denominators of $f^{x}$, there exists a $\sigma$-monomial
$M^{x}$ in $\Y$ such that $M^{x}f^{x}\in\overline{\I} \cap \F\{\Y\}=
\I$. Since $\I$ is reflexive, $Mf\in \I$ and hence $f\in
\overline{\I}$, that is, $\overline{\I}$ is reflexive. The results
about radical and perfect $\sigma$-ideals can be proved similarly.

We now show that $\I$ is prime if and only if $\overline{\I}$ is
prime.
If $\overline{\I}$ is prime, it is clear that  $\I$ is also prime.
For the other side, let $fg\in \overline{\I}$. Then  there exist
$\sigma$-monomials $N_1,N_2$ such that $N_1f\in\F\{\Y\}$,
$N_2g\in\F\{\Y\}$, and hence $N_1f N_2g\in \I$. Since $\I$ is prime,
$N_1f$ or $N_2g$ is in $\I$ that is $f$ or $g$ is in
$\overline{\I}$.
The result about well-mixed $\sigma$-ideals can be proved similarly.\qedd

Given a partial
character $\rho$ on $\Z[x]^{n}$, we define the following binomial
$\sigma$-ideal in $\F\{\Y\}$
 \begin{equation}\label{eq-I+}
 \I^{+}(\rho)=[{\Y^{\f^{+}}-\rho(\f)\Y^{\f^{-}}\,|\,\f\in
 L_{\rho}}].\end{equation}
We will show that any normal binomial $\sigma$-ideal can be written
as the form \bref{eq-I+}.
%


\begin{lem}\label{lm-nl1}
Let $\rho$ be a partial character on $\Zxn$ and $\I(\rho)$ defined
in \bref{eq-I}. Then $\I^{+}(\rho)=\I(\rho)\cap \F\{\Y\}$. As a
consequence, $\I^{+}(\rho)$ is proper and normal.
\end{lem}
\proof It is clear that  $\I^{+}(\rho)\subset \I(\rho)\cap
\F\{\Y\}$.
If $f\in \I(\rho) \cap \F\{\Y\}$, then $f=\sum_{i=1}^s f_i
M_i(\Y^{\f_i}-\rho(\f_i))$ where $f_i\in\F$, $\f_i\in L_{\rho}$, and
$M_i$ are Laurent $\sigma$-monomials in $\Y$.
There exists a $\sigma$-monomial $M$ in $\Y$ such that
 \begin{equation}\label{eq-nl1}
 Mf=\sum_{i=1}^s f_i N_i(\Y^{\f_i^{+}}-\rho(\f_i)\Y^{\f_i^{-}})\in \I^{+}(\rho),\end{equation}
where $N_i$ is a $\sigma$-monomial in $\Y$. We will prove $f\in
\I^{+}(\rho)$ from the above equation.
Without loss of generality, we may assume that $M=y_c^{x^{o}}$ for
some $c$ and $o\in\N$. Note that \bref{eq-nl1} is an algebraic
identity in $y_i^{x^{k}}, i=1,\ldots,n, k\in\N$.
If $N_i$ contains $y_c^{x^{o}}$ as a factor, we move
$F_i=f_iN_i(\Y^{\f_i^{+}}-\rho(\f_i)\Y^{\f_i^{-}})$ to the left hand
side of \bref{eq-nl1} and let $f_1=f-F_i/y_c^{x^{o}}$. Then
$f\in\I^{+}(\rho)$ if and only if $f_1\in\I^{+}(\rho)$.
Repeat the above procedure until no $N_i$ contains $y_c^{x^{o}}$ as
a factor.

If $s=0$ in \bref{eq-nl1}, then $f=0$ and the lemma is proved.
Since $\gcd(\Y^{\f_i^{+}},\Y^{\f_i^{-}})=1$, $y_c^{x^{o}}$ cannot be
a factor of both $\Y^{\f_i^{+}}$ and $\Y^{\f_i^{-}}$.
Let $\Y^{\f_i^{+}}$ be the largest $\sigma$-monomial in
\bref{eq-nl1} not containing $y_c^{x^{o}}$ under a given
$\sigma$-monomial total order . If $\Y^{\f_i^{-}}$ is the largest
$\sigma$-monomial in \bref{eq-nl1} not containing $y_c^{x^{o}}$, the
proving process is similar.
There must exists another $\sigma$-binomial
$f_jN_j(\Y^{\f_j^{+}}-\rho(\f_j)\Y^{\f_j^{-}})$ such that
$N_i\Y^{\f_i^{+}}=N_j\Y^{\f_j^{-}}$.
Let $N_i=\Y^{\p_i}, N_j=\Y^{\p_j}$. Then
$\Y^{\f_i^{+}+\p_i}=\Y^{\f_j^{-}+\p_j}$ and
$\f_i^{+}+\p_i=\f_j^{-}+\p_j$.
We have $p=f_iN_i (\Y^{\f_i^{+}}-\rho(\f_i)\Y^{\f_i^{-}})+
 f_jN_j(\Y^{\f_j^{+}}-\rho(\f_j)\Y^{\f_j^{-}})
  =\frac{f_i}{\rho(\f_j)}(\Y^{\f_j^{+}+\p_j}-\rho(\f_i)\rho(\f_j)\Y^{\f_i^{-}+\p_i})
 +(f_j-\frac{f_i}{\rho(\f_j)})N_j(\Y^{\f_j^{+}}-\rho(\f_j)\Y^{\f_j^{-}})$.
%
%
Since
$\f=\f_j^{+}+\p_j-(\f_i^{-}+\p_i)=\f_i^{+}-\f_i^{-}+\f_j^{+}-\f_j^{-}=\f_i+\f_j\in
L_{\rho}$, we have
$\Y^{\f_j^{+}+\p_j}-\rho(\f_i)\rho(\f_j)\Y^{\f_i^{-}+\p_i}
 =N(\Y^{\f^+}-\rho(\f)\Y^{\f^{-}})\in \I^{+}(\rho)$, where $N$ is a
 $\sigma$-monomial.
As a consequence, $p\in \I^{+}(\rho)$.
%
%
If $N$ contains $y_c^{x^{o}}$, move the term
$\frac{f_i}{\rho(\f_j)}N(\Y^{\f^{+}}-\rho(\f)\Y^{\f^{-}})$ to the left hand
side of \bref{eq-nl1} as we did in the first phase of the proof.
After the above procedure, equation \bref{eq-nl1} is still valid.
Furthermore, the number of $\sigma$-binomials in \bref{eq-nl1} does
not increase, no $N_i$ contains $y_c^{x^{o}}$, and the largest
$\sigma$-monomial $\Y^{\f_i^{+}}$ or $\Y^{\f_i^{-}}$ not containing
$y_c^{x^{o}}$ becomes smaller.
The above procedure will stop after a finite number of steps, which
means $s=0$ in \bref{eq-nl1} and hence $y_c^{x^{o}}f =0$ which means
the original $f$  is in $\I^{+}(\rho)$. Then
$\I^{+}(\rho)=\I(\rho)\cap \F\{\Y\}$.

$\I^{+}(\rho)=\I(\rho)\cap \F\{\Y\}$ is proper. For otherwise
$\I(\rho)=[1]$,  contradicting to Lemma \ref{lm-l1}.
Note that $\I^{+}(\rho)\F\{\Y^{\pm}\}=\I(\rho)$. Then
$\I^{+}(\rho)=\I(\rho)\cap \F\{\Y\}=\I^{+}(\rho)\F\{\Y^{\pm}\}\cap
\F\{\Y\}=\I^{+}(\rho):\mb$, and $\I^{+}(\rho)$ is normal.\qedd

\begin{lem}  \label{lm-nl2}
Let $\rho$ be a partial character over $\Zxn$. Then
$\Y^{\f^+}-c\Y^{\f^-} \in \I^{+}(\rho)$ if and only if $\f\in
L_{\rho}$ and $c=\rho(\f)$.
\end{lem}
\proof By Lemma \ref{lm-nl1}, $\Y^{\f^+}-c\Y^{\f^-} \in
\I^{+}(\rho)$ if and only if $\Y^{\f}-c \in\I(\rho)$ which is
equivalent to $\f\in L_{\rho}$ and $c=\rho(\f)$ by Lemma
\ref{lm-l3}. \qedd

\begin{lem}  \label{lm-nl3}
If $\I$ is a normal binomial $\sigma$-ideal,  then there exists a
unique partial character $\rho$ on $\Z[x]^{n}$ such that
$\I=\I^{+}(\rho)$ and $L_{\rho}=\{\f\in\Zxn\,|\,
\Y^{\f^+}-\rho(\f)\Y^{\f^-} \in \I\}$ which is called the {\em
support lattice} of $\I$.
\end{lem}
\proof We have $\I\cdot \F\{\Y^{\pm}\}\cap \F\{\Y\} =\I:\mb$.
By Theorem \ref{th-l1},
 there exists a partial character $\rho$ such that
$\I\cdot \F\{\Y^{\pm}\}=\I(\rho)$. Then by Lemma \ref{lm-nl1},
$\I=(\I:\mb)=\I\cdot \F\{\Y^{\pm}\}\cap \F\{\Y\}=\I(\rho)\cap
\F\{\Y\}=\I^{+}(\rho)$.
By Lemma \ref{lm-nl2}, we have $L_{\rho}=\{\f\in\Zxn\,|\,
\Y^{\f^+}-\rho(\f)\Y^{\f^-} \in \I=\I^{+}(\rho)\}$. The uniqueness
of $\rho$ comes from the fact that $L_{\rho}$ is uniquely determined
by $\I$.\qedd


By Lemmas \ref{lm-nl1} and  \ref{lm-nl3}, we have
\begin{thm}\label{th-nbi2}
The map $\I(\rho) \Rightarrow \I^{+}(\rho)$ gives a one to one
correspondence between Laurent binomial $\sigma$-ideals in $\F\{\Y^{\pm}\}$ and normal
binomial $\sigma$-ideals in $\F\{\Y\}$.
\end{thm}

Due to Lemma \ref{lm-nl1} and Theorem \ref{th-nbi2}, most properties
of Laurent binomial $\sigma$-ideals can be extended to normal
binomial $\sigma$-ideals.
As a consequence of Corollary \ref{cor-rad}, Lemma \ref{lm-gni1},
and Lemma \ref{lm-nl1}, we have
\begin{cor}\label{cor-rad1}
A normal binomial $\sigma$-ideal is radical.
\end{cor}

As a consequence of Theorem \ref{lm-ld0}, Lemma \ref{lm-gni1}, and
Theorem \ref{th-nbi2}.
\begin{cor}\label{cor-ld01}
If $\F$ is inversive, then the reflexive closure of $\I^+(\rho)$ is also a normal
binomial $\sigma$-ideal whose support lattice is the $x$-saturation
of $L_\rho$.
\end{cor}

%
\begin{cor}\label{th-nl1}
If $\F$ is
algebraically closed and inversive,  then
\begin{description}
\item[(a)] $L_\rho$ is $\Z$-saturated if and only if $\I^+(\rho)$ is prime;
\item[(b)] $L_\rho$ is $x$-saturated if and only if $\I^+(\rho)$ is reflexive;
\item[(c)] $L_\rho$ is saturated if and only if $\I^+(\rho)$ is reflexive prime.
\end{description}
\end{cor}
\proof It is easy to show that
$\I(\rho)=\I^{+}(\rho)\F\{\Y^{\pm}\}$. Then the corollary is a
consequence of Theorem $\ref{th-pr1}$, Lemma \ref{lm-gni1}, and
Lemma \ref{lm-nl1}.\qedd


For properties related with perfect $\sigma$-ideals, it becomes more
complicated. Direct extension of Theorems \ref{th-l2},
and \ref{th-per1} to the normal binomial case is not
correct as shown by the following example.
\begin{exmp}\label{ex-bi-per}
Let $\I = [y_1^x - y_1, y_2^2 - y_1^2, y_2^x + y_2]$ which is a
normal binomial $\sigma$-ideal whose representation matrix is
$L=\left [\begin{array}{ccc}
x-1  & -2    & 0   \\
0    & 2     & x-1 \\
\end{array}\right]$.
Since $o_2=1$, $L$ is $P$-saturated. Also, $L_s = \sat(L) =
 \left [\begin{array}{ccc}
  x-1  & -1 \\
 0     & 1 \\ \end{array}\right]$.
We have $\{\I\} = \{\I,y_2-y_1\}\cap\{\I,y_2+y_1\} = [y_1,y_2]$.
Then $\{\I\}\ne[1]$ and $\I$ is not perfect and hence Theorems
\ref{th-per1} are not correct. Theorem \ref{th-l2}
is also not correct, since the supporting lattice of the prime
component of $\I$ is not $L_s$.
This example also shows that the perfect closure of a normal binomial
$\sigma$-ideal is not necessarily normal.
\end{exmp}

It can be seen that the problem  is due to the occurrence of
$\sigma$-monomials. For any partial character $\rho$, it can be
shown that
 \begin{equation}\label{eq-per1-1}
  \{\I^{+}(\rho)\}:\mb=\{\I(\rho)\}\cap \F\{\Y\}.\end{equation}
We thus have the following modifications for Theorems \ref{th-per1} and \ref{th-l2}.
\begin{cor}\label{th-nl1p}
Let $\F$ be an inversive and algebraically closed $\sigma$-field.
If $\I^{+}(\rho)$ is perfect, then $L_\rho$ is $P$-saturated. Conversely, if
$L_\rho$ is $P$-saturated, then either
$\{\I\}:\mb=[1]$ or $\I$ is perfect.
For any $\rho$, either $\{\I^{+}(\rho)\}:\mb=[1]$ or $\{\I^{+}(\rho)\}:\mb$ is a binomial $\sigma$-ideal whose support lattice is the $P$-saturation of
$L_\rho$.
\end{cor}
\proof If $\I$ is perfect, by Lemma \ref{lm-gni1}, $\I(\rho) =
\I\F\{\Y^{\pm}\}$ is also perfect. By Theorem \ref{th-per1},
$L_\rho$ is $P$-saturated.
If $L_\rho$ is $P$-saturated and $x$-saturated, by Theorem
\ref{th-per1}, either $\I(\rho)=[1]$ or $\I(\rho)$ is perfect. If
$\I(\rho)=[1]$, by \bref{eq-per1-1}, $\{\I\}:\mb=[1]$. If $\I(\rho)$
is perfect, by Lemma \ref{lm-gni1}, $\I=\I^+(\rho)$ is also
perfect.\qedd

Similar results hold for normal well-mixed $\sigma$-ideals.

%
%
%

In the rest of this section, we give decomposition theorems for
perfect binomial $\sigma$-ideals. We first consider normal binomial
$\sigma$-ideals. By Corollary \ref{cor-rad1} and Example
\ref{ex-311}, a normal binomial $\sigma$-ideal is radical but may not
be perfect.
\begin{thm}\label{th-nl2}
Let $\I=\I^{+}(\rho)$ be a normal binomial $\sigma$-ideal and $\F$
an inversive and algebraically closed $\sigma$-field. Then
$\{\I\}:\mb$ is either $[1]$ or can be written as the intersection
of reflexive prime binomial $\sigma$-ideals whose support lattice is
the saturation lattice of $L_\rho$.
\end{thm}
\proof By Theorem \ref{th-l2}, either $\{\I(\rho)\}=[1]$ or
$\{\I(\rho)\}=\bigcap^{s}_{i=1}\I(\rho_i)$, where $\I(\rho_i)$ are
reflexive prime $\sigma$-ideals whose support lattices are
$\sat(L_\rho)$.
By \bref{eq-per1-1} and Lemma \ref{lm-nl1}, either
$\{\I^{+}(\rho)\}:\mb=[1]$ or $\{\I^{+}(\rho)\}:\mb=\{\I(\rho)\}\cap
\F\{\Y\}=(\bigcap^{s}_{i=1}\I(\rho_i))\cap \F\{\Y\}
=\bigcap^{s}_{i=1}(\I(\rho_i)\cap
\F\{\Y\})=\bigcap^{s}_{i=1}\I^{+}(\rho_i)$. By Corollary
\ref{th-nl1}, $ \I^{+}(\rho_i)$ is reflexive and prime whose support
lattices are the saturation of $L_\rho$. \qedd


Now, consider general binomial $\sigma$-ideals.
\begin{lem}\label{lm-nl5}
$\I\subset\F\{\Y\}$ is a reflexive prime binomial $\sigma$-ideal if
and only if $\I=[y_{i_{1}},\ldots,y_{i_{s}}]$ $+\I_1$, where
$\{y_{i_{1}},\ldots,y_{i_{s}}\}=\Y\cap \I$, $\{z_1,\ldots,z_t\}=\Y
\backslash \I$, and $\I_1$ is a normal binomial reflexive prime
$\sigma$-ideal in $\F\{z_1,\ldots,z_t\}$.
\end{lem}
\proof If $\I$ is reflexive and prime, then $(y_i^{x^j})^d\in\I$ if
and only if $y_i\in\I$. Let $\I_1 = \I\cap\F\{z_1,\ldots,z_t\}$.
Then $\I=[y_{i_{1}},\ldots,y_{i_{s}}]+\I_1$. $\I_1$ is clearly
reflexive and prime. We still need to show that $\I_1$ is normal.
Let $N f\in\I_1$ for a $\sigma$-monomial $N$ in $\{z_1,\ldots,z_t\}$
and $f\in\F\{z_1,\ldots,z_t\}$. $N$ cannot be in $\I_1$. Otherwise,
some $z_i$ is in $\I_1$ since $\I_1$ is reflexive and prime, which
contradicts to $\{z_1,\ldots,z_t\}=\Y \backslash \I$. Therefore,
$f\in\I_1$ and $\I_1$ is normal.
The other direction is trivial.\qedd

The $\sigma$-ideal $\I$ in Lemma \ref{lm-nl5} is said to be {\em
quasi-normal}. The following result can be proved similarly to
Theorem \ref{th-l2}.
\begin{thm}\label{th-nl5}
Let $\I$ be a binomial $\sigma$-ideal. If $\F$ is algebraically
closed and inversive, then the perfect $\sigma$-ideal $\{\I\}$ is
either $[1]$ or the intersection of quasi-normal reflexive prime
binomial $\sigma$-ideals.
\end{thm}
\proof We prove the theorem by induction on $n$.
Let $\I_1=\{\I\}:\mb$. Then $\{\I\}=   \I_1 \cap \cap_{i=1}^n
\{\I,y_i\}$. It is easy to check  $\I_1=\{\I:\mb\}:\mb$. Since
$\I:\mb$ is normal, by Theorem \ref{th-nl2}, $\I_1$ is either $[1]$
or intersection of normal reflexive prime $\sigma$-ideals.
If $n=1$, then $\{\I,y_i\}$ must be either $[y_1]$ or $[1]$. Then
the theorem is proved for $n=1$. Suppose the theorem is valid for
$n=1,\ldots,k-1$.
Still use $\{\I\}=   \I_1 \cap \cap_{i=1}^n \{\I,y_i\}$.
Let $\I_i$ be the $\sigma$-ideal obtained by setting $y_i$ to $0$ in
$\I$. By the induction hypothesis, $\I_i$ can be written as
intersection of quasi-normal reflexive prime $\sigma$-ideals in
$\F\{\Y\setminus\{y_i\}\}$. So the theorem is also valid for
$\{\I,y_i\}=\{\I_i,y_i\}$. The theorem is proved.\qedd

\subsection{Characteristic set for normal binomial $\sigma$-ideal}
The theory of characteristic set given in Section \ref{sec-lbi2} will
be extended to the normal $\sigma$-binomial case.

Let $\rho$ be a partial character over $\Zxn$, $L_{\rho}
=(\f_1,\ldots,\f_s)$ where $\fb=\{\f_1,\ldots,\f_s\}$ is a reduced
Gr\"obner basis, and
 \begin{equation}\label{eq-A+}
 \A^{+}(\rho): \Y^{\f_1^+}-\rho(\f_1)\Y^{\f_1^-},
      \ldots,
      \Y^{\f_s^+}-\rho(\f_s)\Y^{\f_s^-}.\end{equation}

We have the following canonical representation for normal binomial
$\sigma$-ideals.
\begin{thm}\label{th-nl3}
Use the notations in \bref{eq-A+}.
Then $\I^{+}(\rho)=\sat(\A^{+}(\rho))$. Furthermore,
$\A^{+}(\rho)$ is a regular and coherent $\sigma$-chain and hence is
a characteristic set of $\I^{+}(\rho)$.
\end{thm}
\proof
Let $\I_1=[\A^{+}(\rho)] :\mb$. We claim $\I_1=\sat(\A^{+}(\rho))$.
It is clear that $\sat(\A^{+}(\rho))\subset
[\A^{+}(\rho)]:\mb=\I_1$.
For the other direction, let $p\in\I_1$ and
$p_1=\prem(p,\A^{+}(\rho))$ which is reduced w.r.t. $\A^{+}(\rho)$.
By \bref{eq-prem}, $p_1\in\I_1$. As a consequence,
$p_1\in[\A(\rho)]$ as Laurent $\sigma$-polynomials in
$\F\{\Y^{\pm}\}$. By Lemma \ref{lm-l1}, $\A(\rho)$ is a
characteristic set of $[\A(\rho)]$. Since $p_1$ is reduced w.r.t.
$\A^+(\rho)$, it is also reduced w.r.t. $\A(\rho)$. Then $p_1=0$ and
hence the claim is proved.

We now prove $\I^+(\rho) = \sat(\A^+(\rho))$.
By the above claim, Lemma \ref{lm-l1}, and Lemma \ref{lm-nl1},
$\sat(\A^{+}(\rho))=[\A^{+}(\rho)]
:\mb=[\A^{+}(\rho)]\F\{\Y^{\pm}\}\cap \F\{\Y\}=[\A(\rho)]\cap
\F\{\Y\}=\I(\rho)\cap \F\{\Y\}=\I^{+}(\rho) $.

It remains to prove that $\A^{+}(\rho)$ is a characteristic set of
$\I_1=[\A^{+}(\rho)] :\mb$. By definition, it suffices to show that
if $p\in\I_1$ is reduced w.r.t. $\A^{+}(\rho)$ then $p=0$. Let
$A_i=\Y^{\f_i}-\rho(\f_i)$ and
$A_i^+=\Y^{\f_i^+}-\rho(\f_i)\Y^{\f_i^-}$. Since $p\in\I_1$, there
exist a $\sigma$-monomial $M$ and $f_{i,j}\in\F\{\Y\}$ such that
$Mp=\sum_{i,j}f_{i,j} (A_{i}^+)^{x^j}$. Then in $\F\{\Y^{\pm}\}$, we
have $p=\sum_{i,j}g_{i,j} A_{i}^{x^j}\in[\A(\rho)]$, where
$g_{i,j}\in\F\{\Y^{\pm}\}$. Since $p$ is reduced w.r.t.
$\A^+(\rho)$, it is also reduced w.r.t. $\A(\rho)$. By Lemma
\ref{lm-l1}, $\A(\rho)$ is a characteristic set of $[\A(\rho)]$ and
hence $p=0$. The claim is proved.

Since $\I_1=\sat(\A^+(\rho))$, $\A^{+}(\rho)$ is also a
characteristic set of $\sat(\A^+(\rho))$. By Theorem \ref{th-rp},
 $\A^{+}(\rho)$ is regular and coherent.\qedd

\begin{exmp}\label{ex-ig}
Let $L=([1-x,x-1]^\tau)$ be a $\Zx$-module and $\rho$ the trivial
partial character on $L$, that is, $\rho(\f)=1$ for $\f\in$L.
By Theorem \ref{th-nl3}, $\I^{+}(\rho)=\sat[y_1y_2^{x}-y_1^{x}y_2]
\subseteq \Q\{y_1, y_2\}$. By Theorem \ref{lm-nl1}, $\I^{+}(\rho)$
is a reflexive prime $\sigma$-ideal.
We can show that
$\I^{+}(\rho)=[y_1^{x^{i}}y_2^{x^{j}}-y_1^{x^{j}}y_2^{x^{i}}\,|\, 0
\leq i < j \leq m]$, which is an infinitely generated
$\sigma$-ideal.
\end{exmp}

As a consequence of Theorem \ref{th-nbi2}, Theorem \ref{th-nl3}, and
Lemma \ref{lm-l1}, we have
\begin{cor}\label{cor-ascr}
Let $\A(\rho)$ and  $\A^{+}(\rho)$ be defined in \bref{eq-Arho} and
\bref{eq-A+}, respectively. Then\newline
$([\A(\rho)]\F\{\Y^{\pm}\})\cap\F\{\Y\}=\sat(\A^{+}(\rho))$.
\end{cor}

As a consequence of Theorem \ref{th-pr1}, Corollary \ref{th-nl1},
and Theorem \ref{th-nl3}.
\begin{cor}\label{cor-asc2}
$[\A(\rho)]$ is a reflexive (prime) $\sigma$-ideal in $\F\{\Y^{\pm}\}$ if
and only if $\sat(\A^+(\rho))$ is a reflexive (prime) $\sigma$-ideal in
$\F\{\Y\}$.
\end{cor}


We now prove the converse of Theorem \ref{th-nl3}.
Let $\f_i\in\Zxn$ and $c_i\in\F^*$, $i=1,\ldots,s$. Consider the
following $\sigma$-chains
\begin{eqnarray}
 \A&:&\Y^{\f_{1}}-c_{1}, \ldots, \Y^{\f_{s}}-c_{s}\label{eq-aa}\\
 \A^{+}&:&\Y^{\f_{1}^+}-c_{1}\Y^{\f_{1}^-}, \ldots,
 \Y^{\Y^{\f_{s}^+}}-c_{s}\Y^{\f_{s}^-}\nonumber
 \end{eqnarray}
in $\F\{\Y^{\pm}\}$ and $\F\{\Y\}$, respectively.
Notice that, when talking about $\A^{+}$ (or $\A$),
all operations are performed in $\F\{\Y\}$ (or $\F\{\Y^{\pm}\}$).
Since $\f_i$ are
assumed to be normal, $\A^+$ is a $\sigma$-chain if and only if $\A$
is a Laurent $\sigma$-chain.

\begin{lem}\label{lm-asc1}
Use the notations in \bref{eq-aa}. Let
$p=a\Y^{\a}+b\Y^{\b}=aN(\Y^{\f}-c)\in\F\{\Y\}$, where
$\a,\b\in\N[x]^n$, $\f\in\Zxn$, $N$ is a $\sigma$-monomial,
$c\in\F^*$.
If $\A^+$ is coherent and regular, then $\prem(p,\A^+)=0$ implies
$\prem(\Y^{\f}-c,\A)=0$.
\end{lem}
\proof
%
%
%
Since $\prem(p,\A^+)=0$, there exists a $\sigma$-monomial $M_1$ such
that $M_1p\in[\A^+]$.
Let $p_1 = \Y^{\f}-c$. Since $r_1=\prem(p_1,\A)=\Y^\g-c_\g$, by
Lemma \ref{lm-rem1}, there exists a $c_1\in\F^*$ such that
$r_1-c_1p_1\in[\A]$. Then, there exists a $\sigma$-monomial $M_2$
such that $M_2Nr_1, M_2Np_1\in\F\{\Y\}$ and
$M_2N(r_1-c_1p_1)\in[\A^+]$ and hence
$M_2M_1N(r_1-c_1p_1)=M_2M_1Nr_1-\frac{c_1}{a}M_2M_1p\in[\A^+]$. Let
$M=M_1M_2N$. From $M_1p\in[\A^+]$, we have
$Mr_1\in[\A^+]\subset\sat(A^+)$.

Suppose $A_i= \Y^{\f_{i}^+}-c\Y^{\f_{i}^-}=
 I_i^+ y_{c_i}^{d_ix^o_i}-cI_i^-$, where $y_{c_i}$ is the leading variable of $A_i$.
A variable like $y_{c_i}^{x^{o_i+k}}$ for $k\in\N$ is called a {\em
main variable} of $\A^+$. A variable $y_{i}^{x^j}$ is called a {\em
parameter} of $\A^+$ is it is not a main variable.
If $M$ contains a main variable of $\A^+$ as a factor. Then let
$z=y_{c_i}^{x^{o_i+k}}$ be the largest one appearing in $M$ under
the variable ordering induced by the lexicographical of the index
$(c_i,o_i+k)$. Let $s=\deg(M,z)$ and $M_1=M/(z^s)$.
We may assume that $d_i$ is a factor of $s$. Otherwise, let
$s_1=\lfloor \frac{s}{d_i}\rfloor$, $s_0 = s-s_1d_i$, and
$M=Mz^{d_i-s_0}=M_1z^{d_i(s_1+1)}$. We still have
$Mr_1\in\sat(A^+)$.
We may use $A_i=0$ to eliminate $z$ from $M$ as follows:
 $M_1z^{s-d_i}(cI_i^-)^{x^k} r_1 =
 M_1z^{s-d_i}(I_i^+ y_{c_i}^{d_ix^o_i}-A_i)^{x^k} r_1=
 M(I_i^+)^{x^k}r_1-M_1z^{s-d_i}(A_i)^{x^k} r_1 \in\sat(A^+)$.
Note that $\deg(M_1z^{s-d_i}(cI_i^-)^{x^k},z)=s-d_i$. Repeat the
above procedure, we may find a $\sigma$-monomial $N$ such that $Nr_1
\in\sat(A^+)$, $N$ does not contain $z$ as a factor, and any
variable $y_i^{x^j}$ in $M$ is smaller than $z$ in the given
variable ordering.
Repeat the procedure, we may finally obtain a $\sigma$-monomial $L$
such that $L$ does not contain main variables of $\A^+$ as factors
and $Lr_1\in\sat(A^+)$. Since $L$ contains only parameters of $\A^+$
and $r_1$ is reduced w.r.t. $\A^+$, $Lr_1$ is also reduced w.r.t.
$\A^+$. Since $\A^+$ is regular and coherent, by Lemma \ref{lm-21},
it is the characteristic set of $\sat(A^+)$. Therefore, $Lr_1=0$,
and $r_1=0$. \qedd

The following example shows that if $\prem(p,\A^+)\ne0$ then the
relation between $\prem(p,$ $\A^+)$ and $\prem(\Y^{\f}-c,\A)$ may be
complicated, where $p=a\Y^{\a}+b\Y^{\b}=aN(\Y^{\f}-c)$.
\begin{exmp}\label{ex-asc2}
Let $p=y_2(y_2-1)$, $A_1=y_1^{-1}y_2^2-1$, and $A_1^+=y_2^2-y_1$.
Then $\prem(p,A_1^+)=y_1-y_2$ in $\F\{y_2\}$. But in
$\F\{y_2^{\pm}\}$, $p$ is represented as $\widetilde{p}=y_2-1$ and
$\prem(\widetilde{p},A_1)=y_2-1$.
\end{exmp}

\begin{lem}\label{lm-asc2}
Use the notations in \bref{eq-aa}. $\A$ is a regular and coherent
$\sigma$-chain in $\F\{\Y^{\pm}\}$ if and only if $\A^{+}$ is a
regular and coherent $\sigma$-chain in $\F\{\Y\}$.
\end{lem}
\proof
If $\A$ is regular and coherent, by Theorem \ref{th-l1} and Corollary \ref{cor-bic}, there exists a partial character $\rho$ over $\Zxn$ such that
$L_\rho=(\f_1,\ldots,\f_s)$, $\rho(\f_i)=c_i$, and $\I(\rho)=[\A]$.
By Theorem \ref{th-nl3}, $\A^+=\A^+(\rho)$ is regular and coherent.

Assume that $\A^+$ is regular and coherent. We first show that
$[\A]\ne[1]$ in $\F\{\Y^{\pm}\}$. It suffices to show that
$\sat(\A^+)$ does not contain a $\sigma$-monomial. Suppose the
contrary, there is a $\sigma$-monomial $M\in\sat(\A^+)$. Since
$\A^+$ is a regular and coherent chain, we have $\prem(M,\A^+)=0$.
Now consider the procedure of $\prem$, it can be shown that the
pseudo-remainder of a nonzero $\sigma$-monomial w.r.t. a binomial
$\sigma$-chain is still a nonzero  $\sigma$-monomial, a
contradiction.

%
Note that $\A$ is always regular since $\sigma$-monomials are
invertible in $\F\{\Y^{\pm}\}$.  Then, it suffices to prove that
$\A$ is coherent.

Let $A_i=\Y^{\f_{i}}-c_{i}$ and
$A_i^+=\Y^{\f_{i}^+}-c_{i}\Y^{\f_{i}^-}$.
Assume $A_i^+$ and $A_j^+$ ($i<j$) have the same leading variable
$y_l$, and
    $A_i^+=I_i^+ y_l^{d_i x^{o_i}}-c_iI_i^-$,
    $A_j^+=I_j^+ y_l^{d_j x^{o_j}}-c_jI_j^-$, where $I_i^-=\Y^{\f_{i}^-}$.
From Definition \ref{def-ghf},  we have $o_i < o_j$ and $d_i| d_j$.
Let $d_i = t d_j$ where $t\in\N$.
From \bref{eq-delta},
$$\Delta(A_i^+,A_j^+)
=\prem((A_i^+)^{x^{o_j-o_i}},A_j^+) =
       c_j^t (I_j^-)^t (I_i)^{x^{o_j-o_i}} -
       (I_j^+)^t  (c_iI_i^+)^{x^{o_j-o_i}}.$$
Comparing to \bref{eq-delta1}, if $\Delta(A_i,A_j)=\Y^\h-c_f$, then
$\Delta(A_i^+,A_j^+)=c_j^tM(\Y^{\h^+}-c_f\Y^{\h^-})$, where $M$ is a
$\sigma$-monomial.
Since $\A^+$ is coherent, $\prem(\Delta(A_i^+,A_j^+),\A^{+})$ $=0$.
By Lemma \ref{lm-asc1}, $\prem(\Delta(A_i,A_j),\A)=0$ which implies
that $\A$ is coherent.\qedd

We now prove the converse of Theorem \ref{th-nl3}.
\begin{thm}\label{th-nl4}
Use the notations in \bref{eq-aa}.
%
%
If $\A^+$ is a regular and coherent $\sigma$-chain, then there is a
partial character $\rho$ over $\Zxn$ such that
$L_\rho=(\f_1,\ldots,\f_s)$, $\rho(\f_i)=c_i$, $\I(\rho)=[\A]$, and
$\I^{+}(\rho)=\sat(\A^{+})$
\end{thm}
\proof
By Lemma \ref{lm-asc2}, $\A$ is regular and coherent.
By Theorem \ref{th-t4}, $\f$ is a reduced Gr\"obner basis for a
$\Zx$-lattice and $[\A]\subset\F\{\Y^{\pm}\}$ is proper.
By Theorem \ref{th-l1} and Corollary \ref{cor-bic},
there exists a partial character $\rho$
such that $L_\rho=(\f_1,\ldots,\f_s)$, $\rho(\f_i)=c_i$, and
$\I(\rho)=[\A]$. By Theorem \ref{th-nl3},
$\I^{+}(\rho)=\sat(\A^{+}(\rho))=\sat(\A^{+})$.\qedd

As a consequence of Theorem \ref{th-nl4} and Lemma \ref{lm-nl1}, we
have the following canonical representation for a normal binomial $\sigma$-ideal.
\begin{cor}\label{cor-bi41}
$\I$ is a normal binomial $\sigma$-ideal if and only if
$\I=\sat(\A^{+})$, where $\A^+$ is a regular and coherent chain
given in \bref{eq-aa}.
\end{cor}

\subsection{Perfect closure of binomial $\sigma$-ideal and binomial $\sigma$-variety}
In this section, we will show that the perfect closure of a binomial
$\sigma$-ideal is also binomial. We will also give a geometric
description of the zero set of a binomial $\sigma$-ideal.
%
%
For the perfect closure of a binomial $\sigma$-ideal, we have
\begin{thm}\label{diff-perfect1}
Let $\F$ be an algebraically closed and inversive $\sigma$-field.
Then the perfect closure of a binomial $\sigma$-ideal $\I$ is
binomial.
\end{thm}

We first remark that it is not known wether the well-mixed closure of a binomial $\sigma$-ideal is still binomial.
Before proving Theorem \ref{diff-perfect1},
we first prove several lemmas. In the rest of this section, we
assume that $\I\subseteq S=\F\{\Y\}$ and $\mb$ the set of
$\sigma$-monomials in $S$.
\begin{lem}\label{diff-anybi}
If $\I$ is a binomial $\sigma$-ideal, then
$\{\I\}:\mb$ is either $[1]$ or a binomial  $\sigma$-ideal.
\end{lem}
\proof It is easy to check
$\{\I\}\F\{\Y^{\pm}\}=\{\I\F\{\Y^{\pm}\}\}$.
By \bref{eq-nbi1},
$\{\I\}:\mb=\{\I\}\F\{\Y^{\pm}\}\cap\F\{\Y\}=\{\I\F\{\Y^{\pm}\}\}\cap\F\{\Y\}$.
Now the lemma follows from  Theorem \ref{th-per1}.\qedd

\begin{lem}\label{diff-radical}
If $\I$ is a $\sigma$-ideal in $\F\{\Y\}$, then
\begin{equation}\label{eq-radical}
  \{\I\}=\{\I\}:\mb\cap \{\I+y_1\}\cap \cdots \cap \{\I+y_n\}
\end{equation}
\end{lem}
\proof The right hand side of \bref{eq-radical} clearly contains
$\{\I\}$. It suffices to show that every reflexive prime $P$
containing  $\I$ contains one of the $\sigma$-ideals on the
right-hand side of \bref{eq-radical}. If $\{\I\}:\mb \subseteq P$,
we are done. Otherwise, there exists an element $f\in
(\{\I\}:\mb)\setminus P$ which implies that there exists a
$\sigma$-monomial $M$ such that $Mf \in \{\I\}\subseteq P$. This
implies $y_i\in P$ for some $i$. Thus, $P$ contains $\{\I+y_i\}$ as
required.\qedd

\begin{lem}\label{diff-bm1}
Let $\I$ be a binomial $\sigma$-ideal in $S=\F\{\Y\}$ and
$S'=\F\{y_1,\ldots,y_{n-1}\}$. If $\I'=\I\cap S'$, then $[\I+y_n]$
is the sum of $[\I'S+y_n]$ and a monomial $\sigma$-ideal in $S'$.
\end{lem}
\proof  Every $\sigma$-binomial involving $y^{x^{k}}_n$ is either
contained in $[y_n]$ or is congruent modulo $[y_n]$ to a
$\sigma$-monomial in $S'$. Thus, all generators of $\I$ which are
not in $\I'$ may be replaced by $\sigma$-monomials in $S'$ when
forming a generating set for $[\I+y_n]$.\qedd

\begin{lem}\label{diff-perfect}
Let $\I$ be a perfect binomial $\sigma$-ideal in $S=\F\{\Y\}$. If
$\MI$ is a $\sigma$-monomial $\sigma$-ideal, then
$\{\I+\MI\}=[\I+\MI_1]$ for some monomial $\sigma$-ideal $\MI_1$.
\end{lem}
\proof
If $1\in\MI$, then the lemma is obviously valid. Otherwise,
$[\I+\MI]:\mb=[1]$. Lemma \ref{diff-radical} yields $\{\I+\MI
\}=\bigcap_{i=1}^{i=n}\{\I+\MI+y_i\}$.
By Corollary \ref{cor-tttt1}, we need only to show that
$\{\I+\MI+y_i\}$ is the sum of $\I$ and a monomial $\sigma$-ideal.
For simplicity, let $i=n$ and write
$S'=\F\{y_1,y_2,\ldots,y_{n-1}\}$. Since $\I$ is perfect, the
$\sigma$-ideal $\I'=\I\cap S'$ is perfect as well. By Lemma
\ref{diff-bm1}, $[\I+\MI+y_n]=[\I'S+\MI'S+y_n]$ where $\MI'$ is a
monomial $\sigma$-ideal in $S'$. By induction on $n$, the perfect
closure of $\I'+\MI'$ in $S'$ has the form $\I'+\MI_1'$, where
$\MI_1'$ is a monomial $\sigma$-ideal of $S'$. Putting this
together, we have
\[
 \begin{array}{llll}
 \{\I+\MI+y_n\} & =& \{\I'S+\MI'S+y_n\}
 = [\I'S+\MI_1'S+y_n]\\
 &\subseteq & [\I+\MI_1'S+y_n]
 \subseteq  \{\I+\MI+y_n\}.
\end{array}\]
So $\{\I+\MI+y_n\}=[\I+\MI_1'S+y_n]$ is $\I$ plus a monomial
$\sigma$-ideal, as required.\qedd

%
\vskip5pt{\noindent\em Proof of Theorem \ref{diff-perfect1}:} We
will prove the theorem by induction on $n$.
By Lemma \ref{diff-anybi}, $\I_1=\{\I\}:\mb$ is binomial.
For $n=1$, by Lemma \ref{diff-radical}, $ \{\I\}=\I_1\cap
\{\I+y_1\}$. If $\{\I+y_1\}=1$ then $ \{\I\}=\I_1$ is binomial.
Otherwise $\{\I+y\}=[y]$ and hence $\I\subset [y]$. Since $\I\subset
\I_1$, $\{\I\}=\I_1\cap [y]=[\I+\I_1]\cap [\I+y]$ is binomial by
Lemma \ref{cor-gbi8}.
Suppose the lemma is valid for $n-1$ variables and let $\I$ be a
binomial $\sigma$-ideal in $S=\F\{\Y\}$. Let $\I_j:=\I \cap S_j$,
where $S_j=\F\{y_1,\ldots,y_{j-1},y_{j+1},\ldots,y_n\}$. By the
induction hypothesis, we may assume that the perfect closure of each
$\I_j$ is binomial. Adding these binomial $\sigma$-ideals to $\I$,
we may assume that each $\I_j$ is perfect begin with.
By Lemma \ref{diff-anybi}, $\I_1=\{\I\}:\mb$ is binomial. Then there
exists a binomial $\sigma$-ideal $\I'$, say $\I'=\I_1$,  such that
$\I_1=[\I+\I']$. By Lemma \ref{diff-bm1}, $[\I+y_j]=[\I_jS+\J_j
S+y_j]$, where $\J_j$ is a monomial $\sigma$-ideal in $S_j$. Since
$\I_j$ is perfect, the $\sigma$-ideal $\I_jS$ is perfect, so we can
apply Lemma \ref{diff-perfect} with $\MI=[\J_j S+y_i]$ to see that
there exists a monomial $\sigma$-ideal $\MI_j$ in $S$ such that
$\{\I+y_j\}=\{\I_jS+\J_j S+y_j\}=[\I_jS+\MI_j]=[\I+\MI_j]$. By Lemma
\ref{diff-radical} and Corollary \ref{cor-gbi8},
$\{\I\} = [\I+\I']\bigcap\cap_{j=1}^n[\I+\MI_j]$ is binomial.\qedd

\begin{exmp}\label{ex-perc11}
Let $p=y_2^{2}-y_1^{2}$. Following the proof of Theorem
\ref{diff-perfect1},  $\{p\} = (\{p\}:\mb)\cap[y_1,y_2]$.
By Example \ref{ex-perc1} and Corollary \ref{cor-ascr},
 $\I_1=\{p\}:\mb=\sat[y_2^{2}-y_1^{2},y_1y_2^{x}-y_1^{x}y_2]= [y_1y_2^{x^i}-y_1^{x^i}y_2,y_2^{1+x^j}-y_1^{1+x^j}\,|\,i,j\in\N]$.
Thus, $\{p\} = \I_1\cap[y_1,y_2]=\I_1$.
%
%
%
\end{exmp}

%
%
%


In the rest of this section, we give a geometric description of the
zero set of a binomial $\sigma$-ideal, which is a generalization of
Theorem 4.1 in \cite{es-bi} to the difference case. The basic idea
of the proof also follows \cite{es-bi}, except we need to consider
the distinction between the perfect $\sigma$-ideals and radical
ideals.


We decompose the affine $n$-space $(\AH)^{n}$ into the union of
$2^{n}$ $\sigma$-coordinate flats:
$$ (\AH^{*})^{\Omega}:=\{(a_1,a_2,\ldots,a_n)\,|\,a_i\neq 0 , i\in \Omega ; a_i=0, i \notin \Omega  \}$$
where $\Omega$ runs over all subsets of $\{1,2,\ldots,n\}$. The Cohn
closure of $ (\AH^{*})^{\Omega}$ in $(\AH)^{n}$ is defined by the
$\sigma$-ideal
 $$M(\Omega):=[y_i|i\notin \Omega]\subset\F\{\Y\}.$$
 The
$\sigma$-coordinate ring of $ (\AH^{*})^{\Omega}$  is the Laurent
polynomial $\sigma$-ring
$\F\{\Omega^{\pm}\}:=\F\{y_i,y_i^{-1},i\in\Omega\}$. We can define a
coordinate projection $ (\AH^{*})^{\Omega'}\longrightarrow
(\AH^{*})^{\Omega}$ whenever $\Omega\subseteq \Omega'\subseteq
\{1,2,\ldots,n\}$ by setting all those coordinates not in $\Omega$
to zero.

 If $X$ is any $\sigma$-variety of $(\AH)^{n}$ and
 $\I=\IH(X)\subseteq \F\{\Y\}$, then the Cohn closure of the
 intersection of $X$ with $(\AH^{*})^{\Omega}$ corresponds to the
 $\sigma$-ideal
 $$\I_{\Omega}:=[\I+M(\Omega)]:\mb_{\Omega}\subset \F\{\Y\}$$
  where $
 \mb_{\Omega}=\{\prod_{i\in \Omega}y_i^{m_i(x)}|m_i(x)\in \N[x]\}$.
%
%
Since $\I$ is perfect, by the difference Nullstellsatz
\cite[p.87]{cohn}
 $$\I=\bigcap_{\Omega}\{\I_{\Omega}\}.$$
 If $\I$ is binomial, then by
 Corollary \ref{diff-quotient} the $\sigma$-ideal $\I_{\Omega}$ is
 also binomial.

\begin{lem}\label{diff-4}
Let $R:=\F\{z_1,z_1^{-1},\ldots,z_t,z_t^{-1}\}\subset
R':=\F\{z_1,z_1^{-1},\ldots,z_t,z_t^{-1},y_1,\ldots,y_s\}$ be a
Laurent polynomial $\sigma$-ring and a polynomial $\sigma$-ring over
it. If $B\subset R'$ is a binomial $\sigma$-ideal and $M\subset R'$
is a monomial $\sigma$-ideal such that $[B+M]$ is a proper
$\sigma$-ideal in $R'$, then $[B+M]\cap R=B\cap R.$
\end{lem}
\proof  This is a $\sigma$-version of \cite[Lemma 4.2]{es-bi}, which
can be proved similarly.\qedd

We can make a classification of all binomial $\sigma$-varieties $X$
by intersecting $X$ with $(\AH^{*})^{\Omega}$, since by Theorem
\ref{diff-perfect1}, the perfect closure of a binomial
$\sigma$-ideal is still binomial.
\begin{thm}\label{diff-structure}
Let $\F$ be any algebraically closed and inversive $\sigma$-field. A
$\sigma$-variety $X\subset \AH^{n}$ is generated by
$\sigma$-binomials if and only if  the following three conditions
hold.
\begin{description}
\item[(1)] For each $(\AH^{*})^{\Omega}$, the $\sigma$-variety
$X\cap(\AH^{*})^{\Omega}$ is generated by $\sigma$-binomials.

\item[(2)] The family of sets $U=\{\Omega\subseteq
\{1,2,\ldots,n\}|X\cap(\AH^{*})^{\Omega}\neq \emptyset$\} is closed
under taking intersections.

\item[(3)] If $\Omega_1,\Omega_2\in U $ and $\Omega_1 \subset \Omega_2$,
then the coordinate projection $(\AH^{*})^{\Omega_2}\longrightarrow
(\AH^{*})^{\Omega_1} $ maps $X\cap (\AH^{*})^{\Omega_2}$ onto a
subset of $X\cap (\AH^{*})^{\Omega_1}$.
\end{description}

\end{thm}

The above theorem can be reduced to the following algebraic version.
\begin{thm}\label{diff-structure-algebra}
Let $\F$ be any algebraically closed and inversive $\sigma$-field. A
perfect $\sigma$-ideal $\I\subset \F\{\Y\}$ is binomial  if and only
if the following three conditions hold.
\begin{description}
\item[(1)]
 For each $\Omega\subseteq\{1,\ldots,n\}$, $\I_{\Omega}$ is
binomial.

\item[(2)] $U=\{\Omega\subseteq
\{1,2,\ldots,n\}\,|\,\{\I_{\Omega}\}\neq [1]$\} is closed under
taking intersections.

\item[(3)] If $\Omega_1,\Omega_2\in U $ and $\Omega_1 \subset \Omega_2$,
then $\I_{\Omega_1}\cap\F\{\Omega_1\}\subset\I_{\Omega_2}$, where
$\F\{\Omega_1\}=\F\{y_i\,|\, y_i\in\Omega_1\}$.
\end{description}
\end{thm}

\proof
%
Suppose $\I$ is a perfect $\sigma$-ideal in $\F\{\Y\}$.
Since  $\I$ is binomial, by Lemma \ref{diff-quotient} $\I_{\Omega}$
is also binomial and (1) is proved.
To prove (2) by contradiction, assume that for $\Omega_1,\Omega_2\in
U$, $\{\I_{\Omega_1}\}\neq[1], \{\I_{\Omega_2}\}\neq[1],
\{\I_{\Omega_1\cap\Omega_2} \}=[1]$.  We consider two cases.
If $\I_{\Omega_1\cap\Omega_2} =[1]$, then for some $m(x)\in \N[x]$
we have  $(\prod_{i\in\Omega_1\cap\Omega_2
}y_i)^{m(x)}\in[\I+M(\Omega_1)+M(\Omega_2)]$. By Corollary
\ref{cor-tttt1},  $(\prod_{i\in\Omega_1\cap\Omega_2 }y_i)^{m(x)}$ is
either in $[\I+M(\Omega_1)]$ or $[\I+M(\Omega_2)]$, so
$\I_{\Omega_1}$ or $\I_{\Omega_2}$ is $[1]$.
For the second case, we have $\I_{\Omega_1\cap\Omega_2} \neq[1]$ and
$\{\I_{\Omega_1\cap\Omega_2}\} =[1]$. Then there exist a finite
number of proper $\sigma$-binomials $B_1,\ldots,B_s$ and
$\sigma$-monomials $m_1,\ldots,m_s$ in $\F\{\Omega_1\cap\Omega_2\}$
such that $m_iB_i\in \I$ and $\{B_1,\ldots,B_s,
y_i,i\notin\Omega_1\cap\Omega_2\}=[1]$. We thus have $
\{B_1,\ldots,B_s\}=[1]$. Since $m_iB_i\in
\I\cap\F\{\Omega_1\cap\Omega_2\}$, we have $B_i\in\I_{\Omega_1}$ and
$B_i\in\I_{\Omega_2}$ and thus
$\{\I_{\Omega_1}\}=\{\I_{\Omega_2}\}=[1]$.
To prove (3), given $\Omega_1,\Omega_2\in U$ and
$\Omega_1\subset\Omega_2$, we have
$\I_{\Omega_2}=[\I_{\Omega_2}:\mb_{\Omega_1} ]$. Set
$R'=\F\{\Omega_1^{\pm}\}\{\{y_i\}_{i\notin \Omega_1}\}$, then
 $$[\I+M(\Omega_1)]R'\cap \F\{\Omega_1^{\pm}\}\subseteq\I_{\Omega_2}R'.$$
Since $\Omega_1\in U$, the $\sigma$-ideal $[\I+M(\Omega_1)]R^{'}$ is
proper. By Lemma \ref{diff-4}, we have $[\I+M(\Omega_1)]R'\cap
\F\{\Omega_1^{\pm}\}=\I R'\cap\F\{\Omega_1^{\pm}\}\subset
\I_{\Omega_2}R'\cap\F\{\Omega_1^{\pm}\}$. So
$\I_{\Omega_1}\cap\F\{\Omega_1\}\subset\I_{\Omega_2}$.

To prove the other driection, let $\I$ be a perfect $\sigma$-ideal
satisfying the three conditions. By the difference Nullstellensatz,
$\I=\cap_{\Omega\in U}\{\I_{\Omega}\}$. By condition (2), $U$ is a
partially ordered set under the inclusion for subsets of
$\{1,\ldots,n\}$. For each $\Omega\in U$, we set
$\J(\Omega)=[\I_{\Omega}\cap \F\{\Omega\}]\F\{\Y\}$ with the
properties that if $\Omega_1\subset \Omega_2$,
$\{\J(\Omega_1)\}\subset\{\J(\Omega_2)\}$. Note that
$[M_{\Omega_1\cap\Omega_2}]\subset[M_{\Omega_1}+M_{\Omega_2}]$. Then
we have
 $$ \I=\cap_{\Omega\in U}\{\I_{\Omega}\}=\cap_{\Omega\in
 U}\{\J(\Omega)+M(\Omega)\}.$$
Now we will prove that
 \begin{equation}\label{eq-pc3}
\cap_{\Omega\in U}\{\J(\Omega)+M(\Omega)\}=\{\cap_{\Omega\in
U}M(\Omega)+\sum_{\Omega\in U}\{\J(\Omega)\cap (\cap_{\Omega_\eta
\nsupseteq \Omega}M(\Omega_\eta))\}\}.  \end{equation}
If $\Omega_2\supseteq\Omega_1$, then
$\{\J(\Omega_2)+M(\Omega_2)\}\supseteq\{\J(\Omega_2)\}\supseteq\{\J(\Omega_1)\}\supseteq
\{\J(\Omega_1)\cap \cap_{\Omega_\eta \nsupseteq
\Omega_1}M(\Omega_\eta)\}$.
If  $\Omega_2\nsupseteq\Omega_1$, we have
$\{\J(\Omega_2)+M(\Omega_2)\}\supseteq M\{\Omega_2\}\supseteq
\{\J(\Omega_1)\cap \cap_{\Omega_\eta \nsupseteq
\Omega_1}M(\Omega_\eta)\}$. So the left hand side contains the right
hand side of \bref{eq-pc3}.
For the other direction, consider a reflexive prime $\sigma$-ideal
$P\supseteq[\cap_{\Omega\in U}M(\Omega)+\sum_{\Omega\in
U}\{\J(\Omega)\cap \cap_{\Omega_\eta \nsupseteq
\Omega}M(\Omega_\eta)\}]$ and set $V=\{\Omega\in U|M(\Omega)\subset
P\}$. Then $V$ is a finite partially order set and nonempty since
$P\supseteq\cap_{\Omega\in U}M(\Omega)$ and
$\{M_{\Omega_1\cap\Omega_2}\}\subset\{M_{\Omega_1}+M_{\Omega_2}\}$.
Let $\Omega_{0}$ be the smallest element of $V$ such that
$P\supseteq M_{\Omega_{0}}$. At the same time,
$P\supset\J(\Omega_0)\cap \cap_{\Omega_\eta \nsupseteq
\Omega_0}M(\Omega_\eta)$, then $P\supseteq \J(\Omega_0)$. Therefore,
$P\supseteq \J(\Omega_0)+M(\Omega_0)$ and $P$ contains the left hand
side of \bref{eq-pc3} and \bref{eq-pc3} is proved. Since
$\cap_{\Omega\in U}M(\Omega)+\sum_{\Omega\in U}\{\J(\Omega)\cap
\cap_{\Omega_\eta \nsupseteq \Omega}M(\Omega_\eta)\}$ is binomial,
the theorem follows from \bref{eq-pc3}.\qedd
%

\section{Algorithms}
\label{sec-alg}

In this section, we give algorithms for most of the results in the preceding sections.
%
%
%
In particular, we  give an algorithm to decompose a finitely generated
perfect binomial $\sigma$-ideal as the intersection of reflexive
and prime binomial $\sigma$-ideals.
The following basic algorithms will be used.
\vskip5pt
\begin{itemize}
\item {\bf Algorithm GHNF}. Let $\fb$ be a finite set of $\Zxn$.
The algorithm computes the generalized Hermite normal form of $[\fb]$, or equivalently,
the reduced Gr\"obner basis of the $\Zx$-module
$(\fb)$ \cite[p. 197]{cox-1998}. A polynomial-time algorithm is given in \cite{GHNF-alg}.

\vskip3pt
\item {\bf Algorithm GKER}. For a matrix $M\in\Zx^{n\times s}$,
 compute a set of generators of the $\Zx$-lattice: $\ker_{\Zx}(M) = \{X\in\Zx^s\,|\, MX=0\}$.
 This can be done by combining Algorithm {\bf{GHNF}} and Theorem \ref{lm-free}.
\end{itemize}

\vskip5pt
Let $\D$ be $\Z$, $\Q[x]$, $\Z_p[x]$, or $\Q[x]/(q(x))$, where
$q(x)$ is an irreducible polynomial in $\Q[x]$. Then $\D$ is either
a PID or a field. The following algorithms will be used.

\vskip5pt
\begin{itemize}
\item {\bf Algorithm HNF}. For  $M\in\D^{n\times s}$, compute the Hermite normal form of
$M$ \cite[p.68]{cohen}.

\vskip3pt
\item {\bf Algorithm KER}. For a matrix $M\in\D^{n\times s}$, compute a basis for the $\D$-module: $\ker_{\D}(M) = \{X\in\D^s\,|\, MX=0\}$ \cite[p.74]{cohen}.
\end{itemize}

\subsection{$x$-saturation of $\Zx$-lattice}
\label{sec-alg1}

In this section, we give algorithms to check whether a $\Zx$-lattice
$L$ is $x$-saturated and in the negative case to compute the
$x$-saturation of $L$.

Let $\f_1,\ldots,\f_s\in\Zxn$ and $L=(\f_1,\ldots,\f_s)$. If $L$ is
not $x$-saturated, then there exist $g_i\in\Zx$ such that
 $\sum_{i=1}^s g_i \f_i = x \h$
and $\h\not\in L$. Setting $g_i(x) = g_i(0) + x\widetilde{g}_i(x)$
and $\widetilde{\h} = \h - \sum_{i=1}^s \widetilde{g}_i(x)\f_i$, we
have
  \begin{equation}\label{eq-x11}
 \sum_{i=1}^s g_i(0) \f_i = x \widetilde{\h}
 \end{equation}
where $\widetilde{\h}\not\in L$. Setting $x=0$ in the above equation, we
have
  $$\sum_{i=1}^s g_i(0) \f_i(0) = 0,$$
that is, $G=(g_1(0),\ldots,g_s(0))^\tau$ is in the kernel of the
matrix $F=[\f_1(0),\ldots,\f_s(0)]\in\Z^{n\times s}$, which can be
computed efficiently \cite[page 74]{cohen}. From $G$ and
\bref{eq-x11}, we may compute $\widetilde{\h}$. This observation
leads to the following algorithm.

\begin{algorithm}[H]\label{alg-satX}
  \caption{\bf --- XFactor$([\f_1,\ldots,\f_s])$} \smallskip
  \Inp{A generalized Hermite normal form $[\f_1,\ldots,\f_s]\subset\Zx^{n\times s}$.}\\
  \Outp{$\emptyset$, if the $\Zx$-lattice $L = (\f_1,\ldots,\f_s)$ is $x$-saturated;
    otherwise, a finite set
    $\{(\h_i,\e_i)\,|\, i=1,\ldots,r\}$ such that
    $\e_i=(e_{i1},\ldots,e_{is})^\tau\in\Z^s$, $\h_i\not\in L$, and
     $x \h_i = \sum_{l=1}^s e_{il} \f_l \in L$,
    $i=1,\ldots,r$.}\medskip

  \noindent
  1. Set $F=[\f_1(0),\ldots,\f_s(0)]\in\Z^{n\times s}$.\\
  2. Compute a basis $E\subset\Z^s$ of the $\Z$-module $\ker_{\Z}(F)$ with Algorithm {\bf KER}.\\
  3. Set $H = \emptyset$.\\
  4. While $E \not= \emptyset$\\
   \SPC 4.1. Let $\e=(e_1,\ldots,e_s)^\tau\in E$ and $E = E\setminus \{\e\}$.\\
   \SPC 4.2. Let $\h=(e_1\f_1+\cdots+e_s\f_s)/x$.\\
   \SPC 4.3. If $\grem(\h,\{\f_1,\ldots,\f_s\})\ne0$,
   then add $(\h,\e)$ to $H$.\\
  5. Return $H$.
\smallskip
\end{algorithm}

We now give the algorithm to compute the $x$-saturation of a
$\Zx$-lattice.
\begin{algorithm}[H]\label{alg-satX1}
  \caption{\bf --- SatX$(\f_1,\ldots,\f_s)$} \smallskip
  \Inp{A finite set $\fb=\{\f_1,\ldots,\f_s\}\subset \Zxn$.}\\
  \Outp{A set of generators of $\sat_{x}(\f_1,\ldots,\f_s)$ .}\medskip

  \noindent
  1. Compute the generalized Hermite normal form $\gb$ of $\fb$ with Algorithm {\bf{GHNF}}. \\
  2. Set $H=$ {\bf XFactor}$(\gb)$.\\
  3. If $H=\emptyset$, then output $\gb$;
      otherwise set $\fb = \col(\gb)\cup \{\h\,|\, (\h,\f)\in H\}$ and goto step 1.  \\
\smallskip

Note. $\col(\gb)$ is the set of columns of $\gb$.
\end{algorithm}

\begin{exmp}
Let $\C$ be the following generalized Hermite normal form.
\[ \C =[\f_1,\f_2,\f_3]
 = \left[  \begin{array}{llllll}
    -x+2               & 1      & 1\\
    3x+2               & 1      & 2x+1 \\
    0                  & 2x     & x^2 \\
\end{array} \right]. \]
In {\bf XFactor}$(\C)$, the kernel of the following matrix
$
 [\f_1(0), \f_2(0), \f_3(0)] = \left[  \begin{array}{llllll}
  2               & 1      & 1\\
  2               & 1      & 1 \\
  0               & 0     & 0 \\
\end{array} \right]
$
is generated by $\e_1 = (0, -1,1)^\tau$ and $\e_2 = (1,-2,0)^\tau$.
In step 4.2 of {\bf XFactor}, we have  $ \h = (-\f_2+\f_3)/x = (0,2,x-2)^\tau$. One can check that
$(0,2,x-2)^\tau\not\in (\C)$.
In {\bf SatX}, computing the generalized Hermite normal form of
$\C\cup \{ (0,2,x-2)^\tau \} $, we have
\[ \C_1
 = \left[  \begin{array}{llllll}
    -x+2               & 1      & 0\\
    3x+2               & -3      & 2 \\
    0                  & 4     & x-2 \\
\end{array} \right]. \]
{\bf XFactor}$(\C_1)$ returns $\emptyset$. So, $(\C_1)$ is
 the $x$-saturation of $(C)$.
%
\end{exmp}

\begin{prop}
Algorithms~{\bf SatX} and {\bf XFactor} are correct.
\end{prop}
\proof From the output of Algorithm~{\bf XFactor}, in step 3 of {\bf
SatX}, we have $(\gb)\subsetneq(\gb\cup \{\h\,|\, (\h,\f)\in H\})\subseteq\sat_x(\fb)$.
Since $\Z[x]^n$ is a Noetherian $\Zx$-module, {\bf SatX} will
terminate and return the $x$-saturation of $(\fb)$. So, it suffices to
show the correctness of Algorithm {\bf XFactor}.

We first explain step 4.2 of Algorithm {\bf XFactor}. Since
$\e\in\ker_{\Z}(F)$, $\h(0)=[\f_1(0),\ldots,$ $\f_s(0)]\e =
[0,\ldots,0]^\tau$. Therefore, $x$ is a factor of $e_1\f_1+\cdots
+e_s\f_s$ and thus $\h=(e_1\f_1+\cdots +e_s\f_s)/x\in\Zxn$.

To prove the correctness of Algorithm {\bf XFactor}, it suffices to
show that $L=\sat_x (L)$ if and only if for each $\e\in E$,
$e_1\f_1+\cdots +e_s\f_s = x\h$ implies $\h\in L$.

Let $E=\{\e_1,\ldots,\e_k\}$ where $\e_i\in\Z^s$. If $L=\sat_x (L)$,
then it is clear that $(\f_1,\ldots, \f_s)\e_i = x\h_i$ implies
$\h_i\in L$. To prove the other direction, let $[\f_1,\ldots,
\f_s]\e_i = x\h_i$ for $1\le i\le k$, where $\h_i\in L$.
Let $x\f\in L$. Then $x\f = \sum_{i=1}^s c_i(x)\f_i$, where
$c_i(x)\in\Z[x]$. If for each $i$, $x | c_i(x)$, then we have $\f =
\sum_{i=1}^s (c_i(x)/x) \f_i \in L$, and the lemma is proved.
Otherwise, set $x=0$ in $x\f = \sum_{i=1}^s c_i(x)\f_i$, we obtain
$\sum_{i=1}^s c_i(0)\f_i(0) = 0$. Hence
$Q=[c_1(0),\ldots,c_s(0)]^\tau \in \ker_{\Z}(F)$ and hence there exist
$a_i\in\Z, i=1,\ldots,k$ such that $Q = \sum_{i=1}^k a_i\e_i$. Then,
$
 \begin{array}{llll}
 [\f_1,\ldots, \f_s]Q & =& \sum_{i=1}^k a_i [\f_1,\ldots, \f_s]\e_i
  =  \sum_{i=1}^k a_i x\h_i = x\widetilde{\h}, \\
\end{array}
$
where $\widetilde{\h}=\sum_{i=1}^k a_i \h_i\in L$. Then,
\[
 \begin{array}{llll}
 x\f & =& \sum_{i=1}^s c_i(x)\f_i
 = \sum_{i=1}^s c_i(0)\f_i + \sum_{i=1}^s x\overline{c}_i(x)\f_i \\
 &=& [\f_1,\ldots, \f_s]Q + x\sum_{i=1}^s \overline{c}_i(x)\f_i
 = x\widetilde{\h} + x\sum_{i=1}^s \overline{c}_i(x)\f_i,
\end{array}\]
where $\overline{c}_i(x) = (c_i(x)-c_i(0))/x \in \Z[x]$. Hence,  $\f
=\widetilde{\h} + \sum_{i=1}^s \overline{c}_i(x)\f_i \in L$ and the
lemma is proved. \qedd



\subsection{$\Z$-saturation of $\Zx$-lattice}
\label{sec-alg2}

The key idea to compute $\sat_\Z(L)$ for a $\Zx$-lattice
$L\in\Z[x]^n$ is as follows. Let $\fb=\{\f_1,\ldots,\f_s\}$.
Then $(\fb)$ is not $\Z$-saturated if and only if a linear
combination of $\f_i$ contains a nontrivial prime factor in $\Z$,
that is, $\sum_i g_i\f_i = p \f$, where $p$ is a prime number and
$\f\not\in (\fb)$.
%
%
%
Furthermore, $\sum_i g_i\f_i = p \f$ with $g_i\ne 0~\mod ~p$ is
valid if and only if $\f_1,\ldots,\f_s$ are linear dependent over
$\Z_p[x]$. The fact that $\Z_p[x]$ is a PID allows us to compute
such linear relations using methods of Hermite normal
forms~\cite{cohen}. The following algorithm is based on this
observation.

\begin{algorithm}[H]\label{alg-satZ1}
  \caption{\bf --- ZFactor} \smallskip
  \Inp{A generalized Hermite normal form $\C=[\c_1,\ldots,\c_s]\subset \Zxn$ of form \bref{ghf}.}\\
  \Outp{$\emptyset$, if $L = (\C)$ is  $\Z$-saturated;
    otherwise, a finite set $\{(\h_i,k_i,\e_i)\,|\,i=1,\ldots,r\}$, such that
    $\h_i\in\Zxn$, $k_i\in\N$, $\e_i=(e_{i1},\ldots,e_{is})^\tau\in\Zx^s$,
    $\h_i\not\in L$ and $k_i \h_i = \sum_{l=1}^s e_{il} \c_l \in L$ for $i=1,\ldots,r$.}\medskip

  \noindent
  1. Read the numbers $t,r_i,l_i, c_{r_i,1,0},i=1,\ldots,t$ from \bref{ghf}.\\
  2. Set $ q = \prod_{i=1}^t c_{r_i,1,0}\in\N $.\\
  3. For any prime factor $p$ of $q$ do\\
    \SPC 3.1.    Set $F=[ \c_{r_1,l_1},\c_{r_2,l_2},\ldots,\c_{r_t,l_t}]\in \Z_p[x]^{n\times t}$.\\
    \SPC 3.2.  Compute a basis $G\subset\Z_p[x]^s$ of the $\Z_p[x]$-module $\ker_{\Z_p[x]}(F)$\\
    \SPC\SPC             with Algorithm {\bf{KER}}.\\
    \SPC 3.3.  If $G\ne\emptyset$, for each $\g=[g_1,\ldots,g_t]^\tau\in G$, let
              $\sum_{i=1}^t g_i\c_{r_i,l_i} = p\h$ in $\Zxn$. \\
    \SPC\SPC  Return the set of $(\h,p,\e)$, where $\e=[e_1,\ldots,e_s]^\tau\in\Zx^s$ such that\\
    \SPC\SPC  $e_{s_k} =g_k,s_k=\sum_{i=1}^k l_i, k=1,\ldots,t$ and $e_{j}=0$ for other $j$.\\
   \SPC 3.4.  Compute the Hermite normal form
     $\B=\{\b_1,\ldots,\b_t\}$\\
   \SPC\SPC of $\{ \c_{r_1,l_1},\ldots,\c_{r_t,l_t} \}$ in $\Z_p[x]^n$ with Algorithm {\bf{HNF}}.\\
   \SPC 3.5.  Let $\C_{-} = \{\f_1,\ldots,\f_l\}$ be given in \bref{eq-inf} and
              $\widetilde{\f}_i =\grem(\f_i,\B)= \f_i +
                \sum_{k=1}^t a_{i,k} \b_k$,\\
    \SPC\SPC           in $\Z_p[x]^n$,   where
                $a_{i,k}\in\Z_p[x]$.\\
   \SPC 3.6.  If $\widetilde{\f}_i=0$ for some $i$, then
             $\f_i + \sum_{k=1}^t a_{i,k} \b_k = p\h_i$ in $\Zxn$.\\
    \SPC\SPC   Return the set of $(\h_i,p,\e_i)$ where $\e_i$ is a vector in $\Zx^s$
    such that\\
   \SPC\SPC  $(\c_1,\ldots,\c_s)\e_i=\f_i + \sum_{k=1}^t a_{i,k} \b_k= p\h_i$.  \\
\end{algorithm}

\begin{algorithm}[H]\label{alg-satZ11}
    \SPC 3.7. Set $E=[\widetilde{\f}_1,\ldots,\widetilde{\f}_l]\in\Z_p[x]^{n\times l}$.\\
   \SPC 3.8. Compute a basis $D$ of $\{X\in\Z_p^l\,|\, EX=0\}$ as a vector space over $\Z_p$. \\
   \SPC 3.9.  If $D\ne\emptyset$, for each $\b=[b_1,\ldots,b_l]^\tau\in D$,
              $\sum_{i=1}^l b_i\widetilde{\f}_{_i} = p\h$ in $\Zxn$.\\
   \SPC\SPC            Return the set of $(\h,p,\e)$ where $\e$ is a vector in $\Zx^s$
    such that\\
    \SPC\SPC $(\c_1,\ldots,\c_s)\e=\sum_{i=1}^l b_i\widetilde{\f}_{_i} = p\h$.  \\
  4. Return $\emptyset$.
\smallskip
\end{algorithm}

\begin{rem}
In steps 3.6 and 3.9, we need to compute $\e_i$ or $\e$.
Since $\B=\{\b_1,\ldots,\b_t\}$ is the Hermite normal form of $\c=\{
\c_{r_1,l_1},\ldots,\c_{r_t,l_t} \}$ in $\Z_p[x]^n$, there exists an
invertible matrix $M_{t\times t}$ such that $[\b_1,\ldots,\b_t] =
[\c_{r_1,l_1},\ldots,\c_{r_t,l_t}] M$.
In Step 3.6, $\e_i$ can be obtained from the relation $\f_i +
\sum_{k=1}^t a_{i,k} \b_k = p\h_i$ and the relation
$[\b_1,\ldots,\b_t] = [\c_{r_1,l_1},\ldots,\c_{r_t,l_t}] M$. Step
3.9 can be treated similarly.
\end{rem}

\begin{rem}
In step 3.8, we need to compute a basis for the vector space
$\{X\in\Z_p^l\,|\, EX=0\}$ over $\Z_p$. We will show how to do this.
A matrix $F\in\Z_p[x]^{m\times s}$ is said to be in standard form if
$F$ has the structure in \bref{ghf} and $\deg(c_{r_i,k_1},x) <
\deg(c_{r_i,k_2},x)$ for $i=1,\ldots,t$ and $k_1 < k_2$.

The matrix $E\in\Z_p[x]^{n\times l}$ can be transformed into
standard form using the following operations: (1) exchange two
columns and (2) add the multiplication of a column by an element
from $\Z_p$ to another column. Equivalently, there exists an
inversive matrix $U\in\Z_p^{l\times l}$ such that $E\cdot U = S$ is
in standard form. Suppose that the first $k$ columns of $S$ are zero
vectors. Then the first $k$ columns of $U$ constitute a basis for
$\ker(E)$.
This can be proved similarly to that of the algorithm to compute a
basis for the kernel of a matrix over a PID \cite[page 74]{cohen}.
\end{rem}


We now give the algorithm to compute the $\Z$-saturation.
\begin{algorithm}[H]\label{alg-satZ}
  \caption{\bf --- SatZ($\f_0,\ldots,\f_s$)} \smallskip
  \Inp{A set of vectors $\fb=\{\f_0,\ldots,\f_s\}\subset\Zxn$.}\\
  \Outp{A reduced Gr\"obner basis $\gb$ such that $(\gb)=\sat_{\Z} (\fb)$.}\medskip

  \noindent
  1. Compute generalized Hermite normal form $\gb$ of $\fb$.\\
  2. Set $S=${\bf ZFactor}$(\gb)$.\\
  3. If $S=\emptyset$, return $\gb$;
   otherwise set $\fb = \col(\gb)\cup \{\h\,|\, (\h,k,\f)\in S\}$ and goto step 1.
\smallskip
\end{algorithm}

\begin{exmp}
Let $\C$ be the following generalized Hermite normal form:
 \[ \C = \left[
\begin{array}{llllll}
    x^2+2x-2               & x+2      & 1 \\
    0                      &  4       & 2x \\
\end{array} \right]. \]
Then, $t=2, r_1 = 1, l_1 = 1, r_2 = 2, l_2 = 2, q = 4$, $\c_{1,1} =
[x^2+2x-2,0]^\tau$, $\c_{2,1} = [x+2,4]^\tau$, $\c_{2,2} =
[1,2x]^\tau$.
Apply algorithm {\bf ZFactor} to $\C$. We have $p=2$. In steps 3.1
and 3.2, $F =\left[\begin{array}{llllll}
    x^2       & 1\\
    0         & 0\\
\end{array}\right]$ and $\ker(F)$ is generated by
$G=\{[-1,x^2]^\tau\}$.
In step 3.3, $x^2\c_{22}-\c_{11}= 2(1-x,x^3)^\tau$ and return
$(1-x,x^3)^\tau$.

In Algorithm {\bf SatZ},  $(1-x,x^3)^\tau$ is added into $\C$ and
\[ \C_1 = \left[  \begin{array}{llllll}
    x^2+2x-2               & x+2      & 1  & 1-x\\
    0                      &  4       & 2x & x^3 \\
\end{array} \right], \]
which is also a generalized Hermite normal form.

Applying Algorithm {\bf ZFactor} to $\C_1$. We have $p=2$ and $t=2$.
In steps 3.1-3.3, $G=\emptyset$. In step 3.4,
 $\B=\left[\begin{array}{llllll}
    x^2       & 1-x\\
    0         & x^3\\
\end{array}\right]$.
In step 3.5, $\C_{-} =\left[  \begin{array}{lll}
     x +2  & 1 & x \\
        4    & 2x & 2x^2  \\
\end{array} \right]$ and $\widetilde{\f}_i\ne0$ for all $i$.
In step 3.7, $E=
 \left[  \begin{array}{llllll}
       x   & 1 & x \\
       0   & 0 & 0 \\
\end{array} \right]. $
In Step 3.8, $D=\{\b\}$, where $\b=[1,0,-1]^\tau$.
In Step 3.9,  $(x+2,4)^\tau - (x,2x^2)^\tau = 2(1,2-x^2)^\tau$.
Add $(1,2-x^2)^\tau$  into $\C_1$ and
compute the generalized Hermite normal form, we have
\[ \C_2 = \left[  \begin{array}{llllll}
    x^2+2x-2               & x+2      & 1  & -1   \\
    0                      &  4       & 2x & x^2-2 \\
\end{array} \right]. \]
Apply Algorithm {\bf ZFactor} again, it is shown that $\C_2$ is
$\Z$-saturated.
\end{exmp}

We will prove the correctness of the algorithm. We denote by
$\sat_p(L)$ the set $\{\f\in\Zxn\, |\, p\f\in L \}$ where $p\in\Z$ is a
prime number.
An infinite set $S$ is said to be {\em linear independent} over a
ring $R$ if any finite set of $S$ is linear independent over $R$,
that is $\sum_{i=1}^k a_i \g_i=0$ for $a_i\in R$ and $\g_i \in S$
implies $a_i=0, i=1,\ldots, k$.

\begin{lem}\label{lm-satz1}
Let  $\C$ be a generalized Hermite normal
form and $L=(\C)$. Then $\sat_p(L) = L$ if and only if $\C_{\infty}$
is linear independent over $\Z_p$, where $\C_{\infty}$
is defined in \bref{eq-inf}.
\end{lem}
\proof  $``\Rightarrow"$ Assume the contrary, that is, $\C_{\infty}
= \{\h_1,\h_2,\ldots\}$ defined in \bref{eq-inf} is linear dependent
over $\Z_p$. Then there exist $a_i\in\Z_p$ not all zero, such that
$\sum_{i=1}^r a_i\h_i = 0$ in $\Z_p[x]^n$ and hence $\sum_{i=1}^r
a_i\h_i = p\g$ in $\Z[x]^n$.
By Lemma \ref{lm-ext1}, $\h_i$ are linear independent over $\Z_p$
and hence $\g\ne{\bf{0}}$.
Since $\sat_p(L) = L$, we have $\g\in L$.
By Lemma \ref{lm-ext2}, there exist $b_i\in\Z$ such that $\g =
\sum_{i=1}^r b_i\h_i$. Hence $\sum_{i=1}^r(a_i-pb_i)\h_i = 0$ in
$\Zxn$. By Lemma \ref{lm-ext1}, $a_i = pb_i$ and hence $a_i=0$ in
$\Z_p[x]$, a contradiction.

$``\Leftarrow"$ Assume the contrary, that is, there exists a
$\g\in\Z[x]^n$, such that $\g\not\in L$ and $p\g \in L$. By Lemma
\ref{lm-ext2}, $p\g = \sum_{i=1}^r a_i\h_i$, where $a_i\in\Z$. $p$
cannot be  a factor of all $a_i$. Otherwise,  $\g = \sum_{i=1}^r
\frac{a_i}{p}\h_i\in L$. Then some of $a_i$ is not zero in $\Z_p$,
which means $\sum_{i=1}^r a_i\h_i=0$ is nontrivial linear relation
among $\C_i$ over $\Z_p$, a contradiction. \qedd

From the $``\Rightarrow"$  part of the above proof, we have
\begin{cor}\label{cor-satz1}
Let  $\sum_{i=1}^r a_i \h_i=0$ be a nontrivial linear relation among
$\h_i$ in $\Z_p[x]^n$, where $a_i\in \Z_p$. Then, in $\Z[x]^n$,
$\sum_{i=1}^r a_i \h_i=p\h$ and $\h\not\in (\C)$.
\end{cor}

\begin{lem}\label{lm-satz11}
Let  $\C=[\c_1,\ldots, \c_s]$ be a generalized Hermite normal
form and $L=(\C)$. Then $\sat_p(L) = L$ if and only if $\C_{\infty}$
is linear independent over $\Z_p$ for the prime factors of $q$
defined in step 2 of Algorithm {\bf ZFactor}.
\end{lem}
\proof By Definition \ref{def-ghf}, the leading monomial of
$x^k\c_{r_i,j}\in\C_{\infty}$ is of the form
$c_{r_i,j,0}x^{k+d_{r_i,j}}{\beps}_{r_i}$ and $c_{r_i,l_i,0} |\ldots
| c_{r_i,2,0} $ $| c_{r_i,1,0}$.
If  $p$ is coprime with $\prod_{i=1}^t c_{r_i,1,0}$, then
$c_{r_i,j,0}\ne0\,\, \mod \,\, p$ for $1\le j\le l_i$. Therefore,
the leading monomials of the elements of $\C_{\infty}$ are linear
independent over $\Z_p$, and hence $\C_{\infty}$ is linear
independent over $\Z_p$. Therefore, it suffices to consider prime
factors of $\prod_{i=1}^t c_{r_i,1,0}$.\qedd

To check whether $\C_{\infty}$ is linear independent over $\Z_p$, we
first consider a subset of $\C_{\infty}$ in the following lemma.
\begin{lem}\label{lm-satz2}
Let  $\C$ be the generalized Hermite normal form given in
\bref{ghf}. Then $\C^{+}$ defined in \bref{eq-inf} is linear
independent over $\Z_p$ if and only if
$\{\c_{r_1,l_1},\c_{r_2,l_2},\ldots,\c_{r_t,l_t}\}$ are linear
independent over $\Z_p[x]$.
\end{lem}
\proof This is obvious since $\sum_{i} \sum_j a_{i,j} x^j
\c_{r_i,l_i} =\sum_{i} p_i \c_{r_i,l_i}$, where $a_{i,j}\in\Z$ and
$p_i = \sum_j a_{i,j} x^j$.\qedd

\begin{lem}\label{lm-satz3}
Let  $\B$ be a Hermite normal form  in $\Z_p[x]^n$ and
$\gb=\{\g_1,\ldots,\g_r\}\subset\Z_p[x]^n$. Then $\gb\cup
\B_{\infty}$ is linear dependent over $\Z_p$ if and only if
\begin{itemize}
\item either $\widetilde{\g}_i=\grem(\g_i,\B)=0$ in $\Z_p[x]^n$ for some $i$, or
\item the residue set $\{\grem(\g_i,\B)\, |\, i=1,\ldots,r\}$ are linear
dependent over $\Z_p$.
\end{itemize}
\end{lem}
\proof We may assume that $\grem(\g_i,\B)=0$ does not happen, since
it gives a nontrivial linear relation of $\gb\cup \B_{\infty}$.
By Lemma \ref{lm-ext2}, $\widetilde{\g}_i =\g_i$ $\mod \,
\B_{\infty}$.
$\gb\cup \B_{\infty}$ is linear dependent over $\Z_p$ if and only if
there exist $a_i\in\Z_p$ not all zero such that $\sum_i a_i\g_i=0$
\mod \, $\B_{\infty}$ over $\Z_p$, which is valid if and only if
$\sum_i a_i \widetilde{\g_i}=0$ $\mod \,$ $\B_{\infty}$. Since
$\widetilde{\g}_i$ are G-reduced with respect to $\B$, $\sum_i a_i
\widetilde{\g}_i=0$ $\mod \,$ $\B_{\infty}$ if and only if $\sum_i
a_i \widetilde{\g}_i=0$, that is $\widetilde{\g}_i$ are linear
dependent over $\Z_p$.\qedd

\begin{prop}
Algorithm~{\bf SatZ} is correct.
\end{prop}
\proof Since $\Zxn$ is Notherian, the algorithm terminates and it suffices to show that Algorithm
{\bf ZFactor} is correct.
Let $C = [\c_1,\ldots,\c_s]$. By Lemma~\ref{lm-satz1}, to check
whether $\sat_\Z(\c_1,\ldots,\c_s)$ is $\Z$-saturated, it suffices
to check for any prime $p$, $\C_{\infty}$ is linear independent on
$\Z_p$.
Furthermore, by Lemma \ref{lm-satz11}, it suffices to consider prime factors of
$\prod_{i=1}^t c_{r_i,1,0}$ in step 3.
 This explain why  only prime factors of $q$ are considered in Step 3.

In steps 3.1 and 3.2, we check whether $\C^{+}$ in \bref{eq-inf} is
linear independent over $\Z_p$. By Lemma \ref{lm-satz2}, we need
only to consider whether $\C_1 = \{\c_{r_1,l_1},
\c_{r_2,l_2},\ldots, \c_{r_t,l_t}\}$ is linear independent over
$\Z_p[x]$. It is clear that $\C_1$ is linear independent over
$\Z_p[x]$ if and only if $G=\emptyset$, where $G$ is given in step
3.2.

In step 3.3,  $\C_1$ is linear dependent over
$\Z_p$. If $G\ne\emptyset$ for any $\g=[g_1,\ldots,g_t]^\tau\in G$,
$\sum_{i=1}^t g_i\c_{r_i,l_i} =0$ in $\Z_p[x]$. Hence $\sum_{i=1}^t
g_i\c_{r_i,l_i}=p\h$ where $\h\in\Zxn$. By Corollary
\ref{cor-satz1}, $\h\not\in L$. The correctness of Algorithm {\bf
ZFactor} is proved in this case.

In steps 3.4 - 3.10, we handle the case where $\C^{+}$ is linear
independent over $\Z_p$.
In step 3.4, we compute the Hermite normal form of $\C_1$ in
$\Z_p[x]^n$, which is possible because $\Z_p[x]^n$ is a PID
\cite{cohen}. Furthermore, we have \cite{cohen}
 $$[\c_{r_1,l_1},\ldots,\c_{r_t,l_t}] N =[\b_1,\ldots,\b_t]$$
where $\{\b_1,\ldots,\b_t\}$ is a Hermite normal form and $N$ is an
inversive matrix in $\Z_p[x]^{t\times t}$.
Then $\C_{\infty}=\C_{-} \cup\C^{+}$ is linear independent over
$\Z_p$ if and only if
\begin{equation}\label{eq-satzp1}
 \widetilde{\C} = \C_{-} \cup \B_{\infty} = \C_{-} \cup
\cup_{j=0}^{\infty}\{x^j\b_1,\ldots,x^j\b_t\}
 \hbox{ is linear independent over }\Z_p.
\end{equation}
By Lemma \ref{lm-satz3}, property \bref{eq-satzp1} is valid if and
only if $\grem(\c,\B)\ne0$ for all $\c\in\C_{-}$ and the residue set
$\widetilde{\C}_{-}$ is linear independent over $\Z_p$, which are
considered in step 3.7 and steps 3.8-3.10, respectively.
Then we either prove $L$ is $\Z$-saturated or find a nontrivial
linear relation for elements in $\C_{\infty}$ over $\Z_p$. By
Corollary \ref{cor-satz1}, such a relation leads to an
$\h\in\sat_{\Z}(L)\setminus L$. The correctness of the algorithm is
proved.\qedd


As a direct consequence of Lemma \ref{lm-wmc} and Algorithm~{\bf ZFactor},
we have the algorithm to compute the $M$-saturation.
\begin{algorithm}[H]\label{alg-satM}
  \caption{\bf --- SatM($\f_0,\ldots,\f_s$)} \smallskip
  \Inp{A set of vectors $\fb=\{\f_0,\ldots,\f_s\}\subset\Zxn$.}\\
  \Outp{A generalized Hermite normal form $\gb$ such that $\sat_M(\fb)=(\gb)$.}\medskip

  \noindent
  1. Using Algorithm~{\bf ZFactor},
  we can compute $m_i\in\N$ and $\g_i\in\Zxn,i=1,\ldots,s$\\
  \SPC such that
     $\sat_\Z (\fb) = (\g_1,\ldots,\g_s)$  and $m_i\g_i\in (\fb)$.\\
  2. Let $S=\emptyset$ and for $i=1,\ldots,s$, if $m_i\ne1$ then $S=S\cup\{(x-o_{m_i})\g_i\}$.\\
  3. Compute the generalized Hermite normal form $\gb$ of $\fb\cup S$ and return $\gb$.\\
  \smallskip
\end{algorithm}

Notices that if $m_i=1$ then $o_{m_i}=0$ and $\g_i\in(\fb)$.
The numbers $m_i$ need not to be unique for the following reasons.
Suppose $m_i=n_ik$ and $n_i\g_i\in (\fb)$.
Then by Corollary \ref{cor-per2},
$o_{m_i} = o_{n_i} + c n_i$ and hence $(x-o_{n_i})\g_i = (x-o_{m_i})\g_i + cn_i\g_i\in (\fb)$.
That is, we can replace $m_i$ by its factor $n_i$.

\begin{prop}\label{lm-satm1}
Algorithm {\bf{SatM}} is correct.
\end{prop}
\proof
Let $L_1=(\fb)$ and $L_2=(\fb, (x-o_{m_1})\g_1,\ldots,(x-o_{m_s})\g_s)$.
We claim that $\sat_\Z(L_1) = \sat_\Z(L_2)$.
Since $L_1\subset L_2$, $\sat_\Z(L_1) \subset \sat_\Z(L_2)$.
Since $\sat_\Z (L_1) = (\g_1,\ldots,\g_s)$, we have $L_2\subset\sat_\Z (L_1)$
and hence $\sat_\Z(L_2)\subset\sat_\Z (L_1)$. The claim is proved.
Then $\sat_\Z(L_2)=(\g_1,\ldots,\g_s)$ and $m_i\g_i\in L_1\subset L_2$.
Since $(x-o_{m_i})\g_i\in L_2,i=1,\ldots,s$, $L_2$ is $M$-saturated by Lemma \ref{lm-wmc}.\qedd

\subsection{Algorithms for Laurent binomial and binomial $\sigma$-ideals}
\label{sec-alg4}
In this section, we will present several algorithms for Laurent
binomial and binomial $\sigma$-ideals, and in particular a
decomposition algorithm for binomial $\sigma$-ideals.
%
%
We first give an algorithm to compute the characteristic set for a
Laurent binomial $\sigma$-ideal.
%
%
\begin{algorithm}[H]\label{alg-cs}
  \caption{\bf --- CharSet} \smallskip
  \Inp{$F$: a finite set of Laurent $\sigma$-binomials in $\F\{\Y^{\pm}\}$.}\\
  \Outp{$\emptyset$, if $[F]=[1]$;
  otherwise, a regular and coherent Laurent binomial $\sigma$-chain $\A$ such that
    $[\A]=[F]$ and $\A$ is a characteristic set of the $\sigma$-ideal $[F]$.}

 \vskip2pt
 \noindent
 1. Let $F=\{\Y^{\f_1}-c_1,\ldots,\Y^{\f_r}-c_r\}$ and $\fb=\{\f_1,\ldots,\f_r\}$.\\
 2. Compute a set of generators $H\subset\Zx^r$ of $\ker_{\Zx}([\f_1,\ldots,\f_r])$
 with Algorithm {\bf{GKER}}.\\
 3. If for any $\h = (h_{1}, \ldots, h_{r})^\tau\in H$, $\prod_{i=1}^r c_i^{h_{i}}\ne1$,
 then return $\emptyset$.\\
 4. Compute the reduced Gr\"obner basis $\gb$ of $\fb$ with Algorithm {\bf{GHNF}}.\\
 5. Let $\gb=\{\g_1,\ldots\g_s\}$ and $\g_i = \sum_{k=1}^r a_{i,k}
 \f_k$, where $a_{i,k}\in \Zx$, $i=1,\ldots,s$.\\
 6. Return $\A=\{g_1,\ldots,g_s\}$, where $g_i = \Y^{\g_i}-d_i$ and $d_i =
 \prod_{k=1}^r c_k^{a_{i,k}}$, $i=1,\ldots,s$.\\
\end{algorithm}
\begin{prop}\label{th-cs}
Algorithm  {\bf CharSet} is correct.
\end{prop}
\proof Steps 1-3 uses Proposition \ref{lm-bi2} to check whether $[F]=[1]$.
Note that $(\fb)$ and $(\gb)$ are the support
lattices of the binomial $\sigma$-ideals $[F]$ and $[G]$, respectively.
By Corollary \ref{lm-bi3}, $[F]=[G]$.
By Theorem \ref{th-t4}, $\A$ is a regular and coherent
$\sigma$-chain and hence a characteristic set of $[\A]$.\qedd

We now show  how to compute the reflexive closure for a
Laurent binomial $\sigma$-ideal.
\begin{algorithm}[H]\label{alg-ref}
  \caption{\bf --- Reflexive} \smallskip
  \Inp{$P$: a finite set of Laurent $\sigma$-binomials in $\F\{\Y^{\pm}\}$, where $\F$ is inversive.}\\
  \Outp{$\A$: a regular and coherent Laurent binomial $\sigma$-chain such that
    $[\A]$ is the reflexive closure of $[P]$.}

 \vskip3pt
 \noindent
 1. Let $G =$ {\bf{CharSet}}$(P)$. If $G=\emptyset$, return $1$.\\
 2. Let $G=\{g_1,\ldots,g_s\}$, $g_i = \Y^{\g_i}-d_i$, and $\gb=[\g_1,\ldots,\g_s]\in\Z[x]^{n\times s}$.\\
 3. $H=${\bf{XFactor}}$(\gb)$.\\
 4. If $H=\emptyset$, return $G$.\\
 5. Let $H= \{(\h_i,\e_i)\,|\, i=1,\ldots,r\}$ and
 $\e_i=(e_{i1},\ldots,e_{is})^\tau$.\\
 6. Let $P:=G\cup\{\Y^{\h_i}- \sigma^{-1}(\prod_{j=1}^s d_j^{e_{i,j}}),i=1,\ldots,r\}$,
                   and go to step 1.\\
\end{algorithm}
\begin{prop}\label{th-algref}
Algorithm  {\bf Reflexive} is correct.
\end{prop}
\proof The algorithm basically follows the proof of Theorem
\ref{lm-ld0}.  By  {\bf{CharSet}},
$[P]=[G]$ and $G$ is a regular and coherent $\sigma$-chain.
In step 4, if $H=\emptyset$, then $(\gb)$ is $x$-saturated,
and by Theorem \ref{th-pr1}, $[G]$ is reflexive and the theorem is proved.
Otherwise, we execute steps 5 and 6. Let $\I_1=[P]$, $L_1=(\fb)$,
$\I_2 = [G\cup\{\Y^{\h_i}- \sigma^{-k_i}(\prod_{j=1}^s
d_j^{b_{i,j}}),i=1,\ldots,r\}]$, and $L_2 =\L(\I_2)$.  Then, we have
$\I_1\varsubsetneq\I_2\subset \I_x$ and $L_1\varsubsetneq L_2\subset
L_x$, where $L_x=\sat_x(L_1)$ and $\I_x$ is the reflexive closure of
$\I_1$.
Similar to the proof of Theorem \ref{lm-ld0}, the algorithm will
terminate and output the reflexive closure of $[P]$.\qedd

\begin{rem}
Similar to Algorithm {\bf{Reflexive}}, we can give algorithms to check whether a Laurent binomial $\sigma$-ideal $\I$ is well-mixed, perfect, or prime,
and in the negative case to
compute the well-mixed or perfect closures of $\I$.
The details are omitted.
\end{rem}

We give a decomposition algorithm for perfect Laurent binomial
$\sigma$-ideals.
\begin{algorithm}[H]\label{alg-ldec}
  \caption{\bf --- DecLaurent} \smallskip
  \Inp{$P$: a finite set of Laurent $\sigma$-binomials in $\F\{\Y^{\pm}\}$,
  where $\F$ is inversive and algebraically closed.}\\
  \Outp{$\emptyset$, if $\{P\}=[1]$ or
  regular and coherent binomial
  $\sigma$-chains $\C_1,\ldots,\C_t$
  such that $[\C_i]$ are Laurent reflexive prime $\sigma$-ideals and
 $\{P\} = \cap_{i=1}^t [\C_i]$ is a minimal decomposition.}

 \medskip
 \noindent
 1. Let $F=${\bf{Reflexive}}$(P)$. If $F=1$, return $\emptyset$. \\
 2. Set $\RB=\emptyset$ and $\FB=\{F\}$.\\
 3. While $\FB\ne\emptyset$.\\
 \SPC 3.1. Let $F=\{\Y^{\f_1}-c_1,\ldots,\Y^{\f_r}-c_r\}\in\FB$,
  $\FB=\FB\setminus\{F\}$.\\
  \SPC 3.2. Let $G =$ {\bf{CharSet}}$(F)$. If $G=\emptyset$, goto step 3.\\
  \SPC 3.3. Let $G=\{g_1,\ldots,g_s\}$, $g_i = \Y^{\g_i}-d_i$, and $\gb=[\g_1,\ldots,\g_s]\in\Z[x]^{n\times s}$.\\
  \SPC 3.4. $H=${\bf{ZFACTOR}}$(\gb)$.\\
  \SPC 3.5. If $H=\emptyset$, add $G$ to $\RB$.\\
  \SPC 3.6. Let $H=\{(\h_i,k_i,\e_i)\,|\, i=1,\ldots,r\}$ and $\e_i=(e_{i1},\ldots,e_{is})^\tau$.\\
  \SPC 3.7. For $i=1,\ldots,t$,
  let $r_{i,1},\ldots,r_{i,k_i}$ be the $k_i$-th roots of  $\prod_{j=1}^s d_j^{e_{i,j}}$.\\
  \SPC 3.8. For $l_1=1,\ldots,k_1$,\ldots,$l_t=1,\ldots,k_t$,
            add $G\cup\{\Y^{\h_1}- r_{1,l_1},\ldots,\Y^{\h_t}- r_{t,l_t}\}$ to $\FB$.\\
 4.  Return $\RB$.
\end{algorithm}

\begin{exmp}\label{ex-dec20}
Let $P=\{g_1,g_2,g_3\}$,
where $g_1 = y_1^{-2}y_1^{x^k} - 1$, $g_2 = y_2^{-2}y_2^{x^k} - 1$, $g_3 =
y_1y_2^{-x}y_3^{2} - 1$, and $k\ge2$.
$[P]$ is already reflexive, so step 1 does nothing.
$P$ is already a regular and coherent chain, so $G=P$.
Let $\g_1 = [x^k-2,0,0]^\tau,\g_2 = [0,x^k-2,0]^\tau, \g_3 = [1,-x,2]^\tau$
be the supports of $g_1$, $g_2$, and $g_3$, respectively.
Then $[\g_1,\g_2,\g_3]$ is already a generalized Hermite normal form.

In Steps 3.4-3.6, $H=\{(\h_1,k_1,\e_1)\}$, where
$\h_1=[1,-x,x^k]^\tau,k_1=2,\e_1=[-1,x,x^k]^\tau$.
In Step 3.7, $r_{1,1}=1$ and $r_{1,2}=-1$.
In Step 3.8, $P_1=y_1y_2^{-x}y_3^{x^k}-1$ and $P_2=y_1y_2^{-x}y_3^{x^k}+1$
are added to $G$ to obtain $\C_1=\{g_1,g_2,g_3,P_1\}$ and $\C_2=\{g_1,g_2,g_3,P_2\}$.
Both $[\C_1]$ and $[\C_2]$ are reflexive and prime,
and are returned.

To see why the algorithm is correct, from $\e_1=[-1,x,x^k]$, we have
$(g_1+1)^{-1}(g_2+1)^x(g_3+1)^{x^k}= y_1^2y_2^{-2x}y_3^{2x^k}=1\, \mod[P]$.
Hence, $P_1P_2=y_1^2y_2^{-2x}y_3^{2x^k}-1\in[P]$.
\end{exmp}

\begin{prop}
Algorithm {\bf DecLaurent} is correct.
\end{prop}
\proof The algorithm basically follows the proof of Theorem
\ref{th-l2}. The proof is similar to that of Theorem
\ref{th-algref}.
By the proof of Theorem \ref{th-l2}, we obtain a  minimal decomposition.
\qedd

In the rest of this section, we give a decomposition algorithm for
binomial $\sigma$-ideals.
Before giving the main algorithm, we give a sub-algorithm {\bf
DecMono} which treats the $\sigma$-monomials. Basically, it gives
the following decomposition
$$ \V(\prod_{i=1}^n y_i^{f_i}) =
 \V(y_1)\cup\V(y_2/y_1)\cup\cdots\cup\V(y_n/\{y_1,\ldots,y_{n-1}\})$$
 where $0\ne f_i\in\N[x]$ and $\V(y_c/S)$ is the set of zeros of $y_c=0$ not
 vanishing any of the variables in $S$. The correctness of the
 algorithm comes directly from the above formula.
\begin{algorithm}[H]\label{alg-decmono}
  \caption{\bf --- DecMono} \smallskip
  \Inp{$(\Y_0,B,\Y_1)$: $\Y_0,\Y_1$ are disjoint subsets of $\Y$ and
   $B$ a finite set of $\sigma$-binomials or $\sigma$-monomials in $\F\{\Y\}$.}\\
  \Outp{$(\Y_{0i},B_i,\Y_{1i})$: $\Y_{0i},\Y_{1i}$ are disjoint subsets of $\Y$,
   $B_i$ consists of proper $\sigma$-binomials,
   and  $\V(\Y_0\cup B/\Y_1) = \cup_{i=1}^r \V(\Y_{0i}\cup B_i/\Y_{1i})$.}

 \medskip
  \noindent
 1. Set $\RB=\emptyset$ and $\FB=\{(\Y_0,B,\Y_1)\}$.\\
 2. While $\FB\ne\emptyset$.\\
 \SPC 2.1. Let $C=(\Y_0,B,\Y_1)\in\FB$, $\FB=\FB\setminus\{C\}$.\\
 \SPC 2.2. For all $y_c\in\Y_0$, let $B_1=B_{y_c=0}$ (replace $y_c^{x^k}$ by $0$) and delete $0$ from $B_1$.\\
 \SPC 2.3. If $B_1$ contains no $\sigma$-monomials, add $(\Y_0,B_1,\Y_1)$ to $\RB$ and goto step 2.\\
 \SPC 2.4. Let $M=\prod_{i=1}^k y_{c_i}^{f_i}\in B_1$, where $0\ne f_i\in\N[x]$.
           $B_1=B_1\setminus\{M\}$.\\
 \SPC 2.5. Let $\Y_2:=\{y_{c_1},\ldots,y_{c_k}\}\setminus\Y_1$.
 If $\Y_2=\emptyset$, go to step 2; else let $\Y_2=\{y_{t_1},\ldots,y_{t_s}\}$.\\
 \SPC 2.6. For $i=1,\ldots,s$,
    add $(\Y_0\cup\{y_{t_i}\},B_1,\Y_1\cup\{y_{t_1},\ldots,y_{t_{i-1}}\})$ to $\FB$.\\
  3. Return $\RB$.
\smallskip
\end{algorithm}

We now give the main algorithm. The algorithm basically follows the
proof of Theorem \ref{th-nl5}. The main modification is that instead
of the perfect ideal decomposition
$\{F\}= (\{F\}:\mb)\bigcap \cap_{i=1}^n \{F,y_i\},$
we use the following zero decomposition
$$\V(F)= \V(\{F\}:\mb)\bigcup \cup_{i=1}^n
  \V(F\cup\{y_i\}/\{y_1,\ldots,y_{i-1}\}).$$
The purpose of using the later decomposition is that many redundant
components can be easily removed by the following criterion
$\V(F/D)=\emptyset$ if $F\cap D\ne\emptyset$, which is done in step
2.5 of Algorithm {\bf DecMono}.
\begin{algorithm}[H]\label{alg-dec}
  \caption{\bf --- DecBinomial} \smallskip
  \Inp{$F$: a finite set of $\sigma$-binomials in $\F\{\Y\}$.}\\
  \Outp{$\emptyset$, if $\{F\}=[1]$
  or $(\C_1,\Y_1),\ldots,(\C_r,\Y_r)$,
  where $\Y_i\subset\Y$ and
  $\C_i$ are regular and coherent $\sigma$-chains containing $\sigma$-binomials or variables in
  $\Y\setminus\Y_i$
  such that
  $\sat(\C_i)$ are reflexive prime $\sigma$-ideals
 and
 $\{F\} = \cap_{i=1}^r \sat(\C_i)$.}

 \medskip
  \noindent
 1. Set $\RB=\emptyset$ and $\FB=${\bf DecMono}$(\emptyset,F,\emptyset)$.\\
 2. While $\FB\ne\emptyset$.\\
 \SPC 2.1. Let $C=(\Z_0,B,\Z_1)\in\FB$, $\FB=\FB\setminus\{C\}$.\\
 \SPC 2.2. If $B=\emptyset$, add $(\Z_0,\Z_1)$ to $\RB$.\\
 \SPC 2.3. Let $E=$ {\bf DecLaurent}$(B)$ in $\F\{\Z^{\pm}\}$, where $\Z=\Y\setminus\Z_0$ and $m=|\Z|$. \\
 \SPC 2.4. If $E=\emptyset$ goto step2.\\
 \SPC 2.5. Let $E=\{E_1,\ldots,E_l\}$ and $E_{l}=$
    $\{\Z^{\f_{l,1}}-c_{l,1},\ldots, \Z^{\f_{l,s_l}}-c_{l,s_l}\}$,
    where $\f_{l,k}\in\Zx^{m}$.\\
 \SPC 2.7. Add
         $(\{\Z_0,     \Z^{\f_{l,1}^+}-c_{l,1}\Z^{\f_{l,1}^-},
              \ldots, \Z^{\f_{l,s_l}^+}-c_{l,s_l}\Z^{\f_{l,s_l}^-}\},\Z_1)$
              to $\RB$, $l=1,\ldots,k$.\\
 \SPC 2.8. Let $\Z = \{y_{c_1},\ldots,y_{c_s}\}$. For
 $i=1,\ldots,s$, do \\
 \SPC\SPC $\FB=\FB\cup$ {\bf DecMono}$(\Z_0\cup\{y_{c_i}\},B,\Z_1\cup\{y_{c_1},\ldots,y_{c_{i-1}}\})$.\\
  3. Return $\RB$.
\smallskip
\end{algorithm}

\begin{exmp}\label{ex-dec21}
Let $\A=\{A_1,A_2,A_3\}$,
where $A_1 = y_1^{x^k} - y_1^2$, $A_2 = y_2^{x^k} - y_2^2$, $A_3 =
y_1y_3^{2} - y_2^x$, and $k\ge2$.
In Step 1, we have $\FB=\{(\emptyset, \A, \emptyset)\}$.
From Example \ref{ex-dec20}, in Step 2.3, we have
$E=\{\C_1,\C_2\}$, where $\C_1$ and $\C_2$ are given in Example \ref{ex-dec20}.
In Step 2.7, $(E_1,\emptyset)$ and  $(E_2,\emptyset)$ are added to $\RB$,
where  $E_1=\{A_1,A_2,A_3,Q_1\}$, $E_2=\{A_1,A_2,A_3,Q_2\}$,
and $Q_1=y_1y_3^{x^k}-y_2^{x}$, $Q_2=y_1y_3^{x^k}+y_2^{x}$.
In Step 2.8, $\Z=\{y_1,y_2,y_3\}$ and $(\{y_1,y_2\},\emptyset,\emptyset)$,
$(\{y_2,y_3),\{A_1\},\{y_1\})$ are added to $\FB$.
Finally, we have the following decomposition
$\{\A\}=[y_1,y_2]\cap[A_1,y_2,y_3]\cap\sat(\C_1)\cap\sat(\C_2)$, where all components are reflexive and prime.
%
%
\end{exmp}

\begin{rem}
Using the algorithm in \cite{gao-dcs}, the following decomposition is obtained:
$\{\A\}=[y_1,y_2]\cap\sat(\A)$. From Example \ref{ex-dec20}, $\sat(\A)$ is not prime.
Then, Algorithm {\bf DecBinomial} is stronger
than the general decomposition algorithm  given in \cite{gao-dcs}.
%
\end{rem}


\begin{prop}\label{th-decbi}
Algorithm {\bf DecBinomial} is correct.
\end{prop}
\proof
%
In step 1, $\sigma$-monomials in $F$ are treated.
In step 2, we will treat the components of $\FB$ one by one. In step
2.1, the component $(\Z_0,B,\Z_1)$ is taken from $\FB$.
In step 2.3, $\{B\}$ is decomposed as
 $\{B\}=\cap^{k}_{l=1}[E_l]$
 in $\F\{\Z^{\pm}\}$, where $E_l$ are regular and coherent $\sigma$-chains and $[E_l]$
 are reflexive prime ideals.
By \bref{eq-nbi1} and Corollary \ref{cor-ascr}, we have
 \begin{equation}\label{eq-ma01}
 \{B\}:\mb=\{B\}\cap\F\{\Z\}= \cap^{k}_{l=1} ([E_l]\F\{\Z^{\pm}\})\cap \F\{\Y\}
=\cap^{k}_{l=1}\sat(E^+_l),\end{equation}
 where   $E^+_l=\{\Z^{\f_{l,1}^+}-c_{l,1}\Z^{\f_{l,1}^-},
              \ldots, \Z^{\f_{l,s_l}^+}-c_{l,s_l}\Z^{\f_{l,s_l}^-}\}$,
              $l=1,\ldots,k$.
Since $E_{l}$ is regular and coherent, by Lemma \ref{lm-asc2},
$E^+_{l}$ is also regular and coherent.
Since $[E_l]$ is reflexive and prime, by Corollary \ref{cor-asc2},
$\sat(E^+_{l})$  is also reflexive and prime.
Note that $E=\emptyset$ in step 2.4 if and only if $[B]$ contains a $\sigma$-monomial.

Since $B\subset \F\{\Z\}$, we have the following decomposition
$$\V(\Z_0\cup B/\Z_1)= \V(\Z_0\cup(\{B\}:\mb)/\Z_1)
  \cup_{i=1}^s \V(\Z_0\cup B\cup\{y_{c_i}\}/\{y_{c_1},\ldots,y_{c_{i-1}}\}\cup\Z_1),$$
 where $\V(\Z_0\cup B\cup\{y_{c_i}\}/\{y_{c_1},\ldots,y_{c_{i-1}}\}\cup\Z_1)$ is
 further simplified with Algorithm {\bf DecMono} in step 2.8.
From \bref{eq-ma01},
 $$\V(\Z_0\cup\{B\}:\mb/\Z_1)=\cup^{k}_{l=1}\V(\sat(\Z_0,E^+_l)/\Z_1)
  =\cup^{k}_{l=1}\V([\Z_0,\sat(E^+_l)]/\Z_1),$$
where $\{\Z_0,E^+_l\}$ is a regular and coherent $\sigma$-chain
since $E^+_l$ does not contain variables in $\Z_0$. The above
formula explains why $(\{\Z_0,E^+_l\},\Z_1)$ is added to $\RB$ in
steps 2.5-2.7.

Let the algorithm returns $\RB=\{(\C_i,\Y_i);i=1,\ldots,m\}$. From
the above proof, we have $\V(F)=\cup^{k}_{l=1}\V(\sat(\C_i)/\Y_i)$.
Since $\Y_i\cap \C_i=\emptyset$ and $\sat(\C_i)$ is a reflexive and prime
$\sigma$-ideal, the Cohn closure of $\V(\sat(\C_i)/\Y_i)$ is
$\V(\sat(\C_i))$ and hence
 $$\V(F)=\cup^{k}_{l=1}\V(\sat(\C_i)/\Y_i)=\cup^{k}_{l=1}\V(\sat(\C_i)).$$
By the difference Hilbert  Nullstellensatz,
 $\{F\}=\cap^{k}_{l=1}\{\sat(\C_i)\} =\cap^{k}_{l=1}\sat(\C_i).$
The algorithm terminates, since after each execution of step 2, in
the new components $(\Y_{0l},B_l,\Y_{1l})$ added to $\FB$ in step
2.8, $B_l$ contains at least one less variables than $B$. \qedd

\section{Conclusion}
In this paper, we initiate the study of binomial $\sigma$-ideals.
Two basic tools used to study  binomial $\sigma$-ideals are the $\Zx$-lattice
and the characteristic set instead of the $\Z$-lattice and the
Gr\"obner basis used in the algebraic case \cite{es-bi}.
%
%

For Laurent binomial $\sigma$-ideals, two main results are proved.
Canonical representations for proper Laurent binomial
$\sigma$-ideals are given in terms of Gr\"obner basis of $\Zx$-lattices, regular and coherent $\sigma$-chains in $\F\{\Y^{\pm}\}$,
and partial characters over $\Zxn$, respectively. It is also shown that
a Laurent binomial $\sigma$-ideal is radical and dimensionally un-mixed.
We also give criteria for a
Laurent binomial $\sigma$-ideal to be reflexive, well-mixed, perfect, and prime
in terms of its support lattice.
It is shown that the reflexive, well-mixed,
and perfect closures of a Laurent binomial $\sigma$-ideal $\I$ is still
binomial whose support lattices are the $x$-, M-, and P-saturation of
the support lattice of $\I$ .
%

For binomial $\sigma$-ideals, we show that certain
properties of algebraic binomial ideals given in \cite{es-bi}
can be extended to the difference case using the theory
of Gr\"obner basis in the case of infinitely many variables.
It is shown that most properties of Laurent binomial
$\sigma$-ideal can be extended to the normal binomial $\sigma$-ideals.

Algorithms are given for the key results of the paper.
We give algorithms to check whether a given Laurent binomial difference ideal
$\I$ is reflexive, prime, well-mixed, or perfect, and in the
negative case, to compute the reflexive, well-mixed, and perfect
closures of $\I$. An algorithm is given to decompose a finitely
generated perfect binomial difference ideal as the intersection of
reflexive prime binomial difference ideals.

Finally, we make a remark about differential binomial ideals.
The study of binomial differential ideals is more difficult,
because the differentiation of a binomial is generally not a binomial.
Differential toric varieties were defined in \cite{d-sres} and were used to connect the differential Chow form \cite{gao} and differential sparse resultant \cite{d-sres}.
But, contrary to the difference case, the defining ideal for a differential toric variety is generally not binomial.

\end{document}